\documentclass[fleqn,leqno, titlepage]{article}
\usepackage{hypertlabook}

\pdftitle{Auxiliary Variables in TLA+}  
\title{\bf Auxiliary Variables in \tlaplus}  
\file{main}
\author{Leslie Lamport and Stephan Merz}
\date{27 May 2017} 

 {\vspace{-.25em}\mbox{}\hfill\rule{\textwidth}{.6pt}\hfill\mbox{}\end{figure}}

\newcommand{\Sp}{\ensuremath{\mathcal{S}}}
\newcommand{\T}{\ensuremath{\mathcal{T}}}

\makeatletter

\def\rangeref#1#2{page\@ifundefined{r@#1}{s {\bf ?--?}\@warning
   {Reference `#1' on page \thepage \space 
    undefined}}{\@ifundefined{r@#2}{s {\bf ?--?}\@warning
   {Reference `#2' on page \thepage \space 
    undefined}}{\edef\@tempa{\@nameuse{r@#1}}\expandafter
    \@tempcnta\expandafter\@cdr\@tempa\@nil\relax
    \edef\@tempa{\@nameuse{r@#2}}\expandafter
    \@tempcntb\expandafter\@cdr\@tempa\@nil\relax
    \ifnum\@tempcnta=\@tempcntb \textilde\pageref{#1}\else 
      s \pageref{#1}--\pageref{#2}\fi}}}
\makeatother

\newtheorem{definitionx}{Definition}

\newtheorem{theoremx}{Theorem}
\newenvironment{theorem}[1]{\begin{theoremx}[#1]{\rm}}{\end{theoremx}}

\begin{document}

\maketitle

\begin{abstract}
  Auxiliary variables are often needed for verifying that an implementation is
  correct with respect to a higher-level specification. They augment the formal
  description of the implementation without changing its semantics---that is, 
  the
  set of behaviors that it describes. This paper explains rules for adding
  history, prophecy, and stuttering variables to \tlaplus\ specifications,
  ensuring that the augmented specification is equivalent to the original one.
  The rules are explained with toy examples, and they are used to verify
  the correctness of a simplified version of a snapshot algorithm due to 
  Afek et al.
\end{abstract}

\tableofcontents

\newpage

\section{Introduction}

With state-based methods, checking that an implementation satisfies a
higher-level specification requires describing how the higher-level
concepts in the specification are represented by the lower-level data
structures of the implementation.  This approach was first proposed in
the domain of sequential systems by Hoare in 1972~\cite{hoare:proof}.
Hoare called the description an \emph{abstraction function}.  The
generalization to concurrent systems was called a \emph{refinement
mapping} by Abadi and Lamport~\cite{abadi:existence}.  They observed
that constructing a refinement mapping may require adding auxiliary
variables to the implementation---variables that do not alter the
behavior of the actual variables and need not be implemented.

This paper is about adding auxiliary variables to \tlaplus\
specifications.  The ideas we present should be applicable to other
state-based specification methods, but we make no attempt to translate
them into those other methods.  We hope that a future paper will
present the basic ideas in a language-independent way and will contain
soundness and completeness proofs.  Our goal here is to teach
engineers writing \tlaplus\ specifications how to add auxiliary
variables when they need them.  

We assume the reader can understand \tlaplus\
specifications.  A basic understanding of refinement mappings will be
helpful but isn't necessary.  \tlaplus\ and refinement mappings are
explained in the book \emph{Specifying
Systems}~\cite{lamport:tla-book} and in material listed on the TLA web
page~\cite{lamport:tla-webpage}.

This is a long paper, in part because it contains 25 figures
with actual \tlaplus\ specifications.  
The paper contains hyperlinks, and we recommend
reading the pdf version on line.
If you are doing that, you can download the source files for all
the \tlaplus\ specifications described in this paper by
  \hyperref{http://research.microsoft.com/en-us/um/people/lamport/tla/auxiliary/auxiliary.html}{}{}{clicking here}.
Otherwise, you can find the URL in the reference 
 list~\cite{lamport:auxiliary-variables-web}.
We expect that engineers will have to study the specifications
carefully to learn how to add auxiliary variables to their
specifications.  

We explain three kinds of auxiliary variables: history, prophecy, and
stuttering variables.  History variables record information about the
system's past behavior.  They have been used since at least the
1970s~\cite{owicki:verifying}.  They were sometimes called ``ghost''
variables.  Prophecy variables predict the future behavior of the
system.  They were introduced by Abadi and Lamport in
1991~\cite{abadi:existence}.  The need for them was also implicit in
an example presented in Herlihy and Wing's classic paper on
linearizability~\cite{herlihy:axioms}.  We found the original prophecy
variables very difficult to use in practice.  The prophecy variables
described here are new, and our experience with them so far indicates
that they are reasonably easy to use in practice.  Stuttering
variables add ``stuttering'' steps---ones that leave the
specification's actual variables unchanged.  Abadi and Lamport
originally used prophecy variables to add stuttering steps, but we
have found it better to introduce stuttering steps with a separate
kind of variable.

We will mostly ignore liveness and consider only safety
specifications.  The canonical form of a \tlaplus\ specification
consists of a safety specification of the form 
 \tlabox{Init /\ [][Next]_{vars}} 
conjoined with a liveness condition.  An auxiliary variable is added
by modifying the safety specification, but leaving the liveness
condition unchanged.  Liveness therefore poses no problem for auxiliary
variables and is discussed only briefly.

\section{Refinement Mappings}

We will illustrate refinement mappings with a simple, useless example.
A user presents a server with a sequence of integer inputs.  The
server responds to each input value $i$ with one of the following
outputs: $Hi$ if $i$ is the largest number input so far, $Lo$ if it's
the smallest number input so far, $Both$ if it's both, and $None$ if
it's neither.  We declare $Hi$, $Lo$, $Both$, and $None$ in a
\textsc{constants} statement.  They are assumed not to be integers.

\subsection{Specification \emph{MinMax}1}

Our first specification appears in a module named $MinMax1$.  It
describes the interaction of the user and the server with two
variables: a variable $x$ to hold an input or a response, and a
variable $turn$ that indicates whether it's the user's turn to input a
value or the server's turn to respond.  The specification also uses a
variable $y$ to hold the set of values input so far.  The initial
predicate is
\begin{display}
\begin{notla}
Init ==  /\ x = None
         /\ turn = "input" 
         /\ y = {}
\end{notla}
\begin{tlatex}
\@x{ Init \.{\defeq}\@s{4.1} \.{\land} x \.{=}None}%
\@x{\@s{39.80} \.{\land} turn \.{=}\@w{input}}%
\@x{\@s{39.80} \.{\land} y \.{=} \{ \}}%
\end{tlatex}
\end{display}
The next-state relation $Next$ equals $InputNum \/ Respond$ where 
$InputNum$ is the user's input action and $Respond$ is the server's output
action.  The definition of $InputNum$ is simple:
\begin{display}
\begin{notla}
InputNum ==  /\ turn = "input"
             /\ turn' = "output"
             /\ x' \in Int
             /\ y' = y
\end{notla}
\begin{tlatex}
\@x{ InputNum \.{\defeq}\@s{4.1} \.{\land} turn \.{=}\@w{input}}%
\@x{\@s{68.01} \.{\land} turn \.{'} \.{=}\@w{output}}%
\@x{\@s{68.01} \.{\land} x \.{'} \.{\in} Int}%
\@x{\@s{68.01} \.{\land} y \.{'}\@s{0.10} \.{=} y}%
\end{tlatex}
\end{display}
To define the $Respond$ action, we must first define operators
$setMax$ and $setMin$ so that, for any finite nonempty set $S$ of
integers, $setMax(S)$ and $setMin(S)$ are the maximum and minimum
element, respectively, of $S$.  The definitions are:
\begin{display}
\begin{notla}
setMax(S)  ==  CHOOSE t \in S : \A s \in S : t >= s

setMin(S)  ==  CHOOSE t \in S : \A s \in S : t =< s
\end{notla}
\begin{tlatex}
 \@x{ setMax ( S )\@s{4.1} \.{\defeq}\@s{4.1} {\CHOOSE} t \.{\in} S \.{:} \A\,
 s \.{\in} S \.{:} t \.{\geq} s}%
\@pvspace{4.0pt}%
 \@x{ setMin ( S )\@s{5.59} \.{\defeq}\@s{4.09} {\CHOOSE} t \.{\in} S \.{:}
 \A\, s \.{\in} S \.{:} t \.{\leq} s}%
\end{tlatex}
\end{display}
The definition of $Respond$ is:
\begin{display}
\begin{notla}
Respond == /\ turn = "output"
           /\ turn' = "input"
           /\ y' = y \cup {x}
           /\ x' = IF x = setMax(y') 
                     THEN IF x = setMin(y') THEN Both ELSE Hi  
                     ELSE IF x = setMin(y') THEN Lo   ELSE None
\end{notla}
\begin{tlatex}
\@x{ Respond \.{\defeq} \.{\land} turn \.{=}\@w{output}}%
\@x{\@s{55.83} \.{\land} turn \.{'} \.{=}\@w{input}}%
\@x{\@s{55.83} \.{\land} y \.{'}\@s{0.10} \.{=} y \.{\cup} \{ x \}}%
\@x{\@s{55.83} \.{\land} x \.{'} \.{=} {\IF} x \.{=} setMax ( y \.{'} )}%
 \@x{\@s{97.13} \.{\THEN} {\IF} x \.{=} setMin ( y \.{'} ) \.{\THEN} Both
 \.{\ELSE} Hi}%
 \@x{\@s{97.13} \.{\ELSE} {\IF} x \.{=} setMin ( y \.{'} ) \.{\THEN}
 Lo\@s{9.84} \.{\ELSE} None}%
\end{tlatex}
\end{display}
Note that action $InputNum$ is enabled iff $turn$ equals $"input"$,
and action $Respond$ is enabled iff $turn$ equals $"output"$.
The complete specification is the formula
 \[ Spec == Init /\ [][Next]_{vars}\]
where $vars$ is the tuple $<<x, turn, y>>$ of variables.  The module
$MinMax1$ we have written thus far is shown in
\lref{targ:MinMax1}{Figure~\ref{fig:MinMax1}}.

\begin{figure} \target{targ:MinMax1}
\begin{notla}
----------------------------- MODULE MinMax1 -----------------------------
EXTENDS Integers

setMax(S) == CHOOSE t \in S : \A s \in S : t >= s
setMin(S) == CHOOSE t \in S : \A s \in S : t =< s

CONSTANTS Lo, Hi, Both, None
ASSUME {Lo, Hi, Both, None} \cap Int = { }

VARIABLES x, turn, y
vars == <<x, turn, y>>

Init ==  /\ x = None
         /\ turn = "input" 
         /\ y = {}

InputNum ==  /\ turn = "input"
             /\ turn' = "output"
             /\ x' \in Int
             /\ y' = y

Respond == /\ turn = "output"
           /\ turn' = "input"
           /\ y' = y \cup {x}
           /\ x' = IF x = setMax(y') 
                     THEN IF x = setMin(y') THEN Both ELSE Hi  
                     ELSE IF x = setMin(y') THEN Lo   ELSE None
          
Next == InputNum \/ Respond

Spec == Init /\ [][Next]_vars
=============================================================================
\end{notla}
\begin{tlatex}
\@x{}\moduleLeftDash\@xx{ {\MODULE} MinMax1}\moduleRightDash\@xx{}%
\@x{ {\EXTENDS} Integers}%
\@pvspace{8.0pt}%
 \@x{ setMax ( S ) \.{\defeq} {\CHOOSE} t \.{\in} S \.{:} \A\, s \.{\in} S
 \.{:} t \.{\geq} s}%
 \@x{ setMin ( S )\@s{1.49} \.{\defeq} {\CHOOSE} t \.{\in} S \.{:} \A\, s
 \.{\in} S \.{:} t \.{\leq} s}%
\@pvspace{8.0pt}%
\@x{ {\CONSTANTS} Lo ,\, Hi ,\, Both ,\, None}%
\@x{ {\ASSUME} \{ Lo ,\, Hi ,\, Both ,\, None \} \.{\cap} Int \.{=} \{ \}}%
\@pvspace{8.0pt}%
\@x{ {\VARIABLES} x ,\, turn ,\, y}%
\@x{ vars \.{\defeq} {\langle} x ,\, turn ,\, y {\rangle}}%
\@pvspace{8.0pt}%
\@x{ Init\@s{2.02} \.{\defeq}\@s{4.1} \.{\land} x \.{=} None}%
\@x{\@s{41.82} \.{\land} turn \.{=}\@w{input}}%
\@x{\@s{41.82} \.{\land} y \.{=} \{ \}}%
\@pvspace{8.0pt}%
\@x{ InputNum \.{\defeq}\@s{4.1} \.{\land} turn \.{=}\@w{input}}%
\@x{\@s{68.01} \.{\land} turn \.{'} \.{=}\@w{output}}%
\@x{\@s{68.01} \.{\land} x \.{'} \.{\in} Int}%
\@x{\@s{68.01} \.{\land} y \.{'}\@s{0.10} \.{=} y}%
\@pvspace{8.0pt}%
\@x{ Respond \.{\defeq} \.{\land} turn\@s{1.52} \.{=}\@w{output}}%
\@x{\@s{55.83} \.{\land} turn \.{'} \.{=}\@w{input}}%
\@x{\@s{55.83} \.{\land} y \.{'}\@s{0.10} \.{=} y \.{\cup} \{ x \}}%
\@x{\@s{55.83} \.{\land} x \.{'} \.{=} {\IF} x \.{=} setMax ( y \.{'} )}%
 \@x{\@s{97.13} \.{\THEN} {\IF} x \.{=} setMin ( y \.{'} ) \.{\THEN} Both
 \.{\ELSE} Hi}%
 \@x{\@s{97.13} \.{\ELSE} {\IF} x \.{=} setMin ( y \.{'} ) \.{\THEN}
 Lo\@s{9.84} \.{\ELSE} None}%
\@pvspace{8.0pt}%
\@x{ Next \.{\defeq} InputNum \.{\lor} Respond}%
\@pvspace{8.0pt}%
\@x{ Spec\@s{1.46} \.{\defeq} Init \.{\land} {\Box} [ Next ]_{ vars}}%
\@x{}\bottombar\@xx{}%
\end{tlatex}
\caption{Module \emph{MinMax}1.} \label{fig:MinMax1}
\end{figure}

\subsection{The Hiding Operator \protect\EE} 

Recall that a behavior is a sequence of states, where a state is an
assignment of values to all possible variables.  For specification
$Spec$ of module $MinMax1$, the interesting part of the state is the
assignment of values to $x$, $turn$, and $y$.  Our specification
allows all other variables to have any value at any state of any
behavior.

The purpose of this specification is to describe the interaction of
the user and the server.  This interaction is described by the values
of $x$ and $turn$.  The value of $y$ is needed only to describe how
the values of $x$ and $turn$ can change.  We consider $x$ and $turn$
to be the externally visible or observable values of the specification and $y$
to be an internal variable.  A philosophically correct specification of
our user/server system would allow only behaviors in which the values
of $x$ and $turn$ are as specified by $Spec$, but would not constrain
the value of $y$.  We can write such a specification in terms of the
temporal-logic operator~$\EE$.

For any temporal formula $F$ and variable $v$, the formula 
 \tlabox{\EE v : F}
is defined approximately as follows.  A behavior $\sigma$ satisfies
 \tlabox{\EE v : F}
iff there exists a behavior $\tau$ satisfying $F$ such that $\tau$ is
identical to $\sigma$ except for the values its states assign to $v$.
The precise definition is more complicated because a temporal formula
of \tlaplus\ may neither require nor prohibit stuttering steps, but we
will use the approximate definition for now.  The operator $\EE$
is much like the ordinary existential quantifier $\E$ except that
 \tlabox{\EE v : F}
asserts the existence not of a single value for $v$ that makes $F$
true but rather of a sequence of values, one for each state in the
behavior, that makes $F$ true on the behavior.  This temporal existential
quantifier $\EE$ satisfies most of the properties of ordinary quantification.
For example, if the variable $v$ does not occur in formula $F$, then
 \tlabox{\EE v : F} is equivalent to $F$.  We sometimes read the formula
\tlabox{\EE v : F} as ``$F$ with $v$ hidden''.

The philosophically correct specification of the $MinMax1$ system
should consist of formula $Spec$ with $y$ hidden.  The obvious way to
write this specification is \tlabox{\EE y : Spec}.  However, we can't
do that for the following reason.  Suppose a module $M$ defines $exp$
to equal some expression.  \tlaplus\ does not allow the expression
\begin{equation} \label{eq1}
  \{v \in exp : v^{2}>42\}
\end{equation}
to appear at any point in module $M$ where $v$ is already declared or
defined.  Since $exp$ must be defined for the expression to have a
meaning, this means that (\ref{eq1}) is illegal if $v$ is a declared
variable that appears in the definition of $exp$.  Similarly, the formula
\tlabox{\EE y : Spec} is illegal because $y$ appears in the definition
of $Spec$.\footnote{There
 are languages for writing math precisely that allow expression
(\ref{eq1}) even if $v$ is already declared.  In such a language,
\tlabox{\EE y : Spec} would be equivalent to \tlabox{\EE w : Spec} for
any identifier $w$, which means it would be equivalent to $Spec$.}

There are ways to write the formula $Spec$ with $v$ hidden in
\tlaplus.  The most convenient ones involve writing it in another
module that instantiates module $MinMax1$.  Chapter~4 of
\emph{Specifying Systems}~\cite{lamport:tla-book} explains one way to
do this.  However, there's little reason to do it since the \tlaplus\
tools cannot check specifications written with $\EE$.  (The TLAPS
proof system may eventually be able to reason about it.)  Instead, we
take the formula \tlabox{\EE y : Spec} to be an abbreviation for the
formula \tlabox{\EE y : \Mmap{Spec}}, where \Mmap{Spec} is the formula
obtained from $Spec$ by expanding all definitions.  Formula
\Mmap{Spec} contains only: \tlaplus\ primitives; the constants $Hi$,
$Lo$, $Both$, and $None$; and the variables $x$, $turn$, and $y$.
Thus \tlabox{\EE y : Spec} is meaningful in a context in which $x$ and
$turn$ are declared variables.  If used in a context in which $y$
already has a meaning, we interpret \tlabox{\EE y : Spec} to be the
formula obtained from \tlabox{\EE y : \Mmap{Spec}} by replacing $y$
everywhere with a new symbol.

What 
 \label{pageMmap}%
it means to expand all definitions in an expression is not
as simple as it might seem.  Consider the following definition:
\begin{equation}
NotUnique(a) == \E i : i # a \NOTLA\label{eq:subs-1}
\end{equation}
It's clear that the following theorem is true:
\begin{equation}
\THEOREM \tlabox{\A a : NotUnique(a)} \NOTLA\label{eq:subs-2}
\end{equation}
Now suppose we follow the definition of $NotUnique$ with:
\begin{equation}
  \begin{noj}
    \CONSTANT i \V{.2}
    \THEOREM NotUnique(i)
    \end{noj}
\NOTLA\label{eq:subs-3}
\end{equation}
Theorem (\ref{eq:subs-2})
obviously implies the theorem of (\ref{eq:subs-3}).  However, a naive
expansion of the definition of $NotUnique$ tells us that 
\Mmap{NotUnique(i)} equals 
 \tlabox{\E i : i # i},
which equals \FALSE. The problem is clear: the bound identifier $i$ in
the definition of $NotUnique$ is not the same $i$ as the one declared
in the \textsc{constant} declaration.  The following definition of
$NotUnique$ is equivalent to (\ref{eq:subs-1})
  \[ NotUnique(a) == \E jku : jku # a
  \]
and with the naive expansion of this definition, 
\Mmap{NotUnique(i)} equals the true formula
  \tlabox{\E jku : jku # i}
of (\ref{eq:subs-3}).  

The easiest way to define the meaning of expanding all definitions in
an expression is to consider (\ref{eq:subs-1}) to define
$NotUnique(a)$ to equal something like
 \tlabox{\E v\_743 : v\_743 # a},
where $v\_743$ is an identifier that cannot be used anywhere else.  In
general, every bound identifier in a definition is replaced by some
unique identifier.

Recursive definitions are not a problem for complete expansion of
definitions because in \tlaplus, a recursive definition is just an
abbreviation for a non-recursive one.  For example
 \[ f[i \in Nat] == \IF {i=0}\THEN 1 \LSE i*f[i-1]
 \]    
is an abbreviation for
 \[ f == \CHOOSE f : f = [i\in Nat |-> \IF {i=0}\THEN 1 \LSE i*f[i-1]]
 \]    
so the bound identifier $f$ to the right of the ``$\deq$'' is not the
same symbol as the $f$ being defined.  (A recursive operator
definition is an abbreviation for a much more complicated ordinary
definition.)\label{pageMmapx}


\subsection{Specification \emph{MinMax}2}

The specification of our system in module $MinMax1$ uses the variable
$y$ to remember the set of all values that the user has input.  Module
$MinMax2$ specifies the same user/server interaction that
remembers only the smallest and largest values input so far, using the
variables $min$ and $max$.  Representing the initial state, before any
values have been input, is a little tricky.  It would be simpler if
the standard $Integers$ module defined a value $\infty$ such that
$-\infty < i < \infty$ for all integers $i$.  So, we will write the
spec pretending that it did.  Afterwards, we'll describe how to obtain
an actual \tlaplus\ spec.

The initial predicate of the specification is:
\begin{display}
\begin{notla}
Init  ==  /\ x = None
          /\ turn = "input" 
          /\ min = Infinity 
          /\ max = MinusInfinity
\end{notla}
\begin{tlatex}
\@x{ Init\@s{4.1} \.{\defeq}\@s{4.1} \.{\land} x \.{=}None}%
\@x{\@s{43.90} \.{\land} turn \.{=}\@w{input}}%
\@x{\@s{43.90} \.{\land} min\@s{1.49} \.{=} \infty}%
\@x{\@s{43.90} \.{\land} max \.{=} -\infty}%
\end{tlatex}
\end{display}
The user's $InputNum$ action is the same as for the $MinMax1$ specification,
except it leaves $min$ and $max$ rather than $y$ unchanged:
\begin{display}
\begin{notla}
InputNum  ==  /\ turn = "input"
              /\ turn' = "output"
              /\ x' \in Int
              /\ UNCHANGED <<min, max>>
\end{notla}
\begin{tlatex}
\@x{ InputNum\@s{4.1} \.{\defeq}\@s{4.1} \.{\land} turn \.{=}\@w{input}}%
\@x{\@s{72.11} \.{\land} turn \.{'} \.{=}\@w{output}}%
\@x{\@s{72.11} \.{\land} x \.{'} \.{\in} Int}%
\@x{\@s{72.11} \.{\land} {\UNCHANGED} {\langle} min ,\, max {\rangle}}%
\end{tlatex}
\end{display}
Here is the system's $Respond$ action:
\begin{display}
\begin{notla}
Respond  ==  /\ turn = "output"
             /\ turn' = "input"
             /\ min' = IF x =< min THEN x ELSE min 
             /\ max' = IF x >= max THEN x ELSE max
             /\ x' = IF x = max' 
                       THEN IF x = min' THEN Both ELSE Hi  
                       ELSE IF x = min' THEN Lo   ELSE None
\end{notla}
\begin{tlatex}
\@x{ Respond\@s{4.1} \.{\defeq}\@s{4.1} \.{\land} turn \.{=}\@w{output}}%
\@x{\@s{64.03} \.{\land} turn \.{'} \.{=}\@w{input}}%
 \@x{\@s{64.03} \.{\land} min \.{'}\@s{1.49} \.{=} {\IF} x \.{\leq}
 min\@s{1.49} \.{\THEN} x \.{\ELSE} min}%
 \@x{\@s{64.03} \.{\land} max \.{'} \.{=} {\IF} x \.{\geq} max \.{\THEN} x
 \.{\ELSE} max}%
\@x{\@s{64.03} \.{\land} x \.{'} \.{=} {\IF} x \.{=} max \.{'}}%
 \@x{\@s{105.33} \.{\THEN} {\IF} x \.{=} min \.{'} \.{\THEN} Both \.{\ELSE}
 Hi}%
 \@x{\@s{105.33} \.{\ELSE} {\IF} x \.{=} min \.{'} \.{\THEN} Lo\@s{9.84}
 \.{\ELSE} None}%
\end{tlatex}
\end{display}
As usual, the complete specification is 
 \[ Spec == Init \;/\ \; [][Next]_{vars}\]
where this time $vars$ is the tuple $<<x, turn, min, max>>$ of variables.

To turn this into a \tlaplus\ specification, we replace $\infty$ and
$-\infty$ by two constants $Infinity$ and $MinusInfinity$.  In the definition
of $Respond$, we replace $x\leq min$ and $x\geq max$ by
$IsLeq(x, min)$ and $IsGeq(x, max)$, where $IsLeq$ and $IsGeq$ are defined
by
\begin{display}
\begin{notla}
IsLeq(i, j)  ==  (j = Infinity) \/ (i =< j)

IsGeq(i, j)  ==  (j = MinusInfinity) \/ (i >= j) 
\end{notla}
\begin{tlatex}
 \@x{ IsLeq ( i ,\, j )\@s{6.07} \.{\defeq}\@s{4.1} ( j \.{=} Infinity )
 \.{\lor} ( i \.{\leq} j )}%
\@pvspace{4.0pt}%
 \@x{ IsGeq ( i ,\, j )\@s{4.09} \.{\defeq}\@s{4.10} ( j \.{=} MinusInfinity )
 \.{\lor} ( i \.{\geq} j )}%
\end{tlatex}
\end{display}
These definitions must be preceded by declarations or definitions of
$Infinity$ and $MinusInfinity$.  They can equal any values, except that
they must not be equal and neither of them should be an
integer---otherwise, the spec wouldn't mean what we want it to mean.
We could declare $Infinity$ and $MinusInfinity$ in a \textsc{constant}
declaration and then add an \textsc{assume} statement asserting that
they aren't in $Int$.  However, we prefer to define them like this:
\begin{display}
\begin{notla}
Infinity       ==  CHOOSE n : n \notin Int

MinusInfinity  ==  CHOOSE n : n \notin (Int \cup {Infinity})
\end{notla}
\begin{tlatex}
 \@x{ Infinity\@s{31.21} \.{\defeq}\@s{4.1} {\CHOOSE} n \.{:} n \.{\notin}
 Int}%
\@pvspace{4.0pt}%
 \@x{ MinusInfinity\@s{4.10} \.{\defeq}\@s{4.10} {\CHOOSE} n \.{:} n
 \.{\notin} ( Int \.{\cup} \{ Infinity \} )}%
\end{tlatex}
\end{display}
This completes module $MinMax2$, which is shown in its entirety
in \lref{targ:MinMax2}{Figure~\ref{fig:MinMax2}}.
\begin{figure} \target{targ:MinMax2}
\begin{notla}
----------------------------- MODULE MinMax2 -----------------------------
EXTENDS Integers, Sequences

CONSTANTS Lo, Hi, Both, None
ASSUME {Lo, Hi, Both, None} \cap Int = { }

Infinity      == CHOOSE n : n \notin Int
MinusInfinity == CHOOSE n : n \notin (Int \cup {Infinity})

IsLeq(i, j) == (j = Infinity) \/ (i =< j)
IsGeq(i, j) == (j = MinusInfinity) \/ (i >= j) 

VARIABLES x, turn, min, max
vars == <<x, turn, min, max>>

Init ==  /\ x = None
         /\ turn = "input" 
         /\ min = Infinity 
         /\ max = MinusInfinity

InputNum ==  /\ turn = "input"
             /\ turn' = "output"
             /\ x' \in Int
             /\ UNCHANGED <<min, max>>

Respond  ==  /\ turn = "output"
             /\ turn' = "input"
             /\ min' = IF IsLeq(x, min) THEN x ELSE min 
             /\ max' = IF IsGeq(x, max) THEN x ELSE max
             /\ x' = IF x = max' THEN IF x = min' THEN Both ELSE Hi  
                                 ELSE IF x = min' THEN Lo   ELSE None
          
Next == InputNum \/ Respond

Spec == Init /\ [][Next]_vars
=============================================================================
\end{notla}
\begin{tlatex}
\@x{}\moduleLeftDash\@xx{ {\MODULE} MinMax2}\moduleRightDash\@xx{}%
\@x{ {\EXTENDS} Integers ,\, Sequences}%
\@pvspace{8.0pt}%
\@x{ {\CONSTANTS} Lo ,\, Hi ,\, Both ,\, None}%
\@x{ {\ASSUME} \{ Lo ,\, Hi ,\, Both ,\, None \} \.{\cap} Int \.{=} \{ \}}%
\@pvspace{8.0pt}%
\@x{ Infinity\@s{27.11} \.{\defeq} {\CHOOSE} n \.{:} n \.{\notin} Int}%
 \@x{ MinusInfinity \.{\defeq} {\CHOOSE} n \.{:} n \.{\notin} ( Int \.{\cup}
 \{ Infinity \} )}%
\@pvspace{8.0pt}%
 \@x{ IsLeq ( i ,\, j )\@s{1.97} \.{\defeq} ( j \.{=} Infinity ) \.{\lor} ( i
 \.{\leq} j )}%
 \@x{ IsGeq ( i ,\, j ) \.{\defeq} ( j \.{=} MinusInfinity ) \.{\lor} ( i
 \.{\geq} j )}%
\@pvspace{8.0pt}%
\@x{ {\VARIABLES} x ,\, turn ,\, min ,\, max}%
\@x{ vars \.{\defeq} {\langle} x ,\, turn ,\, min ,\, max {\rangle}}%
\@pvspace{8.0pt}%
\@x{ Init\@s{2.02} \.{\defeq}\@s{4.1} \.{\land} x \.{=} None}%
\@x{\@s{41.82} \.{\land} turn \.{=}\@w{input}}%
\@x{\@s{41.82} \.{\land} min\@s{1.49} \.{=} Infinity}%
\@x{\@s{41.82} \.{\land} max \.{=} MinusInfinity}%
\@pvspace{8.0pt}%
\@x{ InputNum \.{\defeq}\@s{4.1} \.{\land} turn \.{=}\@w{input}}%
\@x{\@s{68.01} \.{\land} turn \.{'} \.{=}\@w{output}}%
\@x{\@s{68.01} \.{\land} x \.{'} \.{\in} Int}%
\@x{\@s{68.01} \.{\land} {\UNCHANGED} {\langle} min ,\, max {\rangle}}%
\@pvspace{8.0pt}%
\@x{ Respond\@s{8.07} \.{\defeq}\@s{4.1} \.{\land} turn \.{=}\@w{output}}%
\@x{\@s{68.01} \.{\land} turn \.{'} \.{=}\@w{input}}%
 \@x{\@s{68.01} \.{\land} min \.{'}\@s{1.49} \.{=} {\IF} IsLeq ( x ,\, min
 )\@s{3.47} \.{\THEN} x \.{\ELSE} min}%
 \@x{\@s{68.01} \.{\land} max \.{'} \.{=} {\IF} IsGeq ( x ,\, max ) \.{\THEN}
 x \.{\ELSE} max}%
 \@x{\@s{68.01} \.{\land} x \.{'} \.{=} {\IF} x \.{=} max \.{'} \.{\THEN}
 {\IF} x \.{=} min \.{'} \.{\THEN} Both \.{\ELSE} Hi}%
 \@x{\@s{154.37} \.{\ELSE} {\IF} x \.{=} min \.{'} \.{\THEN} Lo\@s{9.84}
 \.{\ELSE} None}%
\@pvspace{8.0pt}%
\@x{ Next \.{\defeq} InputNum \.{\lor} Respond}%
\@pvspace{8.0pt}%
\@x{ Spec\@s{1.46} \.{\defeq} Init \.{\land} {\Box} [ Next ]_{ vars}}%
\@x{}\bottombar\@xx{}%
\end{tlatex}
\caption{Module \emph{MinMax}2.} \label{fig:MinMax2}

\end{figure}

As before, we consider $x$ and $turn$ to be the externally visible
variables, and $min$ and $max$ to be internal variables.  The
philosophically correct specification, which hides the internal
variables $min$ and $max$, is \tlabox{\EE min, max : Spec}.  Of
course, this is an abbreviation for 
 \tlabox{\EE min : (\EE max : Spec}), 
which is equivalent to
 \tlabox{\EE max : (\EE min : Spec}).

\subsection{The Relation Between the Two Specifications} \label{sec:relation}

Using the standard \tlaplus\ naming convention, we have given the two
specifications the same name $Spec$.  To distinguish them, let
$Spec_{1}$ be the specification $Spec$ of module $MinMax1$ and
$Spec_{2}$ be the $Spec$ of module $MinMax2$.

It should be clear that both specifications describe the same behavior
of the external variables $x$ and $turn$.  This means that if we hide
the internal variable $y$ of $Spec_{1}$ and the internal variables $min$
and $max$ of $Spec_{2}$, we should obtain equivalent specifications.
More precisely, we expect to this to be true:
\begin{equation} \label{eqT1}
  (\EE y : Spec_{1}) \ \equiv \ (\EE min, max : Spec_{2}) 
\end{equation}
This formula is equivalent to the conjunction of these two formulas
\begin{eqnarray}
(\EE y : Spec_{1}) \ => \ (\EE min, max : Spec_{2}) \label{eqT1a} \V{.2}
(\EE min, max : Spec_{2}) \ => \ (\EE y : Spec_{1}) \label{eqT1b}   
\end{eqnarray}
We verify (\ref{eqT1}) by separately verifying
(\ref{eqT1a}) and (\ref{eqT1b}).  We first consider (\ref{eqT1a}).

Formula (\ref{eqT1a}) asserts of a behavior $\sigma$ that if there
exists some way of assigning values to $y$ in the states of $\sigma$
to make it satisfy $Spec_{1}$, then $\sigma$ satisfies 
   \tlabox{\EE min, max : Spec_{2}}.
Since the variable $y$ does not appear in \tlabox{\EE min, max : Spec_{2}},
changing the values of $y$ in the states of $\sigma$ doesn't affect
whether it satisfies that formula.  This implies that to verify
(\ref{eqT1a}), it suffices to show that any behavior $\sigma$ that
satisfies $Spec_{1}$ also satisfies \tlabox{\EE min, max : Spec_{2}}.
In other words, to verify (\ref{eqT1a}), it suffices to verify
\begin{equation} \label{eqT1abis}
Spec_{1}  \ => \ (\EE min, max : Spec_{2})
\end{equation}
To verify (\ref{eqT1abis}), we must show that for any behavior
$\sigma$ that satisfies $Spec_{1}$, there exists a way of assigning
values to the variables $min$ and $max$ in the states of $\sigma$ that
makes the resulting behavior satisfy $Spec_{2}$.  A standard way of
doing that is to find explicit expressions \ov{min} and \ov{max} such
that, if in each state of a behavior we assign to the variables
$min$ and $max$ the values of \ov{min} and \ov{max} in that
state, then the resulting behavior satisfies $Spec_{2}$.  We do this by
showing that any behavior satisfying $Spec_{1}$ satisfies the formula
obtained by substituting \ov{min} for $min$ and \ov{max} for $max$ in
$Spec_{2}$.  Let's write that formula
 $\ov{\Mmap{Spec_{2}}}$, 
emphasizing that we must expand all definitions in $Spec_{2}$ before
substituting \ov{min} for $min$ and \ov{max} for $max$.  So,
we verify (\ref{eqT1abis}) by verifying
\begin{equation} \label{eqT1abis2}
Spec_{1} \ => \ \ov{\Mmap{Spec_{2}}}
\end{equation}
We can write formula $\ov{\Mmap{Spec_{2}}}$ (or more precisely, a
formula equivalent to it) in module $MinMax1$ as
follows.  We first add the statement
 \[ M \ == \ \INSTANCE MinMax2 \WITH min <- \ov{min}, \ max <- \ov{max}\]
For every defined symbol $def$ in module $MinMax2$, this statement
defines $M!def$ to be equivalent to $\ov{\Mmap{def}}$, the formula
whose definition is obtained by substituting \ov{min} for $min$ and
\ov{max} for $max$ in the formula obtained by expanding all
definitions in the definition of $def$ in $MinMax2$.%
  \footnote{Note that the declared constants $Hi$, $Lo$, $Both$,
  and $None$ of module $MinMax2$ have been implicitly instantiated
  by the constants of the same name declared in $MinMax1$.}
This \textsc{instance} statement therefore defines $M!Spec$ to
be equivalent to $\ov{\Mmap{Spec_{2}}}$, allowing us to write 
(\ref{eqT1abis2}) in module $MinMax1$ as the theorem:
 \[ \THEOREM Spec \ => \ M!Spec
 \]
We can write a \tlaplus\ proof of this theorem and check it with the
TLAPS theorem prover.  We can also have TLC check this theorem by
creating a model for module $MinMax1$ with specification $Spec$ that
substitutes a finite set of integers for $Int$ and checks the property
$M!Spec$.  But before we can do that, we have to determine what the
expressions \ov{min} and \ov{max} are in the \textsc{instance}
statement.  

Formula (\ref{eqT1abis2}) asserts that in a behavior $\sigma$
satisfying $Spec_{1}$, if in each state of $\sigma$ we assign to $min$
and $max$ the values of \ov{min} and \ov{max} in that state, then the
resulting behavior satisfies $Spec_{2}$.  One way of thinking about
this is that in a behavior satisfying $Spec_{1}$, the values of
\ov{min} and \ov{max} simulate the values that $Spec_{2}$ requires
$min$ and $max$ to assume.

A little thought reveals that \ov{min} and \ov{max} should be defined
as indicated in this statement:
\begin{display}
\begin{notla}
M == INSTANCE MinMax2 
       WITH min <- IF y = {} THEN Infinity      ELSE setMin(y),
            max <- IF y = {} THEN MinusInfinity ELSE setMax(y)
\end{notla}
\begin{tlatex}
\@x{ M \.{\defeq} {\INSTANCE} MinMax2}%
 \@x{\@s{37.69} {\WITH} min\@s{1.49} \.{\leftarrow} {\IF} y \.{=} \{ \}
 \.{\THEN} Infinity\@s{27.11} \.{\ELSE} setMin ( y ) ,\,}%
 \@x{\@s{68.67} max \.{\leftarrow} {\IF} y \.{=} \{ \} \.{\THEN} MinusInfinity
 \.{\ELSE} setMax ( y )}%
\end{tlatex}
\end{display}
Of course, we need to define $Infinity$ and $MinusInfinity$ before we
can write that statement.  They should be defined to be the same
values as in $MinMax2$, so we just copy the definitions from that
module into module $MinMax1$.  We have added the statements in
\lref{targ:MinMax1a}{Figure~\ref{fig:MinMax1a}} to the bottom of the module in
\lref{targ:MinMax1}{Figure~\ref{fig:MinMax1}}
\begin{figure} \target{targ:MinMax1a}
\vspace*{1em}
\begin{notla}
Infinity      == CHOOSE n : n \notin Int
MinusInfinity == CHOOSE n : n \notin (Int \cup {Infinity})

M == INSTANCE MinMax2 
        WITH min <- IF y = {} THEN Infinity      ELSE setMin(y),
             max <- IF y = {} THEN MinusInfinity ELSE setMax(y)
\end{notla}
\begin{tlatex}
\@x{ Infinity\@s{27.11} \.{\defeq} {\CHOOSE} n \.{:} n \.{\notin} Int}%
 \@x{ MinusInfinity \.{\defeq} {\CHOOSE} n \.{:} n \.{\notin} ( Int \.{\cup}
 \{ Infinity \} )}%
\@pvspace{8.0pt}%
\@x{ M \.{\defeq} {\INSTANCE} MinMax2}%
 \@x{\@s{41.79} {\WITH} min\@s{1.49} \.{\leftarrow} {\IF} y \.{=} \{ \}
 \.{\THEN} Infinity\@s{27.11} \.{\ELSE} setMin ( y ) ,\,}%
 \@x{\@s{72.77} max \.{\leftarrow} {\IF} y \.{=} \{ \} \.{\THEN} MinusInfinity
 \.{\ELSE} setMax ( y )}%
\end{tlatex}
\caption{Additions to module \emph{MinMax}1.} \label{fig:MinMax1a}
\end{figure}


\subsection{Refinement In General}

In general, we have two specs: $Spec_{1}$ with variables 
  $x_{1},\ldots,x_{m}, y_{1},\ldots,y_{n}$,
and $Spec_{2}$ with variables
  $x_{1},\ldots,x_{m}, z_{1},\ldots,z_{p}$.  For compactness
let \textbf{x} denote $x_{1},\ldots,x_{m}$, let \textbf{y} denote
$y_{1},\ldots,y_{n}$ and let \textbf{z} denote $z_{1},\ldots,z_{p}$.
We consider \textbf{x} to be the externally visible variables of both
specifications, and we consider \textbf{y} and \textbf{z} to be internal
variables.  

The specifications with their internal variables hidden are written
 \tlabox{\EE \textbf{y}:Spec_{1}} and \tlabox{\EE \textbf{z}:Spec_{2}}.
To verify that 
 \tlabox{\EE \textbf{y}:Spec_{1}} implements \tlabox{\EE \textbf{z}:Spec_{2}},
we must show that for each behavior satisfying $Spec_{1}$, there is
some way to assign values of the variables \textbf{z} in each state so
that the resulting behavior satisfies $Spec_{2}$.  We do that by
explicitly specifying those values of \textbf{z} in terms of the values
of \textbf{x} and \textbf{y}.  More precisely,
for each $z_{i}$ we define an expression \ov{z_{i}} in terms of the
variables \textbf{x} and \textbf{y} and show that $Spec_{1}$
implements $\ov{\Mmap{Spec_{2}}}$, the specification obtained by
expanding all definitions in $Spec_{2}$ and substituting
$z_{1}<-\ov{z_{1}},\,\ldots,\,z_{p}<-\ov{z_{p}}$ in the resulting
formula.  This substitution is
called a \emph{refinement mapping}; and if  $Spec_{1}$
implements $\ov{\Mmap{Spec_{2}}}$, then we say that 
$Spec_{1}$ implements $Spec_{2}$ under the refinement mapping.

The assertion that $Spec_{1}$ implements $Spec_{2}$ under the
refinement mapping 
 $z_{1}<-\ov{z_{1}},\,\ldots,\,z_{p}<-\ov{z_{p}}$ can be expressed
in \tlaplus\ as follows.  Suppose $Spec_{1}$ is formula $Spec1$ in a
module named $Mod1$, and $Spec_{2}$ is formula $Spec2$ in a module
named $Mod2$.  ($Spec1$ and $Spec2$ can be the same
identifier.)  For some identifier $Id$, we add the following statement
to $Mod1$:\footnote{If $Mod2$ has declared \textsc{constants},
 then the statement must
also specify expressions to be substituted for those constants.  A
substitution of the form $id \leftarrow id$ for an identifier $id$ can be
omitted from the \textsc{with} clause.}
 \[ Id == \INSTANCE Mod2\ \WITH \ 
    z_{1}<-\ov{z_{1}},\,\ldots,\,z_{p}<-\ov{z_{p}} \]
The assertion that $Spec_{1}$ implements $Spec_{2}$ under the
refinement mapping can then be expressed in module $Mod1$ by the
following theorem:
 \[ \THEOREM \ Spec1 => Id!Spec2
 \]
This theorem asserts that in any behavior satisfying $Spec_{1}$, the
values of the expressions $\ov{z_{i}}$ are values that $Spec_{2}$
permits the variables $z_{i}$ to have.  
The shape of the theorem makes it explicit that in \tlaplus, implementation is
implication.
The correctness of the theorem
can be checked (but seldom completely verified) with TLC for module
$Mod1$ having $Spec1$ as the specification and $Id!Spec2$ as the
temporal property to be checked.

As we will see, it is sometimes the case that \tlabox{\EE
\textbf{y}:Spec_{1}} implements \tlabox{\EE \textbf{z}:Spec_{2}} but
there does not exist a refinement mapping under which $Spec_{1}$
implements $Spec_{2}$.  In that case, it is almost always possible to
construct the necessary refinement mapping by adding auxiliary
variables to $Spec_{1}$.  Adding auxiliary variables \textbf{a} to the
specification $Spec_{1}$ means finding a specification
$Spec_{1}^{\mathbf{a}}$ such that \tlabox{\EE \mathbf{a}:Spec_{1}^{\mathbf{a}}}
is equivalent to $Spec_{1}$.  Showing that 
 \tlabox{\EE \mathbf{y},\mathbf{a}:Spec_{1}^{\mathbf{a}}} 
implements \tlabox{\EE z:Spec_{2}} shows that 
 \tlabox{\EE \textbf{y}:Spec_{1}} implements
\tlabox{\EE \textbf{z}:Spec_{2}}, since 
  \tlabox{\EE \mathbf{y},\mathbf{a}:Spec_{1}^{\mathbf{a}}}
equals
  \tlabox{\EE \mathbf{y} : \EE a:Spec_{1}^{\mathbf{a}}}
which is equivalent to \tlabox{\EE \textbf{y}:Spec_{1}}.  Even though
we can't define the expressions \ov{z_{i}} in terms of \textbf{x} and
\textbf{y}, we may be able to define them in terms of \textbf{x},
\textbf{y}, and \textbf{a}.

We will define three kinds of auxiliary variables: history variables
that remember what happened in the past, prophecy variables that
predict what will happen in the future, and stuttering variables
that add stuttering steps (ones that don't change \textbf{x} and \textbf{y}).

%
%

\section{History Variables} \label{sec:history}

\subsection{Equivalence of \emph{MinMax}1 and \emph{MinMax}2}

Let us return to the notation of Section~\ref{sec:relation}, so
$Spec_{1}$ is specification $Spec$ of module $MinMax1$ and $Spec_{2}$
is specification $Spec$ of module $MinMax2$.  We observed that
\tlabox{\EE y : Spec_{1}} and \tlabox{\EE min, max : Spec_{2}} are
equivalent, meaning that each implements (implies) the other.  We
found a refinement mapping under which $Spec_{1}$ implements
$Spec_{2}$.  To prove the converse implication, we want to find a
refinement mapping under which $Spec_{2}$ implements $Spec_{1}$.  This
means defining an expression \ov{y} in terms of the variables $x$,
$turn$, $min$, and $max$ such that the values of $x$, $turn$, and $\ov{y}$ in
any behavior allowed by $Spec_{1}$ are values of $x$, $turn$, and $y$
allowed by $Spec_{2}$.

In a behavior of $Spec_{1}$, the value of $y$ is the set of all values
input by the user.  However, in a behavior of $Spec_{2}$, the
variables $min$ and $max$ record only the smallest and largest input
values.  There is no way to reconstruct the set of all values input
from the variables of $MinMax2$.  So, there is no refinement mapping
under which $Spec_{2}$ implements $Spec_{1}$.  To solve this problem,
we write another spec $Spec_{2}^{h}$ that is the same as $Spec_{2}$,
except that it also constrains the behavior of another variable $h$.
More precisely, if we hide $h$ in $Spec_{2}^{h}$, then we get a
specification that's equivalent to $Spec_{2}$.  Expressed
mathematically, this means \tlabox{\EE h:Spec_{2}^{h}} is equivalent
to $Spec_{2}$.  

The initial predicate and next-state action of $Spec_{2}^{h}$ are the
same as those of $Spec_{2}$, except they also describe the values that
$h$ may assume.  In particular, the value of $h$ records information
about previous values of the variable $x$, but does not affect the
current or future values of $x$ or any of the other variables $turn$,
$min$, and $max$ of $Spec_{2}$.  Thus \tlabox{\EE h:Spec_{2}^{h}} is
equivalent to $Spec_{2}$.  We call $h$ a \emph{history} variable.

We write $Spec_{2}^{h}$ as follows in a \tlaplus\ module $MinMax2H$.
The module begins with the statement
 \[ \EXTENDS MinMax2 \]
that simply imports all the declarations and definitions from
$MinMax$, defining $Spec$ to be the specification we are calling
$Spec_{2}$.  The module declares the variable $h$ and defines the
initial predicate $InitH$ of $Spec_{2}^{h}$ by
 \[ InitH == Init \, /\ \, (h = \{\,\})\]
The next-state action $NextH$ is defined to equal 
  $InputNumH \/ RespondH$
where $InputNumH$ and $RespondH$ are defined as follows:
\begin{display}
\begin{notla}
InputNumH == /\ InputNum
             /\ h' = h

RespondH  == /\ Respond
             /\ h' = h \cup {x}
\end{notla}
\begin{tlatex}
\@x{ InputNumH \.{\defeq} \.{\land} InputNum}%
\@x{\@s{72.21} \.{\land} h \.{'} \.{=} h}%
\@pvspace{8.0pt}%
\@x{ RespondH\@s{8.33} \.{\defeq} \.{\land} Respond}%
\@x{\@s{72.21} \.{\land} h \.{'} \.{=} h \.{\cup} \{ x \}}%
\end{tlatex}
\end{display}
The specification $Spec_{2}^{h}$ is the following formula defined
in the module:
\begin{display}
\begin{notla}
SpecH  ==  InitH  /\  [][NextH]_varsH
\end{notla}
\begin{tlatex}
 \@x{ SpecH\@s{4.1} \.{\defeq}\@s{4.1} InitH\@s{4.1} \.{\land}\@s{4.1} {\Box}
 [ NextH ]_{varsH}}%
\end{tlatex}
\end{display}
where $varsH$ equals $<<vars, h>>$.  (Because $vars$ equals
 $<<x, turn, min, max>>$, we can also
 define $varsH$  to equal $<<x, turn, min, max, h>>$; the two
 definitions give equivalent expressions $\UNCHANGED varsH$.)

It's easy to see that this specification asserts that $h$ is always
equal to the set of all values that the user has input thus far, which
is exactly what $Spec_{1}$ asserts about $y$.  Therefore, 
$Spec_{2}^{h}$ implements $Spec_{1}$ under the refinement mapping
$y <- h$---that is, with $\ov{y}$ equal to the expression $h$.  We
express this in module $MinMax2H$ by
\begin{display}
\begin{notla}
M  ==  INSTANCE MinMax1  WITH  y <- h 

THEOREM  SpecH => M!Spec
\end{notla}
\begin{tlatex}
 \@x{ M\@s{4.1} \.{\defeq}\@s{4.1} {\INSTANCE} MinMax1\@s{4.1} {\WITH}\@s{4.1}
 y \.{\leftarrow} h}%
\@pvspace{4.0pt}%
\@x{ {\THEOREM}\@s{4.1} SpecH \.{\implies} M {\bang} Spec}%
\end{tlatex}
\end{display}
The complete module $MinMax2H$ is in \lref{targ:MinMax2H}{Figure~\ref{fig:MinMax2H}}
\begin{figure} \target{targ:MinMax2H}
\begin{notla}
---------------------------- MODULE MinMax2H ----------------------------
EXTENDS MinMax2

VARIABLE h
varsH == <<vars, h>>

InitH == Init /\ (h = {})

InputNumH ==  /\ InputNum
              /\ h' = h

RespondH  == /\ Respond
             /\ h' = h \cup {x}
          
NextH == InputNumH \/ RespondH

SpecH == InitH /\ [][NextH]_varsH

M == INSTANCE MinMax1 WITH y <- h 

THEOREM SpecH => M!Spec
=============================================================================
\end{notla}
\begin{tlatex}
\@x{}\moduleLeftDash\@xx{ {\MODULE} MinMax2H}\moduleRightDash\@xx{}%
\@x{ {\EXTENDS} MinMax2}%
\@pvspace{8.0pt}%
\@x{ {\VARIABLE} h}%
\@x{ varsH \.{\defeq} {\langle} vars ,\, h {\rangle}}%
\@pvspace{8.0pt}%
\@x{ InitH\@s{2.14} \.{\defeq} Init\@s{9.42} \.{\land} ( h \.{=} \{ \} )}%
\@pvspace{8.0pt}%
\@x{ InputNumH \.{\defeq} \.{\land} InputNum}%
\@x{\@s{72.21} \.{\land} h \.{'} \.{=} h}%
\@pvspace{8.0pt}%
\@x{ RespondH\@s{8.33} \.{\defeq} \.{\land} Respond}%
\@x{\@s{72.21} \.{\land} h \.{'} \.{=} h \.{\cup} \{ x \}}%
\@pvspace{8.0pt}%
\@x{ NextH \.{\defeq} InputNumH \.{\lor} RespondH}%
\@pvspace{8.0pt}%
\@x{ SpecH\@s{1.08} \.{\defeq} InitH \.{\land} {\Box} [ NextH ]_{ varsH}}%
\@pvspace{8.0pt}%
\@x{ M \.{\defeq} {\INSTANCE} MinMax1 {\WITH} y \.{\leftarrow} h}%
\@pvspace{8.0pt}%
\@x{ {\THEOREM} SpecH \.{\implies} M {\bang} Spec}%
\@x{}\bottombar\@xx{}%
\end{tlatex}
\caption{Module \emph{MinMax2H}.} \label{fig:MinMax2H}
\end{figure}

\subsection{Disjunctive Representation}

The generalization from the $MinMax$ example is intuitively clear.  We
add a history variable $h$ to a specification by conjoining
$h=exp_{Init}$ to its initial predicate and $(h'=exp_{A})$ to each
subaction $A$ of its next-state action, where expression $exp_{Init}$
can contain the spec's variables and each expression $exp_{A}$ can
contain the spec's variables, both primed and unprimed, and $h$
(unprimed).  
To make this precise, we have to state exactly what a \emph{subaction}
is.  

In general, there may be many different ways to define the
subactions of a next-state action.  In defining $SpecH$ in module
$MinMax2H$, we took $InputNum$ and $Respond$ to be the subactions of
the next-state action $Next$ of $MinMax2$.  However, we can also
consider $Next$ itself to be a subaction of $Next$.  We can do this
and get an equivalent specification $SpecH$ by defining $NextH$ as
follows:
\begin{display}
\begin{notla}
NextH  ==  /\ Next
           /\ \/ (turn = "input")  /\ (h '= h)
              \/ (turn = "output") /\ (h' = h \cup {x})
\end{notla}
\begin{tlatex}
\@x{ NextH\@s{4.1} \.{\defeq}\@s{4.1} \.{\land} Next}%
 \@x{\@s{56.15} \.{\land} \.{\lor} ( turn \.{=}\@w{input} )\@s{6.22} \.{\land}
 ( h \.{'} \.{=} h )}%
 \@x{\@s{67.26} \.{\lor} ( turn \.{=}\@w{output} ) \.{\land} ( h \.{'} \.{=} h
 \.{\cup} \{ x \} )}%
\end{tlatex}
\end{display}
The two specifications are equivalent because the two definitions of
$NextH$ are equivalent.  Their equivalence is asserted by adding the
following theorem to module $MinMax2H$.  The TLAPS proof system easily
checks its \textsc{by} proof.
\begin{display}
\begin{notla}
THEOREM NextH = /\ InputNum \/ Respond
                /\ \/ (turn = "input")  /\ (h '= h)
                   \/ (turn = "output") /\ (h' = h \cup {x})
BY DEF NextH, Next, InputNumH, RespondH, InputNum, Respond
\end{notla}
\begin{tlatex}
\@x{ {\THEOREM} NextH \.{=} \.{\land} InputNum \.{\lor} Respond}%
 \@x{\@s{89.22} \.{\land} \.{\lor} ( turn \.{=}\@w{input} )\@s{6.22} \.{\land}
 ( h \.{'} \.{=} h )}%
 \@x{\@s{100.33} \.{\lor} ( turn \.{=}\@w{output} ) \.{\land} ( h \.{'} \.{=}
 h \.{\cup} \{ x \} )}%
 \@x{ {\BY} {\DEF} NextH ,\, Next ,\, InputNumH ,\, RespondH ,\, InputNum ,\,
 Respond}%
\end{tlatex}
\end{display}
To define what a subaction is, we introduce the concept of a
\emph{disjunctive representation}.  A disjunctive representation of a
formula $N$ is a way of writing $N$ in terms of subactions
 $A_{1}$, \ldots, $A_{m}$ using only the operators $\/ $ and 
 \tlabox{\E k \in K}, 
for some identifiers $k$ and expressions $K$.  For example,
consider the formula: 
\begin{equation} {\NOTLA\label{eq:dis-rep}}
 B \; \/ \; C \; \/\; D \; \/ \; (\begin{noj}
                                     \E i \in S, j \in T : \\ \s{1}
     (\E q \in U : E) \; \/ \; (\E r \in W : {F}))
                                     \end{noj}
\end{equation}
where $B$, $C$, $D$, $E$, and $F$ can be any formulas.
Here is one of the 36 possible disjunctive representations of formula
(\ref{eq:dis-rep}), where each boxed formula is a 
subaction:
\[  \fbox{$B$} \; \/ \; \fbox{$C \; \/\; D$} \; \/ \; 
      (\begin{noj}
       \E i \in S, j \in T : \\ \s{1}
     \fbox{$(\E q \in U : E)$} \; \/ \; (\E r \in W : \fbox{$F$}\,))
       \end{noj}
\]
In other words, this disjunctive representation of (\ref{eq:dis-rep})
has the four subactions $B$, $C \/ D$,  
 \tlabox{\E q \in U : E}, and $F$.  

Each subaction of a disjunction representation has a \emph{context},
which is a pair $<<\textbf{k};\textbf{K}>>$, where $\mathbf{k}$ is an
$n$-tuple of identifiers and $\textbf{K}$ is an $n$-tuple of
expressions, for some $n$.  The contexts of the subactions in the
disjunctive representation of (\ref{eq:dis-rep}) defined above are:
 \[ \begin{noj2}
    \mbox{\underline{subaction}}\;\; & \mbox{\underline{context}} \V{.2}
     B & << >> \\
     C \lor D & << >> \\
     \E q \in U : E \s{1}& <<i,j;\,S,T>> \\
     F & <<i,j,r;\,S,T,W>>
    \end{noj2}
 \]
We let \mbox{$\E \langle i,j;\,S,T\rangle$} be an abbreviation for
\mbox{$\E i\in S,\,j\in T$} and similarly for
\mbox{$\A \langle i,j;\,S,T\rangle$}. 

The generalization of this example should be clear.
The tuple of identifiers in the context of a subaction $A$ are all the
bound identifiers of existential quantifiers within whose scope $A$
lies.\footnote{Everything we do extends easily to handle unbounded
quantification if we pretend that the unbounded quantifiers $\E v$ and
$\A v$ are written $\E v\in\Omega$ and $\A v\in\Omega$, and we define
$e\in\Omega$ to equal $\TRUE$ for every expression $e$.  Since
unbounded quantification seldom occurs in specifications, we will not
discuss this further.} If $<<\textbf{k};\textbf{K}>>$ is the empty
context $<<;>>$, we let \tlabox{\E <<\textbf{k};\textbf{K}>>: F} and
\tlabox{\A <<\textbf{k};\textbf{K}>>: F} equal $F$.

We can now define precisely what it means to add a history variable to
a specification The definition is contained in the hypothesis of this
theorem:
%
%
%
%
\begin{theorem}{History Variable}
Let $Spec$ equal \tlabox{Init \,/\ \,[][Next]_{vars}} and let
$Spec^{h}$ equal
 \tlabox{Init^{h}\, /\ \,[][Next^{h}]_{vars^{h}}},
where:
\begin{itemize}
\item $Init^{h}$ equals $Init /\ (h=exp_{Init})$, for some expression
 $exp_{Init}$ that may contain the specification's
(unprimed) variables.

\item $Next^{h}$ is obtained from $Next$ by replacing each subaction
$A$ of a disjunctive representation of $Next$ with 
  \tlabox{A /\ (h'=exp_{A})}, 
for some expression $exp_{A}$ that may contain primed and unprimed
specification variables, identifiers in the context of $A$, and
constant parameters.
\item $vars^{h}$ equals $<<vars, h>>$
\end{itemize}
Then $Spec$ is equivalent to \tlabox{\EE h : Spec^{h}}.
\end{theorem}
The hypotheses of this theorem are purely syntactic ones: conditions
on the definitions of $Init^{h}$, $Next^{h}$, and $vars^{h}$ plus
conditions on what variables and identifiers may appear in
$exp_{Init}$ and $exp_{A}$.   By a variable or identifier appearing
in an expression $exp$, we mean that it appears in the expression \Mmap{exp}
obtained by expanding all definitions if $exp$.
(See the discussion of definition expansion on page~\pageref{pageMmap}.)


\subsection{Equivalence of Next-State Actions}

When adding an auxiliary variable, it is often useful to rewrite a
specification $Spec$---that is, to replace $Spec$ with a different but
equivalent formula.  This is most often done by rewriting the
next-state action $Next$, which is done by rewriting one or more of
the subactions in a disjunctive representation of $Next$.  We now consider
when we can replace a subaction $A$ in a disjunctive representation
of $Next$ by the subaction $B$.  

We can obviously replace $A$ by $B$ if $A$ and $B$ are equivalent
formulas.  However, this is often too stringent a requirement.  For
example, the two actions
  \[ \begin{noj}
     (x' = \IF{y\geq 0} \THEN x+y \ELSE x - y) \ /\ \ (y'=y) \V{.2}
     (x' = \IF{y> 0} \THEN x+y \ELSE x - y) \ /\ \ (y'=y) 
     \end{noj}
  \]
are not equivalent.  However, they are equivalent if $y$ is a number.
Thus, we can replace one by the other in the next-state action
if $y\in Int$ is an invariant of the specification.  The generalization
of this observation 
is:
\begin{theorem}{Subaction Equivalence} \label{subactEquiv1}
\s{1}Let $A$ be a subaction with context 
  $<<\mathbf{k}; \mathbf{K}>>$
in a disjunctive representation of the next-state action of a
specification $Spec$ with tuple $vars$ of variables, let $Inv$ be an
invariant of $Spec$, and let $B$ be an action satisfying:
\begin{equation} \label{eqEqv1}
 Inv \; => \; 
         \tlabox{\A <<\mathbf{k}; \mathbf{K}>> : A \equiv B}
\end{equation}
Then $Spec$ is equivalent to the specification obtained by replacing
$A$ with $B$ in the next-state action's disjunctive representation.
\end{theorem}
Formula (\ref{eqEqv1}) is an action formula, so it can be proved with
TLAPS but cannot be checked with TLC\@.  

TLC can check directly that two specifications are equivalent by
checking that each specification implies the other.  To check that
specification $Spec$ implies a specification $SpecB$, just run TLC
with a model having $Spec$ as the behavioral specification and $SpecB$
as the property to be checked.  If one spec is obtained by a simple
modification of the other, it should suffice to use small models.  But
in that case, it should not be hard to prove (\ref{eqEqv1})
with TLAPS, where $Inv$ is a simple type invariant.

\subsection{Discussion of History Variables}

As our example, we showed that specifications $MinMax1$ and $MinMax2$ are
equivalent. In practice, we rarely care about checking equivalence of
specifications. We almost always want to show that a specification {\Sp}
satisfies some property $\mathcal{P}$, which means that {\Sp} implies
$\mathcal{P}$. 
For example, $\Sp => []Inv$ asserts that $Inv$ is an invariant of \Sp.
In (\ref{eqT1a}) and (\ref{eqT1b}), the property $\mathcal{P}$ is, like \Sp, a
complete system specification.

 
The most general form of correctness that TLC can check is that one
specification implies another.  For example, the assertion that $Inv$
is an invariant of a specification {\Sp} with tuple $vars$ of
variables is equivalent to the assertion that {\Sp} implies the
specification
 \[
   Inv \ /\ \ [][Inv'\equiv Inv]_{vars}
 \]
We often want to show that a specification {\Sp} implies a
higher-level, more abstract specification {\T}.  The standard way of
doing this is to find a refinement mapping that expresses the values
of {\T}'s variables as functions of the values of {\Sp}'s variables.
This can't be done if specification {\T} remembers in its state
information that is forgotten by {\Sp}.  In that case, we show that
{\Sp} implies \T\ by adding a history variable $h$ to \Sp\ to obtain
the specification $\Sp^{h}$, and we find a refinement mapping to show
$\Sp^{h}$ implies \T. Since \tlabox{\EE h:\Sp^{h}} is equivalent to
\Sp, this shows that \Sp\ implies \T.

One can argue that \T\ is not a good higher-level specification if it
keeps information about the past that doesn't have to be kept by its
implementation \Sp.  However, sometimes that information about the
past can simplify the higher-level specification.  We may also add a
history variable to a specification \Sp\ so we can state the property
we want to show that it satisfies, even if we aren't explicitly
constructing a refinement mapping.  For example the property that \Sp\
requires one kind of action to occur before another can be expressed
as an invariant if we add a history variable that remembers when
actions have occurred.  Because it's a history variable, it doesn't
have to be implemented in the system being specified---that is,
in the actual hardware or software.  Only the variables of \Sp\
must be implemented.

\subsection{Liveness} \label{sec:history-liveness}

A natural liveness requirement for our $MinMax$ specs is that every
input should produce an output.  This requirement is added to both
specifications by conjoining the fairness requirement
$\WF_{vars}(Respond)$ to the formula $Spec$.  (It is just a
peculiarity of our example that the fairness requirements are exactly
the same in both specifications.)  Let us call the resulting
specifications $LSpec$.

The two specifications are still equivalent when the internal
variables are hidden.  Formula (\ref{eqT1}), and hence formulas
(\ref{eqT1a}) and (\ref{eqT1b}), remain true if we replace $Spec_{1}$
and $Spec_{2}$ by $LSpec_{1}$ and $LSpec_{2}$, respectively---where
$LSpec_{1}$ and $LSpec_{2}$ are formulas $LSpec$ of $MinMax1$ and
$MinMax2$, respectively.  We verify (\ref{eqT1a}) the same as before
by verifying the theorem \tlabox{LSpec => M!LSpec} of module
$MinMax1$.
To verify (\ref{eqT1b}), we need to add a history variable $h$ to
$LSpec_{2}$ rather than to $Spec_{2}$.  We now drop the subscripts;
all the formulas we write will be ones defined in $MinMax2$ or
$MinMax2H$.  

To add a history variable to the specification $LSpec$,
we add a history variable to its safety part and then conjoin the
liveness part of $LSpec$.  The resulting specification is defined in
module $MinMax2H$ by
 \[ HLSpec == HSpec /\ \WF_{vars}(Respond) \]
The equivalence of $LSpec$ and \tlabox{\EE h:HLSpec} follows from
the equivalence of $Spec$ and \tlabox{\EE h:HSpec} by this argument:
\begin{display}

  \pflongnumbers 
  \afterPfSpace{.5em}
  \beforePfSpace{.2em}
\begin{proof}
\step{1}{$Spec /\ \WF_{vars}(Respond) \ \equiv \ 
 (\tlabox{\EE h : HSpec}) /\ \WF_{vars}(Respond) $}
   \begin{proof}
   \pf\ Because $Spec$ is equivalent to \tlabox{\EE h : HSpec}.
   \end{proof}
\step{2}{$\begin{noj}
          (\tlabox{\EE h : HSpec}) \ /\ \ \WF_{vars}(Respond)\V{.2}\s{2}
   \equiv \ \ \EE h : (HSpec /\ \WF_{vars}(Respond))
          \end{noj}$}
   \begin{proof}
   \pf\ For any formulas $F$ and $G$, if $h$ does not occur in $G$,
   then \tlabox{(\EE h:F) /\ G}  is equivalent to \tlabox{\EE h: (F /\ G)}.
   \end{proof}  
\qedstep
   \begin{proof}
   \pf\ By definition of $LSpec$ and $HLSpec$, 1 and 2 imply that 
    $LSpec$ is equivalent to \tlabox{\EE h:HLSpec}.
   \end{proof}
\end{proof}
\end{display}
What we have done for this example generalizes in the obvious way.  
For a specification written in the canonical form
 \[ Init /\ [][Next]_{vars} /\ L
 \]
with $L$ a liveness condition, we add a history variable by adding it
just to the safety part, keeping the same liveness condition.
This method of adding a history variable to a specification with
liveness produces unusual specifications.  In our example, if we
expand the definition of $HSpec$, we see that specification $HLSpec$
equals
 \[  InitH \ /\ \ [][NextH]_{varsH} \ /\ \  \WF_{vars}(Respond) \]
This spec is unusual because it contains a fairness condition on the
action $Respond$ that is not a subaction of the next-state relation
$NextH$.  Such specs can be weird.  However, specs obtained in this
way by adding a history variable are not.  Because (i)~a $Respond$ action
is enabled iff a $RespondH$ action is and (ii)~a $NextH$ step is a
$Respond$ step iff it is a $RespondH$ step, specification $HLSpec$ is
equivalent to the normal specification:
\begin{equation}
InitH \ /\ \ [][NextH]_{varsH} \ /\ \  \WF_{varsH}(RespondH)
\end{equation}
In general, if $Spec^{h}$ is obtained from a specification $Spec$ by
adding a history variable $h$, then replacing a fairness requirement
on a subaction $A$ by the same fairness requirement on $A^{h}$
produces a specification equivalent to $Spec^{h}$.
		
The unusual nature of specification $HLSpec$ affects neither the TLC
model checker nor our ability to reason about the specification.  The
same refinement mapping as before shows that $HLSpec$ implements
$MinMax1$ with the added fairness condition when internal variables
are hidden.

\section{Prophecy Variables}

As we have observed, the fundamental task of verification is to show
that the specification $Spec_{1}$ of an implementation satisfies a
specification $Spec_{2}$ of what the implementation is supposed to do.
A history variable remembers the past.  It is needed to find a
refinement mapping to show that $Spec_{1}$ implements $Spec_{2}$ when
$Spec_{2}$ remembers previous events longer than it has to.  A
prophecy variable is one that predicts the future.  It is needed to
find a refinement mapping to show that $Spec_{1}$ implements
$Spec_{2}$ when $Spec_{2}$ makes decisions before it has to.


\subsection{One-Prediction Prophecy Variables} \label{sec:simple-proph}

We begin by showing how to add a simple prophecy variable that makes a
single prediction at a time.  Suppose a disjunctive representation of
the next-state relation contains a subaction $A$ such that
\begin{equation} 
A => (\tlabox{\E i \in \Pi : Pred_{A}(i)})
\NOTLA \label{eq:Proph1}
\end{equation}
for some expression $Pred_{A}(i)$ and constant set
$\Pi$.  Formula (\ref{eq:Proph1}) is equivalent to
\begin{equation} 
A \ \equiv\  A \, /\ \, (\tlabox{\E i \in \Pi : Pred_{A}(i)})
\NOTLA \label{eq:Proph2}
\end{equation}
which means that any $A$ step is an \tlabox{A /\ Pred_{A}(i)} step for
some $i$ in $\Pi$.  We introduce a one-prediction prophecy variable
$p$ whose value is an $i$ for which the next $A$ step is an
 \tlabox{A /\ Pred_{A}(i)} 
step---if there is a next $A$ step.  (There could be more than one
such $i$, since we don't require $Pred_{A}(i) /\ Pred_{A}(j)$ to equal
\FALSE\ if $i#j$.)  We give $p$ that meaning by replacing the
subaction $A$ with a subaction $A^{p}$ defined by
\begin{equation}
A^{p} == A \, /\  \, Pred_{A}(p) \, /\ \, Setp
\NOTLA \label{eq:gen-Ap}
\end{equation}
where $Setp$ determines the value of $p'$.

To ensure that adding the prophecy variable $p$ allows all the
behaviors of the other variables that the original spec does, we must
ensure that $p$ can always have any value in $\Pi$.  We do this by
initializing $p$ to an arbitrary element of $\Pi$ and changing $p$
only by setting it to any arbitrary element of $\Pi$.  Thus we 
modify the spec's initial predicate $Init$ to equal 
 \tlabox{Init /\ (p\in\Pi)}  
and we let $Setp$ equal $p'\in\Pi$, so
\begin{equation}
 A^{p} == A \, /\ \, Pred_{A}(p) \, /\ \, (p' \in \Pi)
\NOTLA \label{eq:ApDef}
\end{equation}
For another subaction $A$ of the next-state relation whose effect is not
being predicted by $p$, we let $A^{p}$ leave the prediction
unchanged, so it is defined simply as: 
\begin{equation}
    A^{p} == A /\ (p'=p) 
 \NOTLA \label{eq:ApDef2}
\end{equation}

\bigskip

\noindent We illustrate this with a simple example: a system in which
integers are sent and received, where sending an integer $i$ is
represented by setting the variable $x$ to $i$, and receiving a value
is represented by setting $x$ to a value $NotInt$ that is not an
integer.  Our specification $SendInt2$ has the receiving action set an
internal variable $z$ to the next value to be sent.  (The initial
value of $z$ is the first value to be sent.)  This simple
specification is in \lref{targ:SendInt2}{Figure~\ref{fig:SendInt2}}.
\begin{figure} \target{targ:SendInt2}
\begin{notla}
----------------------------- MODULE SendInt2 ----------------------------- 
EXTENDS Integers

NotInt == CHOOSE n : n \notin Int

VARIABLE x, z

Init == /\ x = NotInt
        /\ z \in Int

Send == /\ x = NotInt
        /\ x' = z
        /\ z' = NotInt

Rcv == /\ x \in Int
       /\ x' = NotInt
       /\ z' \in Int

Next == Send \/ Rcv

Spec == Init /\ [][Next]_<<x,z>>
=============================================================================
\end{notla}
\begin{tlatex}
\@x{}\moduleLeftDash\@xx{ {\MODULE} SendInt2}\moduleRightDash\@xx{}%
\@x{ {\EXTENDS} Integers}%
\@pvspace{8.0pt}%
\@x{ NotInt \.{\defeq} {\CHOOSE} n \.{:} n \.{\notin} Int}%
\@pvspace{8.0pt}%
\@x{ {\VARIABLE} x ,\, z}%
\@pvspace{8.0pt}%
\@x{ Init\@s{5.17} \.{\defeq} \.{\land} x \.{=} NotInt}%
\@x{\@s{40.87} \.{\land} z\@s{0.52} \.{\in} Int}%
\@pvspace{8.0pt}%
\@x{ Send \.{\defeq} \.{\land} x \.{=} NotInt}%
\@x{\@s{40.87} \.{\land} x \.{'} \.{=} z}%
\@x{\@s{40.87} \.{\land} z \.{'}\@s{0.52} \.{=} NotInt}%
\@pvspace{8.0pt}%
\@x{ Rcv \.{\defeq} \.{\land} x \.{\in} Int}%
\@x{\@s{35.94} \.{\land} x \.{'} \.{=} NotInt}%
\@x{\@s{35.94} \.{\land} z \.{'}\@s{0.52} \.{\in} Int}%
\@pvspace{8.0pt}%
\@x{ Next \.{\defeq} Send \.{\lor} Rcv}%
\@pvspace{8.0pt}%
 \@x{ Spec\@s{1.46} \.{\defeq} Init \.{\land} {\Box} [ Next ]_{ {\langle} x
 ,\, z {\rangle}}}%
\@x{}\bottombar\@xx{}%
\end{tlatex}
\caption{Specification \emph{SendInt}2.} \label{fig:SendInt2}
\end{figure}

Of course, we can describe the behavior of the variable $x$ even more
simply, without using any internal variable.  Such a specification is
in module $SendInt1$ of \lref{targ:SendInt1}{Figure~\ref{fig:SendInt1}}.
\begin{figure} \target{targ:SendInt1}
\begin{notla}
----------------------------- MODULE SendInt1 -----------------------------
EXTENDS Integers

NotInt == CHOOSE n : n \notin Int

VARIABLE x

Init == x = NotInt

Send == /\ x = NotInt
        /\ x' \in Int

Rcv == /\ x \in Int
       /\ x' = NotInt

Next == Send \/ Rcv

Spec == Init /\ [][Next]_x
=============================================================================
\end{notla}
\begin{tlatex}
\@x{}\moduleLeftDash\@xx{ {\MODULE} SendInt1}\moduleRightDash\@xx{}%
\@x{ {\EXTENDS} Integers}%
\@pvspace{8.0pt}%
\@x{ NotInt \.{\defeq} {\CHOOSE} n \.{:} n \.{\notin} Int}%
\@pvspace{8.0pt}%
\@x{ {\VARIABLE} x}%
\@pvspace{8.0pt}%
\@x{ Init\@s{5.17} \.{\defeq} x \.{=} NotInt}%
\@pvspace{8.0pt}%
\@x{ Send \.{\defeq} \.{\land} x \.{=} NotInt}%
\@x{\@s{40.87} \.{\land} x \.{'} \.{\in} Int}%
\@pvspace{8.0pt}%
\@x{ Rcv \.{\defeq} \.{\land} x \.{\in} Int}%
\@x{\@s{35.94} \.{\land} x \.{'} \.{=} NotInt}%
\@pvspace{8.0pt}%
\@x{ Next \.{\defeq} Send \.{\lor} Rcv}%
\@pvspace{8.0pt}%
\@x{ Spec\@s{1.46} \.{\defeq} Init \.{\land} {\Box} [ Next ]_{ x}}%
\@x{}\bottombar\@xx{}%
\end{tlatex}
\caption{Specification \emph{SendInt1}} \label{fig:SendInt1}
\end{figure}
Let $Spec_{1}$ and $Spec_{2}$ be the formulas $Spec$ of modules
$SendInt1$ and $SendInt2$, respectively.  It should be obvious that
$Spec_{1}$ is equivalent to \tlabox{\EE z:Spec_{2}}.  To verify
\tlabox{(\EE z:Spec_{2}) => Spec_{1}}, we just have to verify
$Spec_{2}=>Spec_{1}$, which TLC easily checks for a model that
substitutes a finite set of integers for
$Int$.  We now show how to verify
\tlabox{Spec_{1} => (\EE z:Spec_{2})}.

There is obviously no refinement mapping under which
$Spec_{1}$ implements $Spec_{2}$.  Such a refinement mapping would be
an expression \ov{z} involving only the variable $x$ so that
\ov{z} is the value of $z$ in any state satisfying $Spec_{2}$.
This is impossible, since $z$ could equal any integer in a state in which
$x$ equals $NotInt$, so there is no way to express its value
as a function of the value of $x$.
The variable $z$ of $SendInt2$ is used to predict the value to be sent
before it actually is sent.  To be able to define the value of \ov{z}
for a refinement mapping, we add a prophecy variable $p$ to $SendInt1$
that predicts what the next value to be sent is.

The prophecy variable $p$ must predict the value sent by action $Send$
of $SendInt1$.  Therefore $SendP$ must have the form of 
(\ref{eq:ApDef}).  A little thought shows that $p$ makes the
right prediction if we take $Pred_{Send}(p)$ to equal
$x'=p$.  Since \tlaplus\ doesn't allow identifiers to have subscripts 
we write $PredSend$ instead of $Pred_{Send}$ and define
 \[ PredSend(i) == x' = i
 \]
Condition (\ref{eq:Proph1}) becomes
 \[ Send => \tlabox{\E i \in \Pi : PredSend(i)}
 \]
which is obviously true by definition of $Send$, if we let $\Pi$ 
equal $Int$.  Writing $Pi$ instead of $\Pi$ and $SendP$ instead of
$Send^{p}$, we make the definitions:
 \[ \begin{noj}
    Pi == Int \V{.2}
    SendP ==  Send \, /\ \, PredSend(p) \, /\ \, (p' \in Pi)
    \end{noj}
 \]
(We could of course simply write $Int$ instead of $Pi$, but writing
$Pi$ will help us understand what's going on.)  

For the receive action, $Rcv^{p}$ should have the form (\ref{eq:ApDef2}).

In a behavior of $SendInt2$, when $x$ equals $NotInt$, the value of
$z$ is the next value sent; and when $x$ equals an integer (the value
sent), then $z$ equals $NotInt$.  Therefore, if $SpecP$ is the
specification obtained from specification $Spec$ of $SendInt1$ by
adding the prophecy variable $p$, then $SpecP$ implements the
specification $Spec$ of $SendInt2$ under this refinement mapping:
 \[ \ov{z} \ <- \ \IF x = NotInt \THEN p \ELSE NotInt 
 \]
Note that $SpecP$ predicts the next value to be sent even before the
$SendInt2$ specification does---when the previous value is sent rather
than when it is received.  Although it's not necessary, we'll see
later how we could defer the prediction until the $Rcv$ action is
executed.

The complete specification is contained in module $SendInt1P$ in
\lref{targ:SendInt1P}{Figure~\ref{fig:SendInt1P}}.  
\begin{figure} \target{targ:SendInt1P}
\begin{notla}
----------------------------- MODULE SendInt1P -----------------------------
EXTENDS SendInt1

Pi == Int

PredSend(i) == x' = i

VARIABLE p

varsP == <<x, p>>

InitP ==  Init /\ (p \in Pi)

SendP ==  Send /\ PredSend(p) /\ (p' \in Pi)

RcvP == Rcv /\ (p' = p)

NextP == SendP \/ RcvP

SpecP == InitP /\ [][NextP]_varsP
-----------------------------------------------------------------------------
SI2 == INSTANCE SendInt2 WITH z <- IF x = NotInt THEN p ELSE NotInt

THEOREM SpecP => SI2!Spec
=============================================================================
\end{notla}
\begin{tlatex}
\@x{}\moduleLeftDash\@xx{ {\MODULE} SendInt1P}\moduleRightDash\@xx{}%
\@x{ {\EXTENDS} SendInt1}%
\@pvspace{8.0pt}%
\@x{ Pi \.{\defeq} Int}%
\@pvspace{8.0pt}%
\@x{ PredSend ( i ) \.{\defeq} x \.{'} \.{=} i}%
\@pvspace{8.0pt}%
\@x{ {\VARIABLE} p}%
\@pvspace{8.0pt}%
\@x{ varsP\@s{2.93} \.{\defeq} {\langle} x ,\, p {\rangle}}%
\@pvspace{8.0pt}%
 \@x{ InitP\@s{5.08} \.{\defeq}\@s{4.1} Init\@s{5.17} \.{\land} ( p \.{\in} Pi
 )}%
\@pvspace{8.0pt}%
 \@x{ SendP \.{\defeq}\@s{4.1} Send \.{\land} PredSend ( p ) \.{\land} ( p
 \.{'} \.{\in} Pi )}%
\@pvspace{8.0pt}%
\@x{ RcvP \.{\defeq} Rcv \.{\land} ( p \.{'} \.{=} p )}%
\@pvspace{8.0pt}%
\@x{ NextP \.{\defeq} SendP \.{\lor} RcvP}%
\@pvspace{8.0pt}%
\@x{ SpecP\@s{1.08} \.{\defeq} InitP \.{\land} {\Box} [ NextP ]_{ varsP}}%
\@x{}\midbar\@xx{}%
 \@x{ SI2 \.{\defeq} {\INSTANCE} SendInt2 {\WITH} z \.{\leftarrow} {\IF} x
 \.{=} NotInt \.{\THEN} p \.{\ELSE} NotInt}%
\@pvspace{8.0pt}%
\@x{ {\THEOREM} SpecP \.{\implies} SI2 {\bang} Spec}%
\@x{}\bottombar\@xx{}%
\end{tlatex}

\caption{Specification \emph{SendInt1P} \label{fig:SendInt1P}}
\end{figure}
Observe that we defined $PredSend$ before the declaration of $p$ to
ensure that $PredSend$ is not defined in terms of $p$.

\begin{sloppypar}
The module's theorem asserts that $SpecP$ implements formula $Spec$ of
$SendInt2$ under the refinement mapping defined above.  It can be
checked with TLC by creating a model with temporal specification $SpecP$
and having it check the temporal property $SI2!Spec$.  The model will
have to substitute a finite set of integers for $Int$.  The
specification is very simple and doesn't depend on any properties of
integers, so substituting a set with a few numbers will ensure that we
didn't make a mistake.
\end{sloppypar}

\subsection{One-Prediction Prophecy Variables in General} \label{sec:simple-general}

We generalize our description of a one-prediction prophecy variable in two 
ways.  First, we can allow a prophecy variable to make predictions
about more than one action by replacing more than one subaction $A$ of
a disjunctive representation by an action $A^{p}$ of the form
(\ref{eq:gen-Ap}).  If we do this for subactions $A_{1}$, $A_{2}$,
\ldots, then the value of $p$ makes a prediction about the next step
to occur that is an $A_{1}$ or $A_{2}$ or \ldots\ step.  We can
express this a little more elegantly by generalizing (\ref{eq:gen-Ap})
to allow $Setp$ to depend on $A$ and then letting each action $A$ of
the disjunctive representation be replaced with an action $A^{p}$ defined by
\begin{equation}
A^{p} == A \, /\  \, Pred_{A}(p) \, /\ \, Setp_{A}
\NOTLA \label{eq:gen-Ap2}
\end{equation}
For an action $A$ about which no prediction is being made,
$Pred_{A}(p)$ is the expression $\TRUE$.  We can then replace
(\ref{eq:ApDef}) and (\ref{eq:ApDef2}) by defining $Setp_{A}$ to be
one of the following:
\begin{equation}
  \begin{noj2}
    \mbox{(a)} & Setp_{A} == p'=p \V{.2}
      \mbox{(b)} & Setp_{A} == p'\in\Pi
  \end{noj2}
 \NOTLA \label{eq:ApDef2bis}
\end{equation}
where possibility (a) is allowed only if $Pred_{A}(p)$ is the
expression $\TRUE$ (so $p$ is making no prediction about $A$).  This
is more general because it allows an action that doesn't use the
prediction made by $p$ to make a new prediction.

Our second generalization is needed to handle subactions of a
disjunctive representation having a nonempty context.  For 
a subaction $A$ with context $<<\textbf{k};\textbf{K}>>$,
condition (\ref{eq:Proph1}) contains the identifiers \textbf{k}.
That condition need only hold for values of those identifiers 
in the corresponding set in \textbf{K}.  Thus (\ref{eq:Proph1})
can be generalized to
\begin{equation} 
\tlabox{\A <<\textbf{k};\textbf{K}>> : A => (\E i \in \Pi : Pred_{A}(i))}
\NOTLA \label{eq:Proph1Gen}
\end{equation}
Condition~(\ref{eq:Proph1Gen}) is a condition on pairs of states.
It needn't hold for all pairs of states, only for pairs of states
that can occur in a behavior satisfying the original
specification $Spec$.  We can therefore
replace (\ref{eq:Proph1Gen}) by the requirement%
 \footnote{Formula (\ref{eq:Proph1GenTLA}) does not imply
that (\ref{eq:Proph1Gen}) is true for stuttering steps that are
allowed by $A$.  It can be shown that this doesn't matter, 
and condition (\ref{eq:Proph1GenTLA}) is strong enough.}
\begin{equation}
Spec\; => \; [][\tlabox{\A <<\textbf{k};\textbf{K}>> : A => (\E i \in \Pi : Pred_{A}(i)})]_{vars}
\NOTLA \label{eq:Proph1GenTLA}
\end{equation}
TLC can check this condition with a model having the temporal formula
$Spec$ as its behavioral spec and 
 \[
 \tlabox{[][\A <<\textbf{k};\textbf{K}>> : A => (\tlabox{\E i \in \Pi : Pred_{A}(i)})]_{vars}}
\]
as a property to be checked.

\subsection{Prophecy Array Variables} \label{sec:proph-array}

Our next example is based on one created by Mart\'{\i}n
Abadi~\cite{abadi:undo}.  It is similar to our $SendInt$
specifications in that a sender sends a value $v$ to a receiver with a
$Send$ action that sets the variable $x$ to $v$, and the receiver
receives the value by resetting $x$.  Instead of sending integers, the
values sent are elements of an unspecified constant set $Data$, and we
let the initial value of $x$ be a value $NonData$ not in $Data$.  A
variable $y$ contains a set of values to be sent.  Those values are
chosen by a $Choose$ action, which adds a new data element to $y$.
The high-level specification is formula $Spec$ in module $SendSet$ of
\lref{targ:SendSet}{Figure~{\ref{fig:SendSet}}}.
\begin{figure} \target{targ:SendSet}
\begin{notla}
------------------------------ MODULE SendSet ------------------------------
CONSTANT Data

NonData == CHOOSE d : d \notin Data

VARIABLES x, y
vars == <<x, y>>

Init == (x = NonData) /\ (y = {})

Choose == /\ \E d \in Data \ y : y' = y \cup {d}
          /\ x' = x
          
Send == /\ x = NonData 
        /\ x' \in y
        /\ y' = y \ {x'}
        
Rcv == /\ x \in Data
       /\ x' = NonData
       /\ y' = y
       
Next == Choose \/ Send \/ Rcv

Spec == Init /\ [][Next]_vars
=============================================================================
\end{notla}
\begin{tlatex}
\@x{}\moduleLeftDash\@xx{ {\MODULE} SendSet}\moduleRightDash\@xx{}%
\@x{ {\CONSTANT} Data}%
\@pvspace{8.0pt}%
\@x{ NonData \.{\defeq} {\CHOOSE} d \.{:} d \.{\notin} Data}%
\@pvspace{8.0pt}%
\@x{ {\VARIABLES} x ,\, y}%
\@x{ vars \.{\defeq} {\langle} x ,\, y {\rangle}}%
\@pvspace{8.0pt}%
 \@x{ Init\@s{2.02} \.{\defeq} ( x \.{=} NonData ) \.{\land} ( y \.{=} \{ \}
 )}%
\@pvspace{8.0pt}%
 \@x{ Choose \.{\defeq} \.{\land} \E\, d \.{\in} Data \.{\,\backslash\,} y
 \.{:} y \.{'} \.{=} y \.{\cup} \{ d \}}%
\@x{\@s{50.30} \.{\land} x \.{'} \.{=} x}%
\@pvspace{8.0pt}%
\@x{ Send \.{\defeq} \.{\land} x \.{=} NonData}%
\@x{\@s{40.87} \.{\land} x \.{'} \.{\in} y}%
 \@x{\@s{40.87} \.{\land} y \.{'}\@s{0.10} \.{=} y \.{\,\backslash\,} \{ x
 \.{'} \}}%
\@pvspace{8.0pt}%
\@x{ Rcv \.{\defeq} \.{\land} x \.{\in} Data}%
\@x{\@s{35.94} \.{\land} x \.{'} \.{=} NonData}%
\@x{\@s{35.94} \.{\land} y \.{'}\@s{0.10} \.{=} y}%
\@pvspace{8.0pt}%
\@x{ Next \.{\defeq} Choose \.{\lor} Send \.{\lor} Rcv}%
\@pvspace{8.0pt}%
\@x{ Spec\@s{1.46} \.{\defeq} Init \.{\land} {\Box} [ Next ]_{ vars}}%
\@x{}\bottombar\@xx{}%
\end{tlatex}
\caption{Specification \emph{SendSet}.\label{fig:SendSet}}
\end{figure}
We consider the variable $x$ to be externally visible and $y$ to be
internal.


Our implementation adds to the specification of $SendSet$ an
\emph{undo} operation that removes elements from $y$.  Abadi reports
that this example is a highly simplified abstraction of a real
system in which the implementation contains an undo operation not
present in the specification.  

The implementation specification is
formula $SpecU$ in module $SendSetUndo$ of
\lref{targ:SendSetUndo}{Figure~\ref{fig:SendSetUndo}}.
\begin{figure} \target{targ:SendSetUndo}
\begin{notla}
---------------------------- MODULE SendSetUndo ----------------------------
EXTENDS SendSet

Undo(S)  ==  /\ y' = y \ S
             /\ x' = x
        
NextU  ==  Next  \/  (\E S \in (SUBSET y) : Undo(S))

SpecU  ==  Init  /\  [][NextU]_vars
============================================================================
\end{notla}
\begin{tlatex}
\@x{}\moduleLeftDash\@xx{ {\MODULE} SendSetUndo}\moduleRightDash\@xx{}%
\@x{ {\EXTENDS} SendSet}%
\@pvspace{8.0pt}%
 \@x{ Undo ( S )\@s{4.1} \.{\defeq}\@s{4.1} \.{\land} y \.{'}\@s{0.10} \.{=} y
 \.{\,\backslash\,} S}%
\@x{\@s{65.59} \.{\land} x \.{'} \.{=} x}%
\@pvspace{8.0pt}%
 \@x{ NextU\@s{4.1} \.{\defeq}\@s{4.1} Next\@s{4.1} \.{\lor}\@s{4.1} ( \E\, S
 \.{\in} ( {\SUBSET} y ) \.{:} Undo ( S ) )}%
\@pvspace{8.0pt}%
 \@x{ SpecU\@s{5.18} \.{\defeq}\@s{4.09} Init\@s{4.1} \.{\land}\@s{4.1} {\Box}
 [ NextU ]_{ vars}}%
\@x{}\bottombar\@xx{}%
\end{tlatex}
\caption{Specification \emph{SendSetUndo}.} \label{fig:SendSetUndo}
\end{figure}
Its initial predicate is the same as the initial predicate of the
specification $Spec$ of module $SendSet$, and its next-state action
$NextU$ is the same as the next-state action $Next$ of that module
except it allows $Undo(S)$ steps that remove from $y$ an arbitrarily
chosen non-empty subset $S$ of $y$.

It's clear that the specifications $Spec$ of module $SendSet$ and
$SpecU$ of module $SendSetUndo$ allow the same behaviors of the
variable $x$.  Hence, \tlabox{\EE y:Spec} and \tlabox{\EE y:SpecU} are
equivalent.  It's easy to show that $Spec$ implements $SpecU$ under the
identity refinement mapping in which \ov{y} is defined to equal $y$,
since $NextU$ allows all steps allowed by $Next$.  This implies that
\tlabox{\EE y:Spec} implies \tlabox{\EE y:SpecU}.  To construct a
refinement mapping under which $SpecU$ implements $Spec$, we must
define \ov{y} so that it contains a data value $d$ iff that value is
going to be sent by a $Send$ step rather than being removed from $y$
by an $Undo$ step.  This involves predicting, when $d$ is added to $y$,
whether it will later be sent or ``undone''.

We add a prophecy array variable $p$ to $SpecU$ that makes this prediction,
setting $p[d]$ to either $"send"$ or $"undo"$ when $d$ is added to
$y$.  So, we define our set $Pi$ of possible predictions by
 \[  Pi == \{"send","undo"\}
 \]
The value of $p$ in every state will be a function with
domain equal to the set $y$, with $p[d]\in Pi$ for all $d\in y$.
In other words
 \tlabox{p \in [y -> Pi]} 
will be an invariant of the spec $SpecUP$ obtained by adding the
prophecy variable $p$.  The variable $p$ is therefore making a
predication $p[d]$ for every $d$ in $y$, so $p$ is making an array of
prophecies.  (This is a ``dynamic'' array, because the value of $y$
can change.)

We now define the specification $SpecUP$.  As with a one-prediction prophecy
variable, we obtain the next-state relation of $SpecUP$ by replacing
each subaction $A$ in a disjunctive representation of the next-state
relation with a new action $A^{p}$.  Instead of defining $A^{p}$ as in
(\ref{eq:gen-Ap2}), we define it to equal
\begin{equation}
A^{p} == A \, /\  \, Pred_{A}(p) \, /\ \, (p' \in NewPSet_{A})
\NOTLA \label{eq:ApDefFcn}
\end{equation}
for suitable expressions $Pred_{A}(p)$ and $NewPSet_{A}$.  We need a
condition corresponding to condition (\ref{eq:Proph1GenTLA}) for a
one-prediction prophecy variable to assert that there is a possible value of
$p$ that makes $Pred_{A}(p)$ true.  With an array prophecy variable,
$p$ is no longer an element of $Pi$ but a function in $[Dom -> Pi]$
for some domain $Dom$ that can change.  In our example, $Dom$ equals
$y$.  To make the generalization to an arbitrary spec easier, we
define $Dom$ to equal $y$ and write $Dom$ instead of $y$ where
appropriate.  For our example, we can replace (\ref{eq:Proph1GenTLA})
with
\begin{equation} 
SpecU\; => 
  \; [][\A <<\textbf{k};\textbf{K}>> : 
           A => (\tlabox{\E f \in [Dom -> \Pi] : Pred_{A}(f)})]_{vars}
\NOTLA \label{eq:ProphFcn}
\end{equation}
where $<<\textbf{k};\textbf{K}>>$ is the context of $A$.  (Remember
that for the empty context $<<\,;\,>>$, we define 
\tlabox{\A <<\,;\,>> : F} to equal $F$, for any formula $F$.)

\begin{sloppypar}
We now define the formulas $Pred_{A}$ and $NewPSet_{A}$ for the
disjunctive representation of $NextU$ with subactions $Choose$,
$Send$, $Rcv$, and $Undo(S)$.  The context of the first three
subactions is empty; the context of $Undo(S)$ is 
  $<<S; \,\SUBSET y>>$.
\end{sloppypar}

The variable $p$ does not make any prediction about the $Choose$
action, so $Pred_{Choose}(p)$ should equal \TRUE. The action adds an
element $d$ to its domain, so $Choose^{p}$ must allow $p'[d]$ to equal
any element of $Pi$.  For any element $d$ in the domain $Dom$ of $p$,
the value of $p[d]$ can be left unchanged.  Our definitions 
of $Pred_{Choose}(p)$ and $NewPSet_{Choose}(p)$
are then
\begin{display}
\begin{notla}
PredChoose(p)    == TRUE
NewPSetChoose(p) == {f \in [Dom' -> Pi]  :  \A d \in Dom : f[d] = p[d]}
\end{notla}
\begin{tlatex}
\@x{ PredChoose ( p )\@s{19.31} \.{\defeq} {\TRUE}}%
 \@x{ NewPSetChoose ( p ) \.{\defeq} \{ \,f \.{\in} [ Dom \.{'} \.{\rightarrow}
 Pi ]\@s{4.1} \.{:}\@s{4.1} \A\, d \.{\in} Dom \.{:} f [ d ] \.{=} p [ d ] \,
 \}}%
\end{tlatex}
\end{display}
The prophecy variable $p$ should predict that if the next action is 
a $Send$ action, then it sends a value $d$ in $Dom$ such that $p[d]="send"$.
The value sent by the action is $x'$, so we define
 \[ PredSend(p) == p[x'] = "send"
 \]
The $Send$ action removes the sent element $d$ from $Dom$, thus erasing
the prediction $p$ made about $d$.  The value of $p[d]$ is left
unchanged for all other elements $d$ in $Dom$.  Thus 
$NewPSet_{Send}(p)$ is defined as follows to be a set consisting of a
single function
\[ NewPSetSend(p) == \{\, [d \in Dom' |-> p[d]] \,\}
 \]
No prediction is made about the $Rcv$ action, and it doesn't change
$Dom$, so we have:
\begin{display}
\begin{notla}
PredRcv(p)    == TRUE
NewPSetRcv(p) == {p}
\end{notla}
\begin{tlatex}
\@x{ PredRcv ( p )\@s{19.31} \.{\defeq} {\TRUE}}%
\@x{ NewPSetRcv ( p ) \.{\defeq} \{ p \}}%
\end{tlatex}
\end{display}
The $Undo(S)^{p}$ action should be enabled only when $p$ has predicted
that all the elements in $S$ will not be sent---in other words,
when \tlabox{\A d \in S : p[d] = "undo"} is true.  Since 
the identifier $S$ appears in this formula, it must be an argument
of the definition of $Undo(S)^{p}$.  Thus, we define
 \[ PredUndo(p, S) == \A d \in S : p[d] = "undo"
 \]
The $Undo(S)$ action removes from $Dom$ all the elements for which
$p$ made a prediction about $Undo(S)$---namely, all the elements
of $S$.  We can define $NewPSet_{Undo(S)}$ the same way we defined
$PSetSend_{Send}$, without explicitly mentioning $S$:
 \[ NewPSetUndo(p) == \{\, [d \in Dom' |-> p[d]] \,\}
 \]
We can now declare the variable $p$ and define the specification
$SpecUP$ by
defining the initial predicate
$InitUP$ and defining the next-state relation $NextUP$ 
in terms of the subactions $A^{p}$, using (\ref{eq:ApDefFcn}).
The complete specification is in module $SendSetUndoP$
shown in \lref{targ:SendSetUndoP}{Figure~\ref{fig:SendSetUndoP}}.
\begin{figure} \target{targ:SendSetUndoP}
\begin{notla}
--------------------------- MODULE SendSetUndoP ---------------------------
EXTENDS SendSetUndo

Pi == {"send", "undo"}
Dom == y

PredChoose(p)    == TRUE
NewPSetChoose(p) == {f \in [Dom' -> Pi]  :  \A d \in Dom : f[d] = p[d]}

PredSend(p)    == p[x'] = "send"
NewPSetSend(p) == { [d \in Dom' |-> p[d]] }

PredRcv(p)    == TRUE
NewPSetRcv(p) == {p}

PredUndo(p, S) == \A d \in S : p[d] = "undo"
NewPSetUndo(p) == { [d \in Dom' |-> p[d]] }

VARIABLE p
varsP == <<vars, p>>

InitUP == Init /\ (p = << >>)

ChooseP  == Choose  /\ PredChoose(p)  /\ (p' \in NewPSetChoose(p))
SendP    == Send    /\ PredSend(p)    /\ (p' \in NewPSetSend(p))
RcvP     == Rcv     /\ PredRcv(p)     /\ (p' \in NewPSetRcv(p))
UndoP(S) == Undo(S) /\ PredUndo(p, S) /\ (p' \in NewPSetUndo(p))

NextUP == ChooseP \/ SendP \/ RcvP  \/  (\E S \in SUBSET y : UndoP(S))

SpecUP == InitUP /\ [][NextUP]_varsP
=============================================================================
\end{notla}
\begin{tlatex}
\@x{}\moduleLeftDash\@xx{ {\MODULE} SendSetUndoP}\moduleRightDash\@xx{}%
\@x{ {\EXTENDS} SendSetUndo}%
\@pvspace{8.0pt}%
\@x{ Pi \.{\defeq} \{\@w{send} ,\,\@w{undo} \}}%
\@x{ Dom \.{\defeq} y}%
\@pvspace{8.0pt}%
\@x{ PredChoose ( p )\@s{19.31} \.{\defeq} {\TRUE}}%
 \@x{ NewPSetChoose ( p ) \.{\defeq} \{ f \.{\in} [ Dom \.{'} \.{\rightarrow}
 Pi ]\@s{4.1} \.{:}\@s{4.1} \A\, d \.{\in} Dom \.{:} f [ d ] \.{=} p [ d ]
 \}}%
\@pvspace{8.0pt}%
\@x{ PredSend ( p )\@s{19.31} \.{\defeq} p [ x \.{'} ] \.{=}\@w{send}}%
 \@x{ NewPSetSend ( p ) \.{\defeq} \{ [ d \.{\in} Dom \.{'} \.{\mapsto} p [ d
 ] ] \}}%
\@pvspace{8.0pt}%
\@x{ PredRcv ( p )\@s{19.31} \.{\defeq} {\TRUE}}%
\@x{ NewPSetRcv ( p ) \.{\defeq} \{ p \}}%
\@pvspace{8.0pt}%
 \@x{ PredUndo ( p ,\, S )\@s{6.38} \.{\defeq} \A\, d\@s{0.55} \.{\in} S \.{:}
 p [ d ] \.{=}\@w{undo}}%
 \@x{ NewPSetUndo ( p ) \.{\defeq} \{ [ d \.{\in} Dom \.{'} \.{\mapsto} p [ d
 ] ] \}}%
\@pvspace{8.0pt}%
\@x{ {\VARIABLE} p}%
\@x{ varsP \.{\defeq} {\langle} vars ,\, p {\rangle}}%
\@pvspace{8.0pt}%
\@x{ InitUP \.{\defeq} Init \.{\land} ( p \.{=} {\langle} {\rangle} )}%
\@pvspace{8.0pt}%
 \@x{ ChooseP\@s{7.20} \.{\defeq} Choose\@s{7.08} \.{\land} PredChoose ( p
 )\@s{5.42} \.{\land} ( p \.{'} \.{\in} NewPSetChoose ( p ) )}%
 \@x{ SendP\@s{16.91} \.{\defeq} Send\@s{16.51} \.{\land} PredSend ( p
 )\@s{14.85} \.{\land} ( p \.{'} \.{\in} NewPSetSend ( p ) )}%
 \@x{ RcvP\@s{21.89} \.{\defeq} Rcv\@s{21.44} \.{\land} PredRcv ( p
 )\@s{19.77} \.{\land} ( p \.{'} \.{\in} NewPSetRcv ( p ) )}%
 \@x{ UndoP ( S ) \.{\defeq} Undo ( S ) \.{\land} PredUndo ( p ,\, S )
 \.{\land} ( p \.{'} \.{\in} NewPSetUndo ( p ) )}%
\@pvspace{8.0pt}%
 \@x{ NextUP \.{\defeq} ChooseP \.{\lor} SendP \.{\lor} RcvP\@s{4.1}
 \.{\lor}\@s{4.1} ( \E\, S \.{\in} {\SUBSET} y \.{:} UndoP ( S ) )}%
\@pvspace{8.0pt}%
\@x{ SpecUP\@s{1.08} \.{\defeq} InitUP \.{\land} {\Box} [ NextUP ]_{ varsP}}%
\@x{}\bottombar\@xx{}%
\end{tlatex}

\caption{Specification \emph{SendSetUndoP}} \label{fig:SendSetUndoP}
\end{figure}
Note that the initial value of $p$ is the unique function whose domain
is the empty set.  We could write that function as
 \tlabox{[d \in \{\} |-> exp]}
for any expression $exp$---for example, 42.  However, it's
easier to write that function as $<< >>$ (the empty sequence).

We can now define the refinement mapping under which $SpecUP$
implements specification $Spec$ of module $SendSet$.  The refinement
mapping defines \ov{y} to equal the set of elements $d$ in $y$ with
$p[d]="send"$.  We thus add to the module
\begin{display}
\begin{notla}
SS == INSTANCE SendSet WITH y <- {d \in y : p[d] = "send"}
THEOREM SpecUP => SS!Spec
\end{notla}
\begin{tlatex}
 \@x{ SS \.{\defeq} {\INSTANCE} SendSet {\WITH} y \.{\leftarrow} \{ d \.{\in}
 y \.{:} p [ d ] \.{=}\@w{send} \}}\vspace{.2em}%
\@x{ {\THEOREM} SpecUP \.{\implies} SS {\bang} Spec}%
\end{tlatex}
\end{display}
We can have TLC check this theorem by creating a model having $SpecUP$
as the behavior spec and checking the property $SS!Spec$.

We should also check that condition~(\ref{eq:ProphFcn}) holds for each
subaction $A$.   To do this, we need to create a model with
specification $SpecU$ and have TLC check the property:
\begin{display}
\begin{notla}
[][  /\ Choose => \E f \in [Dom -> Pi] : PredChoose(f)
     /\ Send   => \E f \in [Dom -> Pi] : PredSend(f)
     /\ Rcv    => \E f \in [Dom -> Pi] : PredRcv(f)
     /\ \A S \in SUBSET y : 
            Undo(S) => \E f \in [Dom -> Pi] : PredUndo(f,S)
  ]_vars
\end{notla}
\begin{tlatex}
 \@x{ {\Box} [\@s{4.1} \.{\land} Choose \.{\implies} \E\, f \.{\in} [ Dom
 \.{\rightarrow} Pi ] \.{:} PredChoose ( f )}%
 \@x{\@s{14.35} \.{\land} Send\@s{9.42} \.{\implies} \E\, f \.{\in} [ Dom
 \.{\rightarrow} Pi ] \.{:} PredSend ( f )}%
 \@x{\@s{14.35} \.{\land} Rcv\@s{14.35} \.{\implies} \E\, f \.{\in} [ Dom
 \.{\rightarrow} Pi ] \.{:} PredRcv ( f )}%
\@x{\@s{14.35} \.{\land} \A\, S \.{\in} {\SUBSET} y \.{:}}%
 \@x{\@s{36.78} Undo ( S ) \.{\implies} \E\, f \.{\in} [ Dom \.{\rightarrow}
 Pi ] \.{:} PredUndo ( f ,\, S )}%
\@x{\@s{7.47} ]_{ vars}}%
\end{tlatex}
\end{display}
However, there is a problem in doing this.  TLC will not allow a model
for module $SendSetUndoP$ to have behavior specification $SpecU$
because that spec doesn't describe the behavior of the variable $p$.
We can solve this problem by modifying the specification---temporarily
inserting ``\verb|======|'' into the module before the declaration of
$p$ and then creating the necessary model.  Alternatively, we can move
all the definitions before the declaration of $p$ into module
$SendSetUndo$ and check the condition in a model for that spec.  This
is inelegant because those definitions aren't part of the
$SendSetUndo$ specification.  The proper solution is to move those
definitions from $SendSetUndoP$ and put them in a new module that
extends $SendSetUndo$ and is extended by $SendSetUndoP$.  We can then
check that the condition is satisfied using a model for that
specification that has behavior spec $SpecU$.  We won't bother doing
this, instead putting them in module $SendSetUndoP$.

\bigskip

\noindent
Recall that in the $SendInt$ example of
Section~\ref{sec:simple-proph}, the one-prediction prophecy variable we used
predicted the next value to be sent when the previous value was sent,
while the specification $SendInt2$ didn't choose the next value to be
sent until the previous value was received.  We can use an array
prophecy variable to defer the prediction until it's needed.  We let
the domain $Dom$ of $p$ initially contain a single element---let's
take that element to be $"on"$.  We let the $Send$ action set $Dom$ to
the empty set, and we let the $Rcv$ action set $Dom$ to $\{"on"\}$.

\subsection{Prophecy Data Structure Variables} \label{sec:proph-data-struct}

It is easy to generalize from the $SendSet$ example to an arbitrary
prophecy-array variable.  However, it is useful to generalize still
further from an array to an arbitrary data structure.  These prophecy
data structure variables are the most general ones that we consider.
We propose them as the standard way of defining prophecy variables in
\tlaplus, and we have created a module with definitions that 
simplify adding these prophecy variables.  Both single-prophecy variables
and prophecy array variables are easily expressed as special cases.

As our example of a prophecy data structure variable, we modify the
specification $SendSet$, in which a set of items to be sent is chosen,
to a specification $SendSeq$ in which a sequence of items is chosen.
The value of the variable $y$ is changed from a set of data items to a
sequence of data items.  The next item to be sent is the one at the
head of $y$, and each value chosen is appended to the tail of $y$.
The specification is in module $SendSeq$, shown in
\lref{targ:SendSeq}{Figure~\ref{fig:SendSeq}}.

\begin{figure} \target{targ:SendSeq}
\begin{notla}
------------------------------ MODULE SendSeq ------------------------------
EXTENDS Sequences, Integers

CONSTANT Data

NonData == CHOOSE v : v \notin Data

VARIABLES x, y
vars == <<x, y>>

Init == (x = NonData) /\ (y = <<>>)

Choose == /\ \E d \in Data : y' = Append(y, d)
          /\ x' = x
          
Send == /\ x = NonData /\ y # << >>
        /\ x' = Head(y)
        /\ y' = Tail(y)
        
Rcv == /\ x \in Data
       /\ x' = NonData
       /\ y' = y
       
Next == Choose \/ Send \/ Rcv

Spec == Init /\ [][Next]_vars
=============================================================================
\end{notla}
\begin{tlatex}
\@x{}\moduleLeftDash\@xx{ {\MODULE} SendSeq}\moduleRightDash\@xx{}%
\@x{ {\EXTENDS} Sequences ,\, Integers}%
\@pvspace{8.0pt}%
\@x{ {\CONSTANT} Data}%
\@pvspace{8.0pt}%
\@x{ NonData \.{\defeq} {\CHOOSE} v \.{:} v \.{\notin} Data}%
\@pvspace{8.0pt}%
\@x{ {\VARIABLES} x ,\, y}%
\@x{ vars \.{\defeq} {\langle} x ,\, y {\rangle}}%
\@pvspace{8.0pt}%
 \@x{ Init\@s{2.02} \.{\defeq} ( x \.{=} NonData ) \.{\land} ( y \.{=}
 {\langle} {\rangle} )}%
\@pvspace{8.0pt}%
 \@x{ Choose \.{\defeq} \.{\land} \E\, d \.{\in} Data \.{:} y \.{'} \.{=}
 Append ( y ,\, d )}%
\@x{\@s{50.30} \.{\land} x \.{'} \.{=} x}%
\@pvspace{8.0pt}%
 \@x{ Send \.{\defeq} \.{\land} x \.{=} NonData \.{\land} y \.{\neq} {\langle}
 {\rangle}}%
\@x{\@s{40.87} \.{\land} x \.{'} \.{=} Head ( y )}%
\@x{\@s{40.87} \.{\land} y \.{'}\@s{0.10} \.{=} Tail ( y )}%
\@pvspace{8.0pt}%
\@x{ Rcv \.{\defeq} \.{\land} x \.{\in} Data}%
\@x{\@s{35.94} \.{\land} x \.{'} \.{=} NonData}%
\@x{\@s{35.94} \.{\land} y \.{'}\@s{0.10} \.{=} y}%
\@pvspace{8.0pt}%
\@x{ Next \.{\defeq} Choose \.{\lor} Send \.{\lor} Rcv}%
\@pvspace{8.0pt}%
\@x{ Spec\@s{1.46} \.{\defeq} Init \.{\land} {\Box} [ Next ]_{ vars}}%
\@x{}\bottombar\@xx{}%
\end{tlatex}
\caption{Specification \emph{SendSeq} \label{fig:SendSeq}}
\end{figure}

For our implementation, we add an undo action that removes an
arbitrary element from the sequence $y$.  The specification
is in module $SendSeqUndo$ of \lref{targ:SendSeqUndo}{Figure~\ref{fig:SendSeqUndo}}.  It
defines $RemoveEltFrom(i, seq)$ to be the sequence obtained from
a sequence $seq$ by removing its $i$\tth\ element,
assuming $1 \leq i \leq Len(seq)$.

\begin{figure} \target{targ:SendSeqUndo}
\begin{notla}
---------------------------- MODULE SendSeqUndo ----------------------------
EXTENDS SendSeq

RemoveEltFrom(i, seq) == [j \in 1..(Len(seq)-1) |-> IF j < i THEN seq[j]
                                                             ELSE seq[j+1]]

Undo(i) == /\ y' = RemoveEltFrom(i, y) 
           /\ x' = x
        
NextU == Next \/ (\E i \in 1..Len(y) : Undo(i))

SpecU == Init /\ [][NextU]_vars
============================================================================
\end{notla}
\begin{tlatex}
\@x{}\moduleLeftDash\@xx{ {\MODULE} SendSeqUndo}\moduleRightDash\@xx{}%
\@x{ {\EXTENDS} SendSeq}%
\@pvspace{8.0pt}%
 \@x{ RemoveEltFrom ( i ,\, seq ) \.{\defeq} [ j \.{\in} 1 \.{\dotdot} ( Len (
 seq ) \.{-} 1 ) \.{\mapsto} {\IF} j \.{<} i \.{\THEN} seq [ j ]}%
\@x{\@s{271.81} \.{\ELSE} seq [ j \.{+} 1 ] ]}%
\@pvspace{8.0pt}%
 \@x{ Undo ( i ) \.{\defeq} \.{\land} y \.{'}\@s{0.10} \.{=} RemoveEltFrom ( i
 ,\, y )}%
\@x{\@s{54.66} \.{\land} x \.{'} \.{=} x}%
\@pvspace{8.0pt}%
 \@x{ NextU \.{\defeq} Next \.{\lor} ( \E\, i \.{\in} 1 \.{\dotdot} Len ( y )
 \.{:} Undo ( i ) )}%
\@pvspace{8.0pt}%
\@x{ SpecU\@s{1.08} \.{\defeq} Init \.{\land} {\Box} [ NextU ]_{ vars}}%
\@x{}\bottombar\@xx{}%
\end{tlatex}
\caption{Specification \emph{SendSeqUndo} \label{fig:SendSeqUndo}}
\end{figure}

As before, we want to show that specification $SendU$ of module
$SendSeqUndo$ implements \tlabox{\EE y : Spec}, where $Spec$ is the
specification of module $SendSeq$.  Again, we need to add a prophecy
variable $p$ that predicts whether each element of $y$ will be
sent or ``undone''.  We do this by having $p$ be an element
of $Seq(\{"send", "undo"\})$ that has the same length as
$y$.  The $Choose^{p}$ action should append either $"send"$
or $"undo"$ to the tail of $p$, the $Send^{p}$ action should remove the
head of $p$, and the $Undo(i)$ action should remove the
$i$\tth\ element of $p$.

As we did for the $SendSet$ example, we write a module $SendSeqUndoP$
that extends $SendSeqUndo$.  In it, we define the formulas
$Pred_{A}(p)$ of (\ref{eq:ApDefFcn}) for each of the subactions
$Choose$, $Send$, $Rcv$, and $Undo(i)$.  The definitions are:
\begin{display}
\begin{notla}
PredChoose(p)  == TRUE
PredSend(p)    == p[1] = "send"
PredRcv(p)     == TRUE
PredUndo(p, i) == p[i] = "undo"
\end{notla}
\begin{tlatex}
\@x{ PredChoose ( p )\@s{4.1} \.{\defeq} {\TRUE}}%
\@x{ PredSend ( p )\@s{13.52} \.{\defeq} p [ 1 ] \.{=}\@w{send}}%
\@x{ PredRcv ( p )\@s{18.45} \.{\defeq} {\TRUE}}%
\@x{ PredUndo ( p ,\, i )\@s{1.41} \.{\defeq} p [ i ] \.{=}\@w{undo}}%
\end{tlatex}
\end{display}
We now need to define the expression $NewPSet_{A}$ of
(\ref{eq:ApDefFcn}) for each of these subactions.

Since a sequence of length $n$ is a function with domain $1\dd n$, the
value of $p$ is a function---just as in $SendSetUndoP$ above.  This
time its domain $Dom$ is the set $1\dd Len(y)$.  However, in that
example, if $d$ is in the domain $Dom$ of $p$ in two successive
states, then $p[d]$ represents the same prediction in both states.
This isn't true in the current example.  If $s->t$ is a $Send$ step
and $Len(p)>1$ in state $s$, then the prediction made by $p[2]$ in
state $s$ is the prediction made by $p[1]$ in state $t$.  If $s->t$ is
an $Undo(i)$ step and $j>i$, the prediction made by $p[j]$ in state
$s$ is the prediction made by $p[j-1]$ in state~$t$.

In general, an action $A$ defines a correspondence between some
elements in the domain $Dom$ of $p$ and elements in the domain $Dom'$
of $p'$.  For example, for the 
$Send$ action, each element $i>1$ in
$Dom$ corresponds to the element $i-1$ of $Dom'$.  The formula $p' \in
NewPSet_{A}$ in (\ref{eq:ApDefFcn}) should ensure that if an element
$d$ of $Dom'$ either corresponds to an element $c$ of $Dom$ that makes
a prediction about $A$ or else does not correspond to any element of
$Dom$, then $p'[d]$ can assume any value in $\Pi$; but if $d$
corresponds to an element $c$ of $Dom$ that make no prediction about
$A$, then $p'[d]$ equals $p[c]$.  Instead of defining the formulas
$NewPSet_{A}$ directly, we will define them in terms of the
correspondence between elements of $Dom$ and $Dom'$ made by $A$ and
the set of elements $d$ in $Dom$ for which $p[d]$ makes a prediction
about $A$.

To express formally a correspondence between elements of $Dom$ and
$Dom'$, we introduce the concept of a partial injection.  A
\emph{partial function} from a set $U$ to a set $V$ is a function from
a subset of $U$ to $V$.  In other words, it is an element of $[D->V]$
for some subset $D$ of $U$.  (Remember that $U$ is a subset of
itself.)  An \emph{injection} is a function that maps different
elements in its domain to different values.  In other words, a
function $f$ is an injection iff for all $a$ and $b$ in $\DOMAIN f$,
if $a#b$ then $f[a]#f[b]$.  The set of all partial injections from
$U$ to $V$ is defined in \tlaplus\ by
\begin{display}
\begin{notla}
PartialInjections(U, V) == 
   LET PartialFcns == UNION { [D -> V] : D \in SUBSET U}
   IN  {f \in PartialFcns : \A a, b \in DOMAIN f : (a # b) => (f[a] # f[b])}
\end{notla}
\begin{tlatex}
\@x{ PartialInjections ( U ,\, V ) \.{\defeq}}%
 \@x{\@s{12.29} \.{\LET} PartialFcns \.{\defeq} {\UNION} \{ [ D
 \.{\rightarrow} V ] \.{:} D \.{\in} {\SUBSET} U \}}%
 \@x{\@s{12.29} \.{\IN} \{ f \.{\in} PartialFcns \.{:} \A\, a ,\, b \.{\in}
 {\DOMAIN} f \.{:} ( a \.{\neq} b ) \.{\implies} ( f [ a ] \.{\neq} f [ b ] )
 \}}%
\end{tlatex}
\end{display}
For each subaction $A$, we define a partial injection $DomInj_{A}$
from $Dom$ to $Dom'$ such that an element $c$ of $Dom$ corresponds to
an element $d$ of $Dom'$ iff $c$ is in the domain of $DomInj_{A}$
and $d=DomInj_{A}[c]$.  Here are the definitions for the four subactions,
which are put in module $SendSeqUndoP$:
\begin{display}
\begin{notla}
DomInjChoose  == [d \in Dom |-> d]
DomInjSend    == [i \in 2..Len(y) |-> i-1]
DomInjRcv     == [d \in Dom |-> d]
DomInjUndo(i) == [j \in 1..Len(y) \ {i} |-> IF j < i THEN j ELSE j-1]
\end{notla}
\begin{tlatex}
 \@x{ DomInjChoose\@s{4.35} \.{\defeq} [ d\@s{2.05} \.{\in} Dom \.{\mapsto} d
 ]}%
 \@x{ DomInjSend\@s{13.78} \.{\defeq} [ i\@s{2.05} \.{\in} 2 \.{\dotdot} Len (
 y ) \.{\mapsto} i \.{-} 1 ]}%
\@x{ DomInjRcv\@s{18.71} \.{\defeq} [ d \.{\in} Dom \.{\mapsto} d ]}%
 \@x{ DomInjUndo ( i ) \.{\defeq} [ j\@s{1.63} \.{\in} 1 \.{\dotdot} Len ( y )
 \.{\,\backslash\,} \{ i \} \.{\mapsto} {\IF} j \.{<} i \.{\THEN} j \.{\ELSE}
 j \.{-} 1 ]}%
\end{tlatex}
\end{display}
For the prophecy array variable described in Section~\ref{sec:proph-array}
above, the function $DomInj_{A}$ maps each element $d$ in $Dom$ that is
also in $Dom'$ to itself.  Thus, for each subaction $A$ used in defining
a prophecy array variable, we can define
 \[ DomInj_{A} == [d \in Dom \cap Dom' |-> d]
 \]
A function $f$ such that $f[x]=x$ for all $x\in\DOMAIN f$ is called an
\emph{identity} function.  For convenience, the $Prophecy$ module
defines $IdFcn(S)$ to be the identify function with domain~$S$.  But
we won't bother using it here.  The $Prophecy$ module also defines
$EmptyFcn$ to be the (unique) function whose domain is the empty set.

Let us return to our prophecy data structure example.  For a subaction
$A$, we can define $NewPSet_{A}$ in terms of $DomInj_{A}$ and the
subset $PredDom_{A}$ of $Dom$, which consists of the elements in $Dom$
such that $p[d]$ makes a prediction about $A$.  We define these sets
$PredDom_{A}$ in module $SendSeqUndoP$ for our four subactions as
follows:
\begin{display}
\begin{notla}
PredDomChoose  == {}
PredDomSend    == {1}
PredDomRcv     == {}
PredDomUndo(i) == {i}
\end{notla}
\begin{tlatex}
\@x{ PredDomChoose\@s{4.35} \.{\defeq} \{ \}}%
\@x{ PredDomSend\@s{13.78} \.{\defeq} \{ 1 \}}%
\@x{ PredDomRcv\@s{18.71} \.{\defeq} \{ \}}%
\@x{ PredDomUndo ( i ) \.{\defeq} \{ i \}}%
\end{tlatex}
\end{display}
(Since $PredDom_{Undo(i)}$ depends on the identifier $i$ in its
context, we must define $PredDomUndo$ to have a parameter.)  

We can define $NewPSet_{A}$ to equal the set of all functions $q$ in
$[Dom'->\Pi]$ such that for every element $d$ in $Dom$ that is not in
$PredDom_{A}$ and has a corresponding element $DomInj_{A}[d]$ in
$Dom'$, the value of $q$ on that corresponding element equals
$p[d]$.  More precisely, $NewPSet_{A}$ equals:
 \[\begin{noj}
 \{\,q \in [Dom' -> \Pi] : \\ \s{2}
       \A d \in (\DOMAIN DomInj_{A}) :\: PredDom_A : 
             q[DomInj_{A}[d]] = p[d]  \,\}
  \end{noj}
 \]
We encapsulate definitions like this in a module $Prophecy$.  We find
it most convenient to make this a constant module with constant
parameters $Pi$, $Dom$, and $DomPrime$.  This module is meant to be
instantiated with the parameter $Pi$ instantiated by $\Pi$, with the
parameter $Dom$ instantiated by the appropriate state function $Dom$,
and with $DomPrime$ instantiated by $Dom'$.  The following definition
from module $Prophecy$ allows us to define $NewPSet_{A}(p)$
to equal
 \tlabox{NewPSet(p, DomInj_{A}, PredDom_{A})}:
\begin{display}
\begin{notla}
NewPSet(p, DomInj, PredDom) ==
  {  q \in [DomPrime -> Pi] :
         \A d \in (DOMAIN DomInj) \ PredDom : q[DomInj[d]] = p[d]  }
\end{notla}
\begin{tlatex}
\@x{ NewPSet ( p ,\, DomInj ,\, PredDom ) \.{\defeq}}%
\@x{\@s{8.2} \{\@s{4.1} q \.{\in} [ DomPrime \.{\rightarrow} Pi ] \.{:}}%
 \@x{\@s{30.98} \A\, d \.{\in} ( {\DOMAIN} DomInj ) \.{\,\backslash\,} PredDom
 \.{:} q [ DomInj [ d ] ] \.{=} p [ d ]\@s{4.1} \}}%
\end{tlatex}
\end{display}
\begin{sloppypar} \noindent
For each action in $A$ in our disjunctive decomposition of $NextU$, we
have written the definitions of $Pred_{A}$, $DomInj_{A}$, and
$PredDom_{A}$.  This allows us to define $NewPSet_{A}$, and therefore,
by (\ref{eq:ApDefFcn}), to define $A^{p}$.  The following operator
$ProphAction$ from the $Prophecy$ module allows us to write $A^{p}$
as
  \tlabox{ProphAction(A, p, p', DomInj_{A}, PredDom_{A}, Pred_{A})}:
\end{sloppypar}
\begin{display}
\begin{notla}
ProphAction(A, p, pPrime, DomInj, PredDom,  Pred(_)) ==
    A  /\  Pred(p)  /\  (pPrime \in NewPSet(p, DomInj, PredDom))
\end{notla}
\begin{tlatex}
 \@x{ ProphAction ( A ,\, p ,\, pPrime ,\, DomInj ,\, PredDom ,\,\@s{4.1} Pred
 ( \_ ) ) \.{\defeq}}%
 \@x{\@s{16.4} A\@s{4.1} \.{\land}\@s{4.1} Pred ( p )\@s{4.1}
 \.{\land}\@s{4.1} ( pPrime \.{\in} NewPSet ( p ,\, DomInj ,\, PredDom ) )}%
\end{tlatex}
\end{display}
In module $SendSeqUndoP$, we can define
\begin{display}
\begin{notla}
ChooseP == ProphAction(Choose, p, p', DomInjChoose, 
                       PredDomChoose, PredChoose)
\end{notla}
\begin{tlatex}
 \@x{ ChooseP \.{\defeq} ProphAction ( Choose ,\, p ,\, p \.{'} ,\,
 DomInjChoose ,\,}%
\@x{\@s{116.48} PredDomChoose ,\, PredChoose )}%
\end{tlatex}
\end{display}
The definitions of $SendP$ and $RcvP$ are similar.  However, there is
a problem with the definition of $Undo(i)^{p}$, which we write as
$UndoP(i)$.  Operator $ProphAction$ requires its last argument, which
represents $Pred_{A}$, to be an operator with a single argument.
However, we defined $PredUndo$ to have two arguments: $p$ and its
context identifier $i$.  Since we are defining $UndoP(i)$, the fifth
argument has to be an operator $Op$ so that $Op(p)$ equals
$PredUndo(p,i)$.  So, we should define:
\begin{display}
\begin{notla}
UndoP(i) == LET Op(j) == PredUndo(j, i)
            IN  ProphAction(Undo(i), p, p', DomInjUndo(i), 
                            PredDomUndo(i), Op)
\end{notla}
\begin{tlatex}
 \@x{ UndoP ( i ) \.{\defeq} \.{\LET} Op ( j ) \.{\defeq} PredUndo ( j ,\, i
 )}%
 \@x{\@s{61.83} \.{\IN} ProphAction ( Undo ( i ) ,\, p ,\, p \.{'} ,\,
 DomInjUndo ( i ) ,\,}%
\@x{\@s{141.36} PredDomUndo ( i ) ,\, Op )}%
\end{tlatex}
\end{display}
Using the \tlaplus\ \textsc{lambda} construct (added since 
\emph{Specifying Systems} was published), this can also be written 
as:
\begin{display}
\begin{notla}
UndoP(i) == 
    ProphAction( Undo(i), p, p', DomInjUndo(i), PredDomUndo(i), 
                 LAMBDA j : PredUndo(j, i))
\end{notla}
\begin{tlatex}
\@x{ UndoP ( i ) \.{\defeq}}%
 \@x{\@s{16.4} ProphAction (\, Undo ( i ) ,\, p ,\, p \.{'} ,\, DomInjUndo ( i )
 ,\, PredDomUndo ( i ) ,\,}%
\@x{\@s{75.52} \, {\LAMBDA} j \.{:} PredUndo ( j ,\, i ) \,)}%
\end{tlatex}
\end{display}
It's now straightforward to complete our definition of specification
$SpecUP$.  Doing so, gathering up the definitions made or implied
so far, and rearranging them a bit, we get the beginning of module
$SendSeqUndoP$ shown in
\lref{targ:SendSeqUndoP}{Figure~\ref{fig:SendSeqUndoP}}.

\begin{figure} \target{targ:SendSeqUndoP}
\begin{notla}
--------------------------- MODULE SendSeqUndoP ---------------------------
EXTENDS SendSeqUndo

Pi == {"send", "undo"}
Dom == DOMAIN y

INSTANCE Prophecy WITH DomPrime <- Dom'

PredDomChoose == {}
DomInjChoose  == [d \in Dom |-> d]
PredChoose(p) == TRUE

PredDomSend == {1}
DomInjSend  == [i \in 2..Len(y) |-> i-1]
PredSend(p) == p[1] = "send"

PredDomRcv == {}
DomInjRcv  == [d \in Dom |-> d]
PredRcv(p) == TRUE

PredDomUndo(i) == {i}
DomInjUndo(i)  == [j \in 1..Len(y) \ {i} |-> IF j < i THEN j ELSE j-1]
PredUndo(p, i) == p[i] = "undo"
-----------------------------------------------------------------------
VARIABLE p
varsP == <<vars, p>>

InitUP == Init /\ (p \in [Dom -> Pi])

ChooseP == ProphAction(Choose, p, p', 
                        DomInjChoose, PredDomChoose, PredChoose)

SendP == ProphAction(Send, p, p', DomInjSend, PredDomSend, PredSend)
      
RcvP == ProphAction(Rcv, p, p', DomInjRcv, PredDomRcv, PredRcv)

UndoP(i) == ProphAction(Undo(i), p, p', DomInjUndo(i), PredDomUndo(i), 
                        LAMBDA j : PredUndo(j, i))

NextUP == ChooseP \/ SendP \/ RcvP \/ (\E i \in 1..Len(y) : UndoP(i))

SpecUP == InitUP /\ [][NextUP]_varsP
\end{notla}
\begin{tlatex}
\@x{}\moduleLeftDash\@xx{ {\MODULE} SendSeqUndoP}\moduleRightDash\@xx{}%
\@x{ {\EXTENDS} SendSeqUndo}%
\@pvspace{8.0pt}%
\@x{ Pi \.{\defeq} \{\@w{send} ,\,\@w{undo} \}}%
\@x{ Dom \.{\defeq} {\DOMAIN} y}%
\@pvspace{8.0pt}%
\@x{ {\INSTANCE} Prophecy {\WITH} DomPrime \.{\leftarrow} Dom \.{'}}%
\@pvspace{8.0pt}%
\@x{ PredDomChoose \.{\defeq} \{ \}}%
\@x{ DomInjChoose\@s{7.14} \.{\defeq} [ d \.{\in} Dom \.{\mapsto} d ]}%
\@x{ PredChoose ( p )\@s{7.31} \.{\defeq} {\TRUE}}%
\@pvspace{8.0pt}%
\@x{ PredDomSend \.{\defeq} \{ 1 \}}%
 \@x{ DomInjSend\@s{7.14} \.{\defeq} [ i \.{\in} 2 \.{\dotdot} Len ( y )
 \.{\mapsto} i \.{-} 1 ]}%
\@x{ PredSend ( p )\@s{7.31} \.{\defeq} p [ 1 ] \.{=}\@w{send}}%
\@pvspace{8.0pt}%
\@x{ PredDomRcv \.{\defeq} \{ \}}%
\@x{ DomInjRcv\@s{7.14} \.{\defeq} [ d \.{\in} Dom \.{\mapsto} d ]}%
\@x{ PredRcv ( p )\@s{7.31} \.{\defeq} {\TRUE}}%
\@pvspace{8.0pt}%
\@x{ PredDomUndo ( i ) \.{\defeq} \{ i \}}%
 \@x{ DomInjUndo ( i )\@s{7.14} \.{\defeq} [ j \.{\in} 1 \.{\dotdot} Len ( y )
 \.{\,\backslash\,} \{ i \} \.{\mapsto} {\IF} j \.{<} i \.{\THEN} j \.{\ELSE}
 j \.{-} 1 ]}%
\@x{ PredUndo ( p ,\, i )\@s{8.98} \.{\defeq} p [ i ] \.{=}\@w{undo}}%
\vspace{-2pt}%
\@x{}\midbar\@xx{}%
\vspace{-2pt}%
\@x{ {\VARIABLE} p}%
\@x{ varsP \.{\defeq} {\langle} vars ,\, p {\rangle}}%
\@pvspace{8.0pt}%
 \@x{ InitUP \.{\defeq} Init \.{\land} ( p \.{\in} [ Dom \.{\rightarrow} Pi ]
 )}%
\@pvspace{8.0pt}%
\@x{ ChooseP \.{\defeq} ProphAction ( Choose ,\, p ,\, p \.{'} ,\,}%
\@x{\@s{120.58} DomInjChoose ,\, PredDomChoose ,\, PredChoose )}%
\@pvspace{8.0pt}%
 \@x{ SendP \.{\defeq} ProphAction ( Send ,\, p ,\, p \.{'} ,\, DomInjSend ,\,
 PredDomSend ,\, PredSend )}%
\@pvspace{8.0pt}%
 \@x{ RcvP \.{\defeq} ProphAction ( Rcv ,\, p ,\, p \.{'} ,\, DomInjRcv ,\,
 PredDomRcv ,\, PredRcv )}%
\@pvspace{8.0pt}%
 \@x{ UndoP ( i ) \.{\defeq} ProphAction ( Undo ( i ) ,\, p ,\, p \.{'} ,\,
 DomInjUndo ( i ) ,\, PredDomUndo ( i ) ,\,}%
\@x{\@s{120.96} {\LAMBDA} j \.{:} PredUndo ( j ,\, i ) )}%
\@pvspace{8.0pt}%
 \@x{ NextUP \.{\defeq} ChooseP \.{\lor} SendP \.{\lor} RcvP \.{\lor} ( \E\, i
 \.{\in} 1 \.{\dotdot} Len ( y ) \.{:} UndoP ( i ) )}%
\@pvspace{8.0pt}%
\@x{ SpecUP\@s{1.08} \.{\defeq} InitUP \.{\land} {\Box} [ NextUP ]_{ varsP}}%
\end{tlatex}
\caption{The specification of \emph{SpecUP}.} \label{fig:SendSeqUndoP}
\end{figure}

Finally, we have to define the refinement mapping under which $SpecUP$
implements specification $Spec$ of module $SendSeq$.  The idea is
simple: we let \ov{y} be the subsequence of $y$ containing only those
elements for which the corresponding element of the sequence $p$
equals $"send"$.  The following formal definition is a bit tricky.  
It uses a local recursive definition of an operator $R$ such that
if $yseq$ is any sequence and $pseq$ is a sequence of the same length,
then $R(yseq, pseq)$ is the subsequence of $yseq$ that contains
$yseq[i]$ iff $pseq[i]$ equals $"send"$.
\begin{display}
\begin{notla}
yBar ==  
   LET RECURSIVE R(_, _)
       R(yseq, pseq) ==
          IF yseq = << >> 
            THEN yseq
            ELSE IF Head(pseq) = "send" 
                   THEN <<Head(yseq)>> \o R(Tail(yseq), Tail(pseq))
                   ELSE  R(Tail(yseq), Tail(pseq))
   IN  R(y, p) 
\end{notla}
\begin{tlatex}
\@x{ yBar \.{\defeq}}%
\@x{\@s{12.29} \.{\LET} {\RECURSIVE} R ( \_ ,\, \_ )}%
\@x{\@s{32.79} R ( yseq ,\, pseq ) \.{\defeq}}
\@x{\@s{52.46} {\IF} yseq \.{=} {\langle}\, {\rangle}}%
\@x{\@s{60.66} \.{\THEN} yseq}%
\@x{\@s{60.66} \.{\ELSE} {\IF} Head ( pseq ) \.{=}\@w{send}}%
 \@x{\@s{100.18} \.{\THEN} {\langle} Head ( yseq ) {\rangle} \.{\circ} R (
 Tail ( yseq ) ,\, Tail ( pseq ) )}%
\@x{\@s{100.18} \.{\ELSE} R ( Tail ( yseq ) ,\, Tail ( pseq ) )}%
\@x{\@s{12.29} \.{\IN} R ( y ,\, p )}%
\end{tlatex}
\end{display}
We then instantiate module $SendSeq$ and state as follows the theorem
asserting that $SpecUP$ implements formula $Spec$ of that model under
the refinement mapping.
\begin{display}
\begin{notla}
SS == INSTANCE SendSeq WITH y <- yBar

THEOREM SpecUP => SS!Spec
\end{notla}
\begin{tlatex}
\@x{ SS \.{\defeq} {\INSTANCE} SendSeq {\WITH} y \.{\leftarrow} yBar}%
\@pvspace{2.0pt}%
\@x{ {\THEOREM} SpecUP \.{\implies} SS {\bang} Spec}%
\end{tlatex}
\end{display}
TLC can check this theorem in the usual way.

\subsection{Checking the Definitions}

We have shown how to define a specification $Spec^{p}$ for an
arbitrary specification $Spec$ by defining a state function $Dom$ and,
for every subaction $A$ of a disjunctive representation of the
next-state action of $Spec$, defining $Pred_{A}$, $DomInj_{A}$, and
$PredDom_{A}$.  These definitions must satisfy certain conditions
to ensure that \tlabox{\EE p:Spec^{p}} is equivalent to $Spec$.
We now state those conditions.

The first condition is (\ref{eq:ProphFcn}).  The $Prophecy$ module
defines this operator:
\begin{display}
\begin{notla}
ExistsGoodProphecy(Pred(_)) == \E q \in [Dom -> Pi] : Pred(q)
\end{notla}
\begin{tlatex}
 \@x{ ExistsGoodProphecy ( Pred ( \_ ) ) \.{\defeq} \E\, q \.{\in} [ Dom
 \.{\rightarrow} Pi ] \.{:} Pred ( q )}%
\end{tlatex}
\end{display}
For a subaction $A$ with an empty context, we can write
(\ref{eq:ProphFcn}) as
 \[ Spec\; => \; [][ 
  A => (ExistsGoodProphecy(Pred_{A})]_{vars}
 \]
(Remember that the $Prophecy$ module will be instantiated with the
appropriate expression substituted for $Dom$.)  
To see how this
definition is used if $A$ has a non-empty context, here is how
condition is (\ref{eq:ProphFcn}) is expressed for the subaction
$UndoP(i)$ of specification $SpecU$ in our $SendSeq$ example:
\begin{display}
\begin{notla}
SpecU  =>  [][\,\A i \in Dom : 
                Undo(i) => 
                  ExistsGoodProphecy( LAMBDA p : PredUndo(p,i))]_vars
\end{notla}
\begin{tlatex}
 \@x{ SpecU\@s{4.1} 
      \.{\implies}\@s{4.1} {\Box} [\, \A\, i \.{\in} Dom \.{:}}%
\@x{\@s{70.18} Undo ( i ) \.{\implies}}%
 \@x{\@s{78.38} ExistsGoodProphecy ( {\LAMBDA} p \.{:} PredUndo ( p ,\, i ) )
\, ]_{ vars}}%
\end{tlatex}
\end{display}
The only condition we require of $DomInj_{A}$ is that it be a partial
function from $Dom$ to $Dom'$.  This is expressed as
$IsDomInj(DomInj_{A})$ using this definition from module
$Prophecy$
\begin{display}
\begin{notla}
IsDomInj(DomInj) == DomInj \in PartialInjections(Dom, DomPrime)
\end{notla}
\begin{tlatex}
 \@x{ IsDomInj ( DomInj ) \.{\defeq} DomInj \.{\in} PartialInjections ( Dom
 ,\, DomPrime )}%
\end{tlatex}
\end{display}
As with the $ExistsGoodProphecy$ condition, it needs to hold only
for $A$ steps in a behavior satisfying the specification $Spec$.
Hence the general requirement on $DomInj_{A}$ for an action $A$ with
context $<<\textbf{k};\textbf{K}>>$ is
 \[ Spec\; => 
  \; [][\A <<\textbf{k};\textbf{K}>> : 
           A => IsDomInj(DomInj_{A})]_{vars}
 \]
Because $IsDomInj$ does not have an operator argument, no local
definition or \textsc{lambda} expression is needed even if the context
is nonempty.  For example, if $A$ is subaction $Undo(i)$ of
specification $SpecU$ of the $SendSeq$ example, this condition is
written:
 \[ \A i \in Dom : Undo(i) => IsDomInj(DomInjUndo(i))
 \]
Finally, we need a condition on $PredDom_{A}$.  Remember that 
$PredDom_{A}$ should equal the set of elements $d$ of $Dom$ such
that $p[d]$ is making predictions about $A$.  Actually, it suffices
that $PredDom_{A}$ contain all such elements.  (It may contain
other elements as well.)   This is equivalent to
the requirement that any element not in $PredDom_{A}$ does not make a
prediction about $A$.  Making a prediction about $A$ means affecting
the value of $Pred_{A}$, so not making a prediction means not affecting
its value.  Thus, $p[d]$ does not make a prediction about 
$A$ iff setting $p[d]$ to any value in $\Pi$ does not change the
value of $Pred_{A}$.  You should be able to convince yourself
that the value of $Pred_{A}$ does not depend
on the value of $p[d]$ for any $d$ not in $PredDom_{A}$ iff the
following formula is true:
  \[ \begin{noj}
     \A q, r \in [Dom -> Pi] : \\ \s{1}
          (\A d \in PredDom_{A} : q[d] = r[d]) => (Pred_{A}(q) = Pred_{A}(r))
     \end{noj}
 \]
In addition to this requirement, to ensure that our formulas make
sense, we also make the obvious requirement that $PredDom_{A}$ is a
subset of $Dom$.  The following definition appears in module $Prophecy$.
\begin{display}
\begin{notla}
IsPredDom(PredDom, Pred(_)) ==
  /\ PredDom \subseteq Dom
  /\ \A q, r \in [Dom -> Pi] : 
          (\A d \in PredDom : q[d] = r[d]) => (Pred(q) = Pred(r))
\end{notla}
\begin{tlatex}
\@x{ IsPredDom ( PredDom ,\, Pred ( \_ ) ) \.{\defeq}}%
\@x{\@s{8.2} \.{\land} PredDom \.{\subseteq} Dom}%
\@x{\@s{8.2} \.{\land} \A\, q ,\, r \.{\in} [ Dom \.{\rightarrow} Pi ] \.{:}}%
 \@x{\@s{36.11} ( \A\, d \.{\in} PredDom \.{:} q [ d ] \.{=} r [ d ] )
 \.{\implies} ( Pred ( q ) \.{=} Pred ( r ) )}%
\end{tlatex}
\end{display}
Remembering that the condition on $PredDom_{A}$ needs to hold only for
$A$ steps in a behavior satisfying the specification, we can express
it for an action $A$ with an empty context as:
 \[ Spec \;=>\; [][A => IsPredDom(PredDom_{A}, Pred_{A})]_{vars}
 \]
Because the second argument of $IsPredDom$ is an operator argument,
we again need to use a local definition or a \textsc{lambda} expression
to express the condition if the subaction $A$ has a nonempty context.
For example, here is how we can express it for the $Undo(i)$
subaction in our $SendSeq$ example using a local definition:
%
\begin{display}
\begin{notla}
[][\A i \in Dom : 
        Undo(i)  =>  LET Op(p) == PredUndo(p, i)
                     IN  IsPredDom(PredDomUndo, Op)]_vars
\end{notla}
\begin{tlatex}
\@x{ {\Box} [\, \A\, i \.{\in} Dom \.{:}}%
 \@x{\@s{17.47} Undo ( i ) \,\.{\implies} \, \.{\LET} Op ( p ) \.{\defeq} PredUndo
 ( p ,\, i )}%
\@x{\@s{68.79} \,\, \.{\IN} IsPredDom ( PredDomUndo ,\, Op ) \,]_{ vars}}%
\end{tlatex}
\end{display}
Here is how it is written with a \textsc{lambda} expression:
\begin{display}
\begin{notla}
[][\A i \in Dom : 
       Undo(i) => IsPredDom(PredDomUndo, 
                            LAMBDA p : PredUndo(p, i))]_vars
\end{notla}
\begin{tlatex}
\@x{ {\Box} [ \,\A\, i \.{\in} Dom \.{:}}%
\@x{\@s{21.57} Undo ( i ) \,\.{\implies}\, IsPredDom (\, PredDomUndo ,\,}%
\@x{\@s{126.02} \,\,\,{\LAMBDA} p \.{:} PredUndo ( p ,\, i ) \,) \,]_{ vars}}%
\end{tlatex}
\end{display}
The following definition from module $Prophecy$ allows us
to combine the three conditions.
\begin{display}
\begin{notla}
ProphCondition(A, DomInj, PredDom,  Pred(_))  ==
     A  =>  /\ ExistsGoodProphecy(Pred)
            /\ IsDomInj(DomInj)
            /\ IsPredDom(PredDom, Pred)
\end{notla}
\begin{tlatex}
 \@x{ ProphCondition ( A ,\, DomInj ,\, PredDom ,\,\@s{4.1} Pred ( \_ )
 )\@s{4.1} \.{\defeq}}%
 \@x{\@s{20.5} A\@s{4.1} \.{\implies}\@s{4.1} \.{\land} ExistsGoodProphecy (
 Pred )}%
\@x{\@s{51.68} \.{\land} IsDomInj ( DomInj )}%
\@x{\@s{51.68} \.{\land} IsPredDom ( PredDom ,\, Pred )}%
\end{tlatex}
\end{display}
Using this definition, the requirements on the definitions are
expressed for specification $SendSeqUndo$ in
\lref{targ:proph-cond}{Figure~\ref{fig:proph-cond}}.%
%
\begin{figure} \target{targ:proph-cond}
\begin{notla}
Condition == 
  /\ ProphCondition(Choose, DomInjChoose, PredDomChoose, PredChoose)
  /\ ProphCondition(Send,   DomInjSend,   PredDomSend,   PredSend)
  /\ ProphCondition(Rcv,    DomInjRcv,    PredDomRcv,    PredRcv)
  /\ \A i \in Dom : 
         ProphCondition(Undo(i), DomInjUndo(i), PredDomUndo(i),  
                        LAMBDA p : PredUndo(p, i))

THEOREM SpecU => [][Condition]_vars
\end{notla}
\begin{tlatex}
\@x{ Condition \.{\defeq}}%
 \@x{\@s{8.2} \.{\land} ProphCondition ( Choose ,\, DomInjChoose ,\,
 PredDomChoose ,\, PredChoose )}%
 \@x{\@s{8.2} \.{\land} ProphCondition ( Send ,\,\@s{9.4} DomInjSend 
 ,\,\@s{9.4} PredDomSend ,\,\@s{9.4} PredSend )}%
 \@x{\@s{8.2} \.{\land} ProphCondition ( Rcv ,\,\@s{14.2} DomInjRcv 
 ,\,\@s{14.3} PredDomRcv ,\,\@s{14.2} PredRcv )}%
\@x{\@s{8.2} \.{\land} \A\, i \.{\in} Dom \.{:}}%
 \@x{\@s{30.63} ProphCondition ( Undo ( i ) ,\, DomInjUndo ( i ) ,\,
 PredDomUndo ( i ) ,\,}%
\@x{\@s{104.29} \,{\LAMBDA} p \.{:} PredUndo ( p ,\, i ) )}%
\@pvspace{8.0pt}%
\@x{ {\THEOREM} SpecU \.{\implies} {\Box} [ Condition ]_{ vars}}%
\end{tlatex}
\caption{Action requirements for specification \emph{SendSeqUndo}.} 
  \label{fig:proph-cond}
\end{figure}

We encounter the same problem here that we encountered in checking
condition (\ref{eq:ProphFcn}) for the $SendSet$ example.  We would
like to put these requirements in module $SendSetUndoP$
(\lref{targ:SendSetUndoP}{Figure~\ref{fig:SendSetUndoP}}), right
before the declaration of the variable $p$.  However, TLC can't check
the theorem in a model for that specification because $SpecU$ does not
specify the values of variable $p$.  We can either move all the
definitions that are now in $SendSetUndoP$ before the declaration of
$p$ into a separate module, put them at the end of module
$SendSetUndo$, or end the module before the declaration of $p$
by adding ``\verb|=====|'' when checking the condition.

You should check these conditions when adding a prophecy variable.
They provide a good way to debug your definitions, before you try
checking that the specification with prophecy variable implements the
desired specification.

\subsection{Liveness}

As with our other auxiliary variables, we add a prophecy variable to
the safety part of a specification, but we keep the liveness part the
same.  As we remarked in Section~\ref{sec:history-liveness}, this
produces unusual specifications in which the liveness property can
assert a fairness condition about an action that isn't a subaction of
the next-state action.

For history variables, although the form of the specifications is
unusual, the specifications are not.  This is not the case for
prophecy variables.  If $Spec$ has a liveness condition, the
specification $Spec^{p}$ obtained from it by adding a prophecy
variable can be weird.  As an example, suppose that in the $SendInt$
specifications of Section~\ref{sec:simple-proph}, instead of taking $Pi$ to
equal $Int$, we let it equal $Int \cup \{\infty\}$ for some value
$\infty\notin Int$.  Everything we did would work exactly as before,
and the theorem at the end of module $SendInt1P$ in
\lref{targ:SendInt1P}{Figure~\ref{fig:SendInt1P}} would still be true.
If a $SendP$ step set $p'$ to $\infty$, predicting that the next value
to be sent is $\infty$, then the system would simply halt (stutter
forever) before the next $SendP$ step because $Send /\ PredSend(\infty)$
equals \FALSE.

Now suppose we add a liveness condition $\WF_{vars}(Next)$ to
our $SendInt$ specifications, requiring that they never halt.
We would then have
 \[ SpecP == InitP /\ [][NextP]_{varsP} /\ \WF_{vars}(Next)\]
This formula $SpecP$ implies that infinitely many $Next$ steps must
occur, so a behavior can't halt.  The weak fairness conjunct therefore
requires that $SpecP$ not set $p'$ to $\infty$.  This is weird.
The liveness property $\WF_{vars}(Next)$ doesn't just require that
something must eventually happen; it also prevents something (setting
$p$ to $\infty$) from ever happening.  The technical term for this
weirdness is that the formula $SpecP$ is \emph{not
machine closed}~\cite{abadi:existence}, which means that its liveness
property affects safety as well as liveness.  

Non-machine closed specs should never be used to describe \emph{how} a
system works.  You can't understand how to implement a system if the
next-state action doesn't describe what it can and cannot do next.  In
\emph{extremely} rare cases, a non-machine closed high-level spec is
the best way to describe \emph{what} a system should do, rather than
how it should do it.  You are very unlikely to encounter such a
situation in practice.

While the non-machine closed spec we get by adding a prophecy variable
to a spec with liveness can be weird, this weirdness causes no
problem.  We don't have to implement $SpecP$.   We use it only to check
the correctness of $Spec$; and the presence of the liveness property
makes no difference in what we do.  With our modified $SendInt$
specifications, we can check that $SpecP$ implies $SI2!Spec$ exactly as
we did before.  

\section{Stuttering Variables}

\subsection{Adding Stuttering Steps to a Simple Action}

Suppose $Spec_{1}$ is a specification of a (24-hour) clock that
displays only the hour---a specification we can write as
\begin{equation}
Spec_{1} == (h = 0) /\ [][h' = (h+1)\,\%\, 24]_{h}
\NOTLA \label{eq:HourClock}
\end{equation}
Let $Spec_{2}$ be this specification of an hour-minute clock:
\begin{equation}
Spec_{2} == \begin{conj}
                (h=0) /\ (m=0) \\
                [][ \;\begin{conj}
                    m' = (m+1)\,\%\,60 \\
                    h' = \IF m'=0 \THEN (h+1)\,\%\,24 \LSE h \;]_{<<h,m>>}
                    \end{conj}
                \end{conj}
\NOTLA \label{eq:HourMinClock}
\end{equation}
If we ignore the variable $m$ in $Spec_{2}$, then we get a clock that
displays only the hour.  Thus, \tlabox{\EE m:Spec_{2}} should be
equivalent to $Spec_{1}$.  (The 59 steps each hour that change only
$m$ are stuttering steps that are allowed by $Spec_{1}$.)  It's easy to
see that $Spec_{2}$ implies $Spec_{1}$, so \tlabox{\EE m:Spec_{2}}
implies $Spec_{1}$.  There is no refinement mapping with which we can
prove that $Spec_{1}$ implies \tlabox{\EE m:Spec_{2}}.  To construct
the necessary refinement mapping we need to add an auxiliary variable
$s$ to $Spec_{1}$ to obtain a specification $Spec_{1}^{s}$  that adds
59 steps that change only $s$ to each step that increments $h$.  Such
an auxiliary variable is called a \emph{stuttering} variable because
it changes $Spec_{1}$ only by requiring it to add steps that leave
the variable $h$ of $Spec_{1}$ unchanged.

To add such a variable $s$ to a specification $Spec$ to form
$Spec^{s}$, we let the next-state action of $Spec^{s}$ take
``normal'' steps that satisfy the next-state action of $Spec$ when $s$
equals $\top$ (usually read ``top''), which is some value that is not
a positive integer.  The value of $s$ in the initial state equals $\top$.
When $s$ is set to a positive integer, the
specification $Spec^{s}$ allows only stuttering steps that decrement
$s$, leaving the variables of $Spec$ unchanged.  When $s$ counts down
to zero, it is set equal to $\top$ again.  We add these stuttering
steps before and/or after steps of some particular subaction of the
next-state action.  Here, we assume that this is a ``simple''
subaction, meaning that its context is empty.

Suppose we want the specification $Spec^{s}$ to add stuttering steps
to each step of a particular subaction.  We replace each
subaction $A$ by the action $A^{s}$ defined as follows.  For each $A$
other than that particular subaction, we define $A^{s}$ by:
\begin{equation}
 A^{s} == (s = \top) \,/\ \, A \,/\ \, (s'=s)
\NOTLA\label{eq:As}
\end{equation}
To add $initVal$ stuttering steps after a step of an
action $A$, for a positive
integer $initVal$ (whose value may depend on the variables of $Spec$),
we define 
 \[
  A^{s} == \begin{noj}
            \IF{s = \top} \\ \s{.5}\begin{noj}
                          \THEN A \,/\ \, (s'=initVal)  \\
              \ELSE \begin{conj}
                    vars'=vars \\
                  s' = \IF{s=1}\THEN \top \LSE s-1
                    \end{conj}
                          \end{noj}
            \end{noj}
 \]
We can generalize
this by replacing the set of natural numbers 
with an arbitrary set $\Sigma$ having
a well-founded partial order $\prec$ with smallest element $\bot$ (read
``bottom'')\footnote{
This means that $\bot \prec \sigma$ for all $\sigma\in\Sigma$,
and any decreasing chain $\sigma_{1} \succ \sigma_{2} \succ \ldots$
of elements in $\Sigma$
must be finite.},
letting $initVal$ be an arbitrary element of $\Sigma$, replacing
$s=1$ with $s=\bot$, and replacing $s-1$ by $decr(s)$ for some
operator $decr$ such that \tlabox{decr(s)\prec s} for
all $s\in \Sigma$.  The generalization is:
   \begin{equation}
  A^{s} == \begin{noj}
            \IF{s = \top} \\ \s{.5}\begin{noj}
                          \THEN A \,/\ \, (s'=initVal)  \\
              \ELSE \begin{conj}
                    vars'=vars \\
                  s' = \IF{s=\bot}\THEN \top \LSE decr(s) 
                    \end{conj}
                          \end{noj}
            \end{noj}
   \NOTLA \label{eq:A0sPost}
  \end{equation}
We can add $initVal$ stuttering steps before an $A$ step, rather than
after it, as follows.  The stuttering steps should only be taken when
they can be followed by an $A$ step, which is the case only when an $A$
step is enabled.  Remembering that $\ENABLED A$ is the state
predicate that is true iff an $A$ step is enabled, the stuttering
steps can be added with this definition of $A^{s}$:
\begin{equation}
    A^{s} ==
     \begin{conj}
     \ENABLED A\\
     \IF{s = \bot} \begin{noj}
                   \THEN A \,/\ \, (s'=\top) \\
                   \ELSE \begin{conj}
                         vars' = vars \\
                         s' = \IF{s=\top}\THEN initVal\LSE decr(s)
                         \end{conj}
                   \end{noj}
     \end{conj}
  \NOTLA \label{eq:A0sPre}
\end{equation}
We could generalize (\ref{eq:A0sPost}) and (\ref{eq:A0sPre}) to allow
putting stuttering steps both before and after an $A$ step.  We
won't bother to do this because it is probably seldom needed, and it
wouldn't be significantly simpler than adding two separate stuttering
variables.

\subsection{Adding Stuttering Steps to Multiple Actions}

To generalize what we did in the preceding section, we first consider
how to add stuttering steps before or after an action $A$ that may
have a nonempty context.  We assume that the set $\Sigma$, its $\bot$
element, and the operator $decr$ do not depend on the values of the
context variables.  (This should be true in most real examples.)
However, $initVal$ may depend on them.  We therefore must make sure
that $initVal$ is evaluated with the values of the context variables
for which the action $A$ is ``executed''.  This is no problem for
stuttering steps added after the $A$ step, where $A^{s}$ is defined by
(\ref{eq:A0sPost}).  (Since the stuttering steps do nothing but
decrement the value of $s$, it makes no difference for which value of
the context variables $A^{s}$ is ``executed'' when $s#\top$.)
However, it is a problem for stuttering steps added before an $A$
step, where $A^{s}$ is defined by (\ref{eq:A0sPre}).

To solve the problem for $A^{s}$ defined by (\ref{eq:A0sPre}), we let
the non-$\top$ values of $s$ be records with a $val$ component that
equals the value of $s$ described in (\ref{eq:A0sPre}), and a $ctxt$
component that equals the tuple of values of the context when
$initVal$ is evaluated.  In other words, $ctxt$ is set by the
\textsc{else} clause in the second conjunct of (\ref{eq:A0sPre}).  The
condition that $s.ctxt$ equals the values of the context variables is
added as a conjunct to the \textsc{then} clause to make sure that $A$
is executed only in that context.  The precise definition is given
below.

We often need to add stuttering steps before or after more than one
subaction of the next-state action.  We could do that by adding
separate stuttering variables, or we could introduce a stuttering
array variable.  However, because the stuttering steps we add to an
action all occur immediately before or after that subaction, we can
add them all with the same stuttering variable~$s$.  Stuttering steps
can be added to each such subaction with its own set $\Sigma$ and
hence its own values of $initVal$ and $\bot$ and of the operator
$decr$.  We just let the value of $s$ indicate the action to which the
stuttering steps are being added.  We do this by adding to the
non-$\top$ values of $s$ an additional $id$ component that identifies
the action for which the stuttering steps are being added.  The
component is set when $s$ is first set to a non-$\top$ value, and
execution of the new subaction $A^{s}$ is enabled when $s#\top$ only
if $s.id$ equals the identifier of $A$.

We write the three definitions (\ref{eq:As}), (\ref{eq:A0sPost}), and
(\ref{eq:A0sPre}) in \tlaplus\ using the three operators $NoStutter$,
$PostStutter$, and $PreStutter$, respectively, shown in
\lref{targ:Stuttering}{Figure~\ref{fig:Stuttering}}.  
\begin{figure} \target{targ:Stuttering}
\begin{notla}
----------------------------- MODULE Stuttering ----------------------------
top == [top |-> "top"]

VARIABLES s, vars

NoStutter(A) == (s = top) /\ A /\ (s' = s)

PostStutter(A, actionId, context, bot, initVal, decr(_)) ==
  IF s = top THEN /\ A 
                  /\ s' = [id |-> actionId,  ctxt |-> context, val |-> initVal]
             ELSE /\ s.id = actionId
                  /\ UNCHANGED vars 
                  /\ s'= IF s.val = bot THEN top 
                                        ELSE [s EXCEPT !.val = decr(s.val)]

PreStutter(A, enabled, actionId, context, bot, initVal, decr(_)) == 
  IF s = top 
    THEN /\ enabled
         /\ UNCHANGED vars 
         /\ s' = [id |-> actionId, ctxt |-> context, val |-> initVal] 
    ELSE /\ s.id = actionId 
         /\ IF s.val = bot THEN /\ s.ctxt = context
                                /\ A 
                                /\ s' = top
                           ELSE /\ UNCHANGED vars  
                                /\ s' = [s EXCEPT !.val = decr(s.val)]  
=============================================================================
\end{notla}
\begin{tlatex}
\@x{}\moduleLeftDash\@xx{ {\MODULE} Stuttering}\moduleRightDash\@xx{}%
\@x{ top \.{\defeq} [ top \.{\mapsto}\@w{top} ]}%
\@pvspace{8.0pt}%
\@x{ {\VARIABLES} s ,\, vars}%
\@pvspace{8.0pt}%
 \@x{ NoStutter ( A ) \.{\defeq} ( s \.{=} top ) \.{\land} A \.{\land} ( s
 \.{'} \.{=} s )}%
\@pvspace{8.0pt}%
 \@x{ PostStutter ( A ,\, actionId ,\, context ,\, bot ,\, initVal ,\, decr (
 \_ ) ) \.{\defeq}}%
\@x{\@s{8.2} {\IF} s \.{=} top \.{\THEN} \.{\land} A}%
 \@x{\@s{84.08} \.{\land} s \.{'} \.{=} [ id \.{\mapsto} actionId ,\,\@s{4.1}
 ctxt \.{\mapsto} context ,\, val \.{\mapsto} initVal ]}%
\@x{\@s{52.77} \.{\ELSE} \.{\land} s . id \.{=} actionId}%
\@x{\@s{84.08} \.{\land} {\UNCHANGED} vars}%
\@x{\@s{84.08} \.{\land} s \.{'} \.{=} {\IF} s . val \.{=} bot \.{\THEN} top}%
 \@x{\@s{176.19} \.{\ELSE} [ s {\EXCEPT} {\bang} . val \.{=} decr ( s . val )
 ]}%
\@pvspace{8.0pt}%
 \@x{ PreStutter ( A ,\, enabled ,\, actionId ,\, context ,\, bot ,\, initVal
 ,\, decr ( \_ ) ) \.{\defeq}}%
\@x{\@s{8.2} {\IF} s \.{=} top}%
\@x{\@s{16.4} \.{\THEN} \.{\land} enabled}%
\@x{\@s{47.71} \.{\land} {\UNCHANGED} vars}%
 \@x{\@s{47.71} \.{\land} s \.{'} \.{=} [ id \.{\mapsto} actionId ,\, ctxt
 \.{\mapsto} context ,\, val \.{\mapsto} initVal ]}%
\@x{\@s{16.4} \.{\ELSE} \.{\land} s . id \.{=} actionId}%
 \@x{\@s{47.71} \.{\land} {\IF} s . val \.{=} bot \.{\THEN} \.{\land} s . ctxt
 \.{=} context}%
\@x{\@s{150.07} \.{\land} A}%
\@x{\@s{150.07} \.{\land} s \.{'} \.{=} top}%
\@x{\@s{118.76} \.{\ELSE} \.{\land} {\UNCHANGED} vars}%
 \@x{\@s{150.07} \.{\land} s \.{'} \.{=} [ s {\EXCEPT} {\bang} . val \.{=}
 decr ( s . val ) ]}%
\end{tlatex}

\caption{The beginning of the \emph{Stuttering} module.} \label{fig:Stuttering}
\end{figure}
The module should be instantiated by substituting the new stuttering
variable for $s$ and the tuple of variables of the original
specification for $vars$.  It defines $\top$, which is written $top$,
to be a value that is different from the values assigned to $s$ by
$PostStutter$ and $PreStutter$ actions (and that TLC knows is
different from those values).  The other values in (\ref{eq:A0sPost})
and (\ref{eq:A0sPre}) are provided by the following arguments to
$PostStutter$ and $PreStutter$.
\begin{describe}{$actionId$}
\item[$A$] The action to which stuttering steps
are being added.

\item[$id$] An identifier to distinguish that action $A$ from other
actions to which stuttering steps are added.  We like to let it be the
name of the action (which is a string).

\item[$bot$] The $\bot$ (smallest) element of $\Sigma$.

\item[$initVal$] The value we have been calling by that name.

\item[$decr$] The operator we have been calling by that name.  
It must take a single argument.

\item[$enabled$] A formula that should be equivalent to $\ENABLED A$.
We can often find such a formula that TLC can evaluate much more
efficiently than $\ENABLED A$.  You can use TLC to check that
$enabled$ is equivalent to $\ENABLED A$ by checking that 
 $enabled \equiv \ENABLED A$ 
is an invariant of the original specification.

\item[$context$] The tuple of context identifiers of $A$.  (You can use
$i$ instead of a 1-tuple $<<i>>$.)  The $context$ argument is used in
the $PostStutter$ action only to set the $ctxt$ component of~$s$.
This component may be used in defining refinement mappings.
\end{describe}
Note that $PostStutter$ and $PreStutter$ add at least one stuttering
step, adding exactly one such step if $initVal=bot$.  It is often more
convenient to use operators that add one fewer stuttering step.  These
are the operators $MayPostStutter$ and $MayPreStutter$ defined in the
$Stuttering$ module as shown in
\lref{targ:Stuttering2}{Figure~\ref{fig:Stuttering2}}.
\begin{figure} \target{targ:Stuttering2}
\begin{notla}
MayPostStutter(A, actionId, context, bot, initVal, decr(_)) ==
  IF s = top THEN /\ A 
                  /\ s' = IF initVal = bot
                            THEN s
                            ELSE [id |-> actionId, ctxt |-> context, 
                                  val |-> initVal]
             ELSE /\ s.id = actionId
                  /\ UNCHANGED vars 
                  /\ s'= IF decr(s.val) = bot 
                           THEN top 
                           ELSE [s EXCEPT !.val = decr(s.val)]

MayPreStutter(A, enabled, actionId, context, bot, initVal, decr(_)) == 
  IF s = top 
    THEN /\ enabled
         /\ IF initVal = bot
              THEN A /\ (s' = s)
              ELSE /\ UNCHANGED vars 
                   /\ s' = [id |-> actionId, ctxt |-> context, 
                            val |-> decr(initVal)] 
    ELSE /\ s.id = actionId 
         /\ IF s.val = bot THEN /\ s.ctxt = context
                                /\ A 
                                /\ s' = top
                           ELSE /\ UNCHANGED vars  
                                /\ s' = [s EXCEPT !.val = decr(s.val)]  
============================================================================
\end{notla}
\begin{tlatex}
 \@x{ MayPostStutter ( A ,\, actionId ,\, context ,\, bot ,\, initVal ,\, decr
 ( \_ ) ) \.{\defeq}}%
\@x{\@s{8.2} {\IF} s \.{=} top \.{\THEN} \.{\land} A}%
\@x{\@s{84.08} \.{\land} s \.{'} \.{=} {\IF} initVal \.{=} bot}%
\@x{\@s{124.44} \.{\THEN} s}%
 \@x{\@s{124.44} \.{\ELSE} [ id \.{\mapsto} actionId ,\, ctxt \.{\mapsto}
 context ,\,}%
\@x{\@s{158.53} val \.{\mapsto} initVal ]}%
\@x{\@s{52.77} \.{\ELSE} \.{\land} s . id \.{=} actionId}%
\@x{\@s{84.08} \.{\land} {\UNCHANGED} vars}%
\@x{\@s{84.08} \.{\land} s \.{'} \.{=} {\IF} decr ( s . val ) \.{=} bot}%
\@x{\@s{124.44} \.{\THEN} top}%
 \@x{\@s{124.44} \.{\ELSE} [ s {\EXCEPT} {\bang} . val \.{=} decr ( s . val )
 ]}%
\@pvspace{8.0pt}%
 \@x{ MayPreStutter ( A ,\, enabled ,\, actionId ,\, context ,\, bot ,\,
 initVal ,\, decr ( \_ ) ) \.{\defeq}}%
\@x{\@s{8.2} {\IF} s \.{=} top}%
\@x{\@s{16.4} \.{\THEN} \.{\land} enabled}%
\@x{\@s{47.71} \.{\land} {\IF} initVal \.{=} bot}%
\@x{\@s{67.02} \.{\THEN} A \.{\land} ( s \.{'} \.{=} s )}%
\@x{\@s{67.02} \.{\ELSE} \.{\land} {\UNCHANGED} vars}%
 \@x{\@s{98.33} \.{\land} s \.{'} \.{=} [ id \.{\mapsto} actionId ,\, ctxt
 \.{\mapsto} context ,\,}%
\@x{\@s{133.27} val \.{\mapsto} decr ( initVal ) ]}%
\@x{\@s{16.4} \.{\ELSE} \.{\land} s . id \.{=} actionId}%
 \@x{\@s{47.71} \.{\land} {\IF} s . val \.{=} bot \.{\THEN} \.{\land} s . ctxt
 \.{=} context}%
\@x{\@s{150.07} \.{\land} A}%
\@x{\@s{150.07} \.{\land} s \.{'} \.{=} top}%
\@x{\@s{118.76} \.{\ELSE} \.{\land} {\UNCHANGED} vars}%
 \@x{\@s{150.07} \.{\land} s \.{'} \.{=} [ s {\EXCEPT} {\bang} . val \.{=}
 decr ( s . val ) ]}%
\@x{}\bottombar\@xx{}%
\end{tlatex}
\caption{The end of the \emph{Stuttering} module.} \label{fig:Stuttering2}
\end{figure}
Unlike the original definitions, the actions they define do not
execute any stuttering step when $initVal$ equals $bot$.

As a simple example, let formula $Spec$ be defined as in
Figure~\ref{fig:hour} to equal the hour clock specification of
(\ref{eq:HourClock}).
\begin{figure}
\begin{notla}
---------------------------- MODULE Hour ----------------------------
EXTENDS Integers

VARIABLE h

Init == h=0

Next == h' = (h + 1)  
      
Spec == Init /\ [][Next]_h
=============================================================================
\end{notla}
\begin{tlatex}
\@x{}\moduleLeftDash\@xx{ {\MODULE} Hour}\moduleRightDash\@xx{}%
\@x{ {\EXTENDS} Integers}%
\@pvspace{8.0pt}%
\@x{ {\VARIABLE} h}%
\@pvspace{8.0pt}%
\@x{ Init\@s{4.12} \.{\defeq} h \.{=} 0}%
\@pvspace{8.0pt}%
\@x{ Next \.{\defeq} h \.{'} \.{=} ( h \.{+} 1 )\@s{4.1} \.{\%}\@s{4.1} 24}%
\@pvspace{8.0pt}%
\@x{ Spec\@s{1.46} \.{\defeq} Init \.{\land} {\Box} [ Next ]_{ h}}%
\@x{}\bottombar\@xx{}%
\end{tlatex}
\caption{The hour clock specification.} \label{fig:hour}
\end{figure}
Let us suppose that a module $HourMin$ defines a formula $Spec$ to
equal the hour-minute clock specification $Spec_{2}$ of
(\ref{eq:HourMinClock}).
To construct a refinement mapping under which the hour clock
specification implements the hour-minute clock specification, we add
59 stuttering steps before each $Next$ step of the hour-clock
specification.  The obvious way to do that is to let $\Sigma$ be the
set $1\dd 59$ ordered by \,$<$\,, with $\bot$ equal to 1 and $initVal$
equal to 59.  However, our refinement mapping becomes simpler if we
use the reverse ordering $>$ of $1\dd 59$, with $\bot$ equal to $59$ and
$initVal$ equal to 1.  The refinement mapping can then define \ov{m}
to equal 0 when $s=\top$ and $s.val$ when $s#\top$.

We use the operator $PreStutter$ of the $Stuttering$ module to define
$Next^{s}$.  We instantiate that module with $vars$ equal to $h$ and
with $s$ equal to the stuttering variable, which we also call $s$.
For the arguments of $PreStutter$, observe that:
\begin{itemize}
\item $Next$ is always enabled, so $\ENABLED Next$ equals \TRUE.

\item Since we are adding stuttering steps to only one action, it doesn't
matter what constant we choose for the $actionId$ argument.

\item $Next$, which is the only subaction in the trivial disjunctive
representation of $Next$, has a null context.  We can therefore let
the $context$ argument be any constant.
\end{itemize}
We therefore add the following to the end of module $Hour$.  
\begin{display}
\begin{notla}
vars == h
VARIABLE s 
INSTANCE Stuttering  

InitS == Init /\ (s = top)

NextS == PreStutter(Next, TRUE, "Next", "", 59, 1, LAMBDA j : j+1) 

SpecS == InitS /\ [][NextS]_<<vars, s>>

HM == INSTANCE HourMin WITH m <- IF s = top THEN 0 ELSE s.val

THEOREM SpecS => HM!Spec
\end{notla}
\begin{tlatex}
\@x{ vars \.{\defeq} h}%
\@x{ {\VARIABLE} s}%
\@x{ {\INSTANCE} Stuttering}%
\@pvspace{8.0pt}%
\@x{ InitS\@s{4.12} \.{\defeq} Init \.{\land} ( s \.{=} top )}%
\@pvspace{8.0pt}%
 \@x{ NextS \.{\defeq} PreStutter ( Next ,\, {\TRUE} ,\,\@w{Next} ,\,\@w{\s{.15}} ,\,
 59 ,\, 1 ,\, {\LAMBDA} j \.{:} j \.{+} 1 )}%
\@pvspace{8.0pt}%
 \@x{ SpecS\@s{1.08} \.{\defeq} InitS \.{\land} {\Box} [ NextS ]_{ {\langle}
 vars ,\, s {\rangle}}}%
\@pvspace{8.0pt}%
 \@x{ HM \.{\defeq} {\INSTANCE} HourMin {\WITH} m \.{\leftarrow} {\IF} s \.{=}
 top \.{\THEN} 0 \.{\ELSE} s . val}%
\@pvspace{8.0pt}%
\@x{ {\THEOREM} SpecS \.{\implies} HM {\bang} Spec}%
\end{tlatex}
\end{display}
TLC can easily check this theorem.

\subsection{Correctness of Adding a Stuttering Variable}

How do we check that adding a stuttering variable using the operators
of the $Stuttering$ module produces a specification $Spec^{s}$ such
that
  \tlabox{\EE s: Spec^{s}}
is equivalent to the original specification $Spec$?  The construction
ensures that each behavior of $Spec^{s}$ is obtained by adding
stuttering steps to a behavior of $Spec$, so \tlabox{\EE s: Spec^{s}}
implies $Spec$.  It can fail to be equivalent to $Spec$ only if it
either adds an infinite sequence of stuttering steps, or if we have
used an incorrect $enabled$ argument for $PreStutter$.  It will be
equivalent to $Spec$ if the following conditions are satisfied for
every use of the $PostStutter$ and $PreStutter$ operators, with
arguments named as above, for some constant set $\Sigma$:
\begin{enumerate}
\item For every $\sigma$ in $\Sigma$, the sequence of values
      $\sigma$, $decr(\sigma)$, $decr(decr(\sigma))$, \ldots\
      is contained in $\Sigma$ and
      eventually reaches $bot$.

\item $initVal$ is in $\Sigma$.

\item $enabled$ is equivalent to $\ENABLED A$ [for $PreStutter$ only]

\end{enumerate}
%
%
%
Condition~1 is a condition only on the constants $\Sigma$, $bot$, and
$decr$.  It can be written as $StutterConstantCondition(\Sigma,bot,decr)$
using the following definition from the $Stuttering$ module:
\begin{display}
\begin{notla}
StutterConstantCondition(Sigma, bot, decr(_)) ==
  LET InverseDecr(S) == {sig \in Sigma \ S : decr(sig) \in S}
      R[n \in Nat] == IF n = 0 THEN {bot}
                               ELSE LET T == R[n-1] 
                                    IN  T \cup InverseDecr(T)
                        
  IN Sigma = UNION {R[n] : n \in Nat}
\end{notla}
\begin{tlatex}
\@x{ StutterConstantCondition ( Sigma ,\, bot ,\, decr ( \_ ) ) \.{\defeq}}%
 \@x{\@s{8.2} \.{\LET} InverseDecr ( S ) \.{\defeq} \{ sig \.{\in} Sigma
 \.{\,\backslash\,} S \.{:} decr ( sig ) \.{\in} S \}}%
 \@x{\@s{28.59} R [ n \.{\in} Nat ] \.{\defeq} {\IF} n \.{=} 0 \.{\THEN} \{
 bot \}}%
\@x{\@s{134.14} \.{\ELSE} \.{\LET} T \.{\defeq} R [ n \.{-} 1 ]}%
\@x{\@s{165.45} \.{\IN} T \.{\cup} InverseDecr ( T )}%
\@pvspace{8.0pt}%
\@x{\@s{8.2} \.{\IN} Sigma \.{=} {\UNION} \{ R [ n ] \.{:} n \.{\in} Nat \}}%
\end{tlatex}
\end{display}
This condition can be checked by TLC by putting it into an \textsc{assume}
statement or else putting it in the \textsf{Evaluate Constant Expression}
field of a model's \emph{Model Checking Results} page.  In either case,
the model must replace $Nat$ by a $0\dd n$ for a
(sufficiently large) integer $n$,
and $\Sigma$ must also be replaced with a finite set if it is infinite.
The $Stuttering$ module defines $AltStutterConstantCondition$ to
be equivalent to $StutterConstantCondition$ if $\Sigma$ is finite,
and it doesn't require redefining $Nat$.

The last two conditions are ones that need only hold for behaviors
of $Spec$.  They can be stated formally as follows, where
$A$ is a subaction with context~\tlabox{<<\mathbf{k}; \mathbf{K}>>}:
 \begin{eqnarray}
    Spec & => & [][\A <<\mathbf{k}; \mathbf{K}>> : 
                       A => (initVal \in \Sigma)]_{vars} 
             {\NOTLA \label{eq:stuttering-cond}} \V{.4}
    Spec & => & [](\A <<\mathbf{k}; \mathbf{K}>> :
                            enabled \equiv \ENABLED A)
    \end{eqnarray}
%
TLC can check them in the obvious way.

\subsection{Adding Infinite Stuttering} \label{sec:infinite-stutter}

The type of stuttering variable we have been describing adds a finite
number of stuttering steps before or after a step of a subaction.
There is another type of stuttering variable that adds an infinite
number of stuttering steps not associated with an action.  You are
unlikely ever to have to use one, but we include it for completeness.

Suppose we want to find a refinement mapping under which a spec
$Spec_{1}$ that allows only halting behaviors implements a spec
$Spec_{2}$ that allows behaviors in which externally visible variables
stop changing, but internal variables keep changing forever.  We
obviously can't do that, because if all the variables of $Spec_{1}$
stop changing, then no expression defined in terms of those variables
can keep changing forever.  None of the methods we have described thus
far for adding an auxiliary variable $a$ to $Spec_{1}$ can help us,
because they all have the property that if every behavior allowed by
$Spec_{1}$ halts, then so does every behavior allowed by
$Spec_{1}^{a}$.

It's hard to devise a practical example in which this problem would
arise.  One possibility is for $Spec_{2}$ to have a server perform
internal actions looking for user input that may never arrive, while
in $Spec_{1}$ the server just waits for input.  Although unlikely to
arise, the problem is easy to solve, so we briefly sketch a solution.

The solution is to add a stuttering variable that is required to
stutter forever.  Let $Spec$ equal $Init /\ [][Next]_{vars}$ and let
$UC$ be the stuttering action \tlabox{\UNCHANGED vars}.  Since
$[Next]_{vars}$ equals $Next \/ UC$, we can write $Spec$ as 
 $Init /\ [][Next \/ UC]_{vars}$.  
We can therefore add the subaction $UC$ to any disjunctive
representation of $Next$.  We add a history variable $s$ as described
in Section~\ref{sec:history}, using a disjunctive representation
containing the subaction $UC$.  This defines an action $A^{s}$ for
every subaction $A$ and produces the specification 
  $Init^{s} /\ [][Next^{s}]_{<<vars, s>>}$.  
(We choose $UC^{s}$ so it implies $s'#s$.)  We then define $Spec^{s}$
to equal
 \[ Init^{s} /\ [][Next^{s}]_{<<vars, s>>} /\ \WF_{<<vars, s>>}(UC^{s})\]
Since \tlabox{\ENABLED UC^{s}} equals $\TRUE$, the fairness
requirement $\WF_{<<vars,s>>}(UC^{s})$ implies that infinitely many
$UC^{s}$ steps occur.  These are steps that leave the variables in
$vars$ unchanged, changing only $s$.  Since we have added $s$ as
a history variable, 
  \tlabox{\EE s : Init^{s} /\ [][Next^{s}]_{<<vars, s>>}}
is equivalent to $Spec$.  Since any \tlaplus\ spec allows stuttering
steps, this implies that
  \tlabox{\EE s: Spec^{s}} 
is equivalent to $Spec$.

\subsection{Liveness}

Liveness poses no problem when adding a stuttering variable.  As with
other auxiliary variables, we obtain $Spec^{s}$ by adding the
stuttering variable to the safety part of $Spec$ and then conjoining
to it the liveness conjunct of $Spec$.  (This is true as well for the
kind of stuttering variable described in
Section~\ref{sec:infinite-stutter}, where $Spec^{s}$ contains a liveness
conjunct.)

Although $Spec^{s}$ may have an unusual form, it isn't weird.  If
$Spec$ is machine closed then $Spec^{s}$ is also machine closed.
However, putting it into a standard form with fairness conditions only
on subactions of $Next^{s}$ is not as simple as it is for history
variables.

\section{The Snapshot Problem}

We now consider an example of using auxiliary variables to show that
an algorithm satisfies its specification.  Our example is based on an
algorithm of Afek et al.~\cite{afek:atomic-snap}.  Their algorithm
implements what they call a \emph{single-writer atomic snapshot
memory}, which will call simply a \emph{snapshot object}.  Their
algorithm implements a snapshot object using an unbounded amount of
storage.  They also present a second algorithm that uses a bounded
amount of storage and implements a more general type of object, but we
restrict ourselves to their first, simpler algorithm.  Moreover, we
consider only a simplified version of this simpler algorithm; their
algorithm can be checked by adding the same auxiliary variables used
for the simplified version.

\subsection{Linearizability}

A snapshot algorithm is used to implement an atomic read of an array
of memory registers, each of which can be written by a different
process.  Its specification is a special case of a linearizable
specification of a data object---a concept introduced by
Herlihy and Wing~\cite{herlihy:axioms}.

A data object, also called a state machine, executes commands from
user processes.  It is described by an initial state of the
object and an operator $Apply$, where $Apply(i, cmd, st)$ describes
the output and new state of the object that results from process $i$
executing command $cmd$ when the object has state $st$.  It is
specified formally by these declared constants:
\begin{display}
\begin{notla}
CONSTANTS  Procs, Commands(_), Outputs(_), InitObj, 
           Apply(_, _, _)
\end{notla}
\begin{tlatex}
 \@x{ {\CONSTANTS}\@s{4.1} Procs ,\, Commands ( \_ ) ,\, Outputs ( \_ ) ,\,
   InitObj ,\,}%
\@x{\@s{58.85} Apply ( \_ ,\, \_ ,\, \_ )}%
\end{tlatex}
\end{display}
They have the following meanings:
\begin{describe}{$Apply(i,cmd,st)$}
\item[$Procs$] The set of processes.

\item[$Commands(i)$] The set of commands that process $i$ can issue.

\item[$Outputs(i)$] The set of outputs the commands issued by process $i$
can produce.

\item[$InitObj$] The initial state of the object.

\item[$Apply(i,cmd,st)$] A record with $output$ and $newState$ fields
describing the result of process $i$ executing command $cmd$ when
the object is in state~$st$.
\end{describe}
A linearizable implementation  of the data object
is one in which the state of the object is internal, the only
externally visible actions being the issuing of the command and the
return of its output.  More precisely, a process $i$ executes a
command $cmd$ with a $BeginOp(i, cmd)$ step, followed by a $DoOp(i)$
step that modifies the state of the object, followed by an $EndOp(i)$
step.  The $BeginOp$ and $EndOp$ steps are externally visible, meaning
that they modify externally visible variables (and perhaps internal
variables), while the $DoOp$ step modifies only internal
variables---including an internal variable describing the state of the
object.

To simplify the specification, we assume that the sets of commands and
of outputs are disjoint.  We can then use a single externally visible
variable $interface$, letting $BeginOp(i, cmd)$ set $interface[i]$ to
the command $cmd$ and letting $EndOp(i)$ set it to the command's
output.  We also introduce an internal variable $istate$ to hold the
internal state of the processes---needed to remember, while a process
is executing a command, whether or not it has performed the $DoOp$
step and, if it has, what output was produced.  We do this by letting
$BeginOp(i, cmd)$ set $istate[i]$ to $cmd$, and letting $DoOp(i)$
set it to the command's output.  Here is the definition of the
next-state action and its subactions.
\begin{display}
\begin{notla}
BeginOp(i, cmd) == /\ interface[i] \in Outputs(i)
                   /\ interface' = [interface EXCEPT ![i] = cmd]
                   /\ istate' = [istate EXCEPT ![i] = cmd]
                   /\ object' = object

DoOp(i) == /\ interface[i] \in Commands(i)
           /\ istate[i] = interface[i]
           /\ LET result == Apply(i, interface[i], object)
              IN  /\ object' = result.newState
                  /\ istate' = [istate EXCEPT ![i] = result.output]
           /\ interface' = interface
           
EndOp(i) == /\ interface[i] \in Commands(i)
            /\ istate[i] \in Outputs(i)
            /\ interface' = [interface EXCEPT ![i] = istate[i]]
            /\ UNCHANGED <<object, istate>> 

Next == \E i \in Procs : \/ \E cmd \in Commands(i) : BeginOp(i, cmd)
                         \/ DoOp(i) 
                         \/ EndOp(i)
\end{notla}
\begin{tlatex}
 \@x{ BeginOp ( i ,\, cmd ) \.{\defeq} \.{\land} interface [ i ] \.{\in}
 Outputs ( i )}%
 \@x{\@s{93.61} \.{\land} interface \.{'} \.{=} [ interface {\EXCEPT} {\bang}
 [ i ] \.{=} cmd ]}%
 \@x{\@s{93.61} \.{\land} istate \.{'}\@s{1.47} \.{=} [ istate {\EXCEPT}
 {\bang} [ i ] \.{=} cmd ]}%
\@x{\@s{93.61} \.{\land} object \.{'} \.{=} object}%
\@pvspace{8.0pt}%
\@x{ DoOp ( i ) \.{\defeq} \.{\land} interface [ i ] \.{\in} Commands ( i )}%
\@x{\@s{56.82} \.{\land} istate [ i ] \.{=} interface [ i ]}%
 \@x{\@s{56.82} \.{\land} \.{\LET} result \.{\defeq} Apply ( i ,\, interface [
 i ] ,\, object )}%
\@x{\@s{67.93} \.{\IN} \.{\land} object \.{'} \.{=} result . newState}%
 \@x{\@s{88.33} \.{\land} istate \.{'}\@s{1.47} \.{=} [ istate {\EXCEPT}
 {\bang} [ i ] \.{=} result . output ]}%
\@x{\@s{56.82} \.{\land} interface \.{'} \.{=} interface}%
\@pvspace{8.0pt}%
\@x{ EndOp ( i ) \.{\defeq} \.{\land} interface [ i ] \.{\in} Commands ( i )}%
\@x{\@s{61.67} \.{\land} istate [ i ] \.{\in} Outputs ( i )}%
 \@x{\@s{61.67} \.{\land} interface \.{'} \.{=} [ interface {\EXCEPT} {\bang}
 [ i ] \.{=} istate [ i ] ]}%
\@x{\@s{61.67} \.{\land} {\UNCHANGED} {\langle} object ,\, istate {\rangle}}%
\@pvspace{8.0pt}%
 \@x{ Next \.{\defeq} \E\, i \.{\in} Procs \.{:} \.{\lor} \E\, cmd \.{\in}
 Commands ( i ) \.{:} BeginOp ( i ,\, cmd )}%
\@x{\@s{97.40} \.{\lor} DoOp ( i )}%
\@x{\@s{97.40} \.{\lor} EndOp ( i )}%
\end{tlatex}
\end{display}
Initially, $interface[i]$ and $istate[i]$ equal some output, for each
$i$.  We let that equal $InitOutput(i)$ for some \textsc{constant}
operator $InitOutput$.  We also add a fairness requirement to imply
that any command that has begun (with a $BeginOp$ step) eventually
completes (with an $EndOp$ step).  The complete specification (with
the action definitions above elided) is in module $Linearizability$,
shown in \lref{targ:Linearizability}{Figure~\ref{fig:Linearizability}}. 
\begin{figure} \target{targ:Linearizability}
\begin{notla}
-------------------------- MODULE Linearizability --------------------------
CONSTANTS  Procs, Commands(_), Outputs(_), InitOutput(_), 
           ObjValues, InitObj, Apply(_, _, _)

ASSUME LinearAssumps == 
         /\ InitObj \in ObjValues
         /\ \A i \in Procs : InitOutput(i) \in Outputs(i)
         /\ \A i \in Procs : Outputs(i) \cap Commands(i) = { }
         /\ \A i \in Procs, obj \in ObjValues : 
              \A cmd \in Commands(i) : 
                  /\ Apply(i, cmd, obj).output \in Outputs(i)
                  /\ Apply(i, cmd, obj).newState \in ObjValues

VARIABLES object, interface, istate
vars == <<object, interface, istate>>

Init == /\ object = InitObj
        /\ interface = [i \in Procs |-> InitOutput(i)]
        /\ istate = [i \in Procs |-> InitOutput(i)]
        
BeginOp(i, cmd) == ...
DoOp(i) == ...
EndOp(i) == ...
Next == ...

SafeSpec == Init /\ [][Next]_vars   

Fairness == \A i \in Procs : WF_vars(DoOp(i)) /\ WF_vars(EndOp(i))  
Spec == Init /\ [][Next]_vars  /\ Fairness
=============================================================================
\end{notla}
\begin{tlatex}
\@x{}\moduleLeftDash\@xx{ {\MODULE} Linearizability}\moduleRightDash\@xx{}%
 \@x{ {\CONSTANTS}\@s{4.1} Procs ,\, Commands ( \_ ) ,\, Outputs ( \_ ) ,\,
 InitOutput ( \_ ) ,\,}%
\@x{\@s{58.85} ObjValues ,\, InitObj ,\, Apply ( \_ ,\, \_ ,\, \_ )}%
\@pvspace{8.0pt}%
\@x{ {\ASSUME} LinearAssumps \.{\defeq}}%
\@x{\@s{46.44} \.{\land} InitObj \.{\in} ObjValues}%
 \@x{\@s{46.44} \.{\land} \A\, i \.{\in} Procs \.{:} InitOutput ( i ) \.{\in}
 Outputs ( i )}%
 \@x{\@s{46.44} \.{\land} \A\, i \.{\in} Procs \.{:} Outputs ( i ) \.{\cap}
 Commands ( i ) \.{=} \{ \}}%
 \@x{\@s{46.44} \.{\land} \A\, i \.{\in} Procs ,\, obj \.{\in} ObjValues
 \.{:}}%
\@x{\@s{65.75} \A\, cmd \.{\in} Commands ( i ) \.{:}}%
 \@x{\@s{77.07} \.{\land}\@s{3.71} Apply ( i ,\, cmd ,\, obj ) . output
 \.{\in} Outputs ( i )}%
 \@x{\@s{77.07} \.{\land}\@s{3.71} Apply ( i ,\, cmd ,\, obj ) . newState
 \.{\in} ObjValues}%
\@pvspace{8.0pt}%
\@x{ {\VARIABLES} object ,\, interface ,\, istate}%
\@x{ vars \.{\defeq} {\langle} object ,\, interface ,\, istate {\rangle}}%
\@pvspace{8.0pt}%
\@x{ Init\@s{2.02} \.{\defeq} \.{\land} object \.{=} InitObj}%
 \@x{\@s{37.72} \.{\land} interface \.{=} [ i \.{\in} Procs \.{\mapsto}
 InitOutput ( i ) ]}%
 \@x{\@s{37.72} \.{\land} istate \.{=} [ i \.{\in} Procs \.{\mapsto}
 InitOutput ( i ) ]}%
\@pvspace{8.0pt}%
\@x{ BeginOp ( i ,\, cmd ) \.{\defeq} \.{\dots}}%
\@x{ DoOp ( i ) \.{\defeq} \.{\dots}}%
\@x{ EndOp ( i ) \.{\defeq} \.{\dots}}%
\@x{ Next \.{\defeq} \.{\dots}}%
\@pvspace{8.0pt}%
\@x{ SafeSpec \.{\defeq} Init \.{\land} {\Box} [ Next ]_{ vars}}%
\@pvspace{8.0pt}%
 \@x{ Fairness\@s{0.49} \.{\defeq} \A\, i \.{\in} Procs \.{:} {\WF}_{ vars} (
 DoOp ( i ) ) \.{\land} {\WF}_{ vars} ( EndOp ( i ) )}%
\@pvspace{2.0pt}%
 \@x{ Spec \.{\defeq} Init \.{\land} {\Box} [ Next ]_{ vars}\@s{4.1} \.{\land}
 Fairness}%
\@x{}\bottombar\@xx{}%
\end{tlatex}
\caption{Module \emph{Linearizability}.} \label{fig:Linearizability}
\end{figure}
Any particular linearizable object can be specified by instantiating
the module with the appropriate constants.  The module includes an
assumption named $LinearAssumps$, to check that the instantiated
constants satisfy the properties they should for the module to specify
a linearizable object.  To state all those properties, the
specification introduces a constant $ObjValues$ to describe the set of
all possible states of the object.  This set could be defined to equal
the following rather complicated expression.  Trying to understand it
provides a good lesson in set theory.
\begin{display}
\begin{notla}
LET ApplyProcTo(i,S) == 
       {Apply(i, cmd, x).newState :  x \in S, cmd \in Commands(i)}

    ApplyTo(S) == UNION {ApplyProcTo(i, S) : i \in Procs}

    ApplyITimes[i \in Nat] == 
        IF i = 0 THEN {InitObj}
                 ELSE ApplyTo(ApplyITimes[i-1])

IN  UNION {ApplyITimes[i] : i \in Nat}
\end{notla}
\begin{tlatex}
\@x{ \.{\LET} ApplyProcTo ( i ,\, S ) \.{\defeq}}%
 \@x{\@s{32.69} \{ Apply ( i ,\, cmd ,\, x ) . newState \.{:}\@s{4.1} x
 \.{\in} S ,\, cmd \.{\in} Commands ( i ) \}}%
\@pvspace{4.0pt}%
 \@x{\@s{20.39} ApplyTo ( S ) \.{\defeq} {\UNION} \{ ApplyProcTo ( i ,\, S )
 \.{:} i \.{\in} Procs \}}%
\@pvspace{4.0pt}%
\@x{\@s{20.19} ApplyITimes [ i \.{\in} Nat ] \.{\defeq}}
\@x{\@s{40.89} {\IF} i \.{=} 0 \.{\THEN} \{ InitObj \}}%
\@x{\@s{75.47} \.{\ELSE} ApplyTo ( ApplyITimes [ i \.{-} 1 ] )}%
\@pvspace{4.0pt}%
\@x{ \.{\IN} {\UNION} \{ ApplyITimes [ i ] \.{:} i \.{\in} Nat \}}%
\end{tlatex}
\end{display}

\subsection{The Linearizable Snapshot Specification}

By a snapshot object, we mean what Afek et al.~\cite{afek:atomic-snap}
called an \emph{atomic snapshot memory}.  In a snapshot object, the
processes are either readers or writers.  Reader and writer should be
thought of as roles; the same physical process can act as both a
reader and a writer.  A snapshot object is an array of registers, one
per writer.  A write operation writes a value to the writer's register
and produces as output some fixed value that is not a possible
register value.  A read operation has a single command that produces
the object's state (an array of register values) as output and leaves
that state unchanged.

The specification declares four constants: the sets $Readers$ and
$Writers$ of reader and writer processes; the set $RegVals$ of
possible register values; and a value $InitRegVal$ in $RegVals$ that
is the initial value of a register.  We call the snapshot object a
\emph{memory} and use different names for some of the parameters of
the $Linearizability$ module, including $MemVals$ and $InitMem$ for
$ObjValues$ and $InitObj$.  We define $NotMemVal$ be the single reader
command and $NotRegVal$ to be the single write command output.  The
complete specification is in module $LinearSnapshot$ of
\lref{targ:LinearSnapshot}{Figure~\ref{fig:LinearSnapshot}}.  (The
\textsc{assume} is added at the end of the module so TLC will check
that the assumption $LinearAssumps$ of module $Linearizability$ is
true under the instantiation.)

\begin{figure} \target{targ:LinearSnapshot}
\begin{notla}
--------------------------- MODULE LinearSnapshot ---------------------------
CONSTANTS Readers, Writers, RegVals, InitRegVal

ASSUME /\ Readers \cap Writers = {}
       /\ InitRegVal \in RegVals

Procs == Readers \cup Writers

MemVals == [Writers -> RegVals]
InitMem == [i \in Writers |-> InitRegVal]

NotMemVal == CHOOSE v : v \notin MemVals
NotRegVal == CHOOSE v : v \notin RegVals

Commands(i) == IF i \in Readers THEN {NotMemVal}
                                ELSE RegVals

Outputs(i) == IF i \in Readers THEN MemVals 
                               ELSE {NotRegVal}

InitOutput(i) == IF i \in Readers THEN InitMem ELSE NotRegVal  

Apply(i, cmd, obj) == IF i \in Readers 
                        THEN [newState |-> obj, output |-> obj]
                        ELSE [newState |-> [obj EXCEPT ![i] = cmd],
                              output   |-> NotRegVal]

VARIABLES mem, interface, istate

INSTANCE Linearizability WITH ObjValues <- MemVals, InitObj <- InitMem,  
                              object <- mem

ASSUME LinearAssumps
============================================================================
\end{notla}
\begin{tlatex}
\@x{}\moduleLeftDash\@xx{ {\MODULE} LinearSnapshot}\moduleRightDash\@xx{}%
\@x{ {\CONSTANTS} Readers ,\, Writers ,\, RegVals ,\, InitRegVal}%
\@pvspace{8.0pt}%
\@x{ {\ASSUME} \.{\land} Readers \.{\cap} Writers \.{=} \{ \}}%
\@x{\@s{38.24} \.{\land} InitRegVal \.{\in} RegVals}%
\@pvspace{8.0pt}%
\@x{ Procs \.{\defeq} Readers \.{\cup} Writers}%
\@pvspace{8.0pt}%
\@x{ MemVals \.{\defeq} [ Writers \.{\rightarrow} RegVals ]}%
 \@x{ InitMem\@s{2.60} \.{\defeq} [ i \.{\in} Writers \.{\mapsto} InitRegVal
 ]}%
\@pvspace{8.0pt}%
\@x{ NotMemVal \.{\defeq} {\CHOOSE} v \.{:} v \.{\notin} MemVals}%
\@x{ NotRegVal\@s{6.27} \.{\defeq} {\CHOOSE} v \.{:} v \.{\notin} RegVals}%
\@pvspace{8.0pt}%
 \@x{ Commands ( i ) \.{\defeq} {\IF} i \.{\in} Readers \.{\THEN} \{ NotMemVal
 \}}%
\@x{\@s{144.52} \.{\ELSE} RegVals}%
\@pvspace{8.0pt}%
\@x{ Outputs ( i ) \.{\defeq} {\IF} i \.{\in} Readers \.{\THEN} MemVals}%
\@x{\@s{130.21} \.{\ELSE} \{ NotRegVal \}}%
\@pvspace{8.0pt}%
 \@x{ InitOutput ( i ) \.{\defeq} {\IF} i \.{\in} Readers \.{\THEN} InitMem
 \.{\ELSE} NotRegVal}%
\@pvspace{8.0pt}%
\@x{ Apply ( i ,\, cmd ,\, obj ) \.{\defeq} {\IF} i \.{\in} Readers}%
 \@x{\@s{110.27} \.{\THEN} [ newState \.{\mapsto} obj ,\, output \.{\mapsto}
 obj ]}%
 \@x{\@s{110.27} \.{\ELSE} [ newState \.{\mapsto} [ obj {\EXCEPT} {\bang} [ i
 ] \.{=} cmd ] ,\,}%
\@x{\@s{144.36} output\@s{11.04} \.{\mapsto} NotRegVal ]}%
\@pvspace{8.0pt}%
\@x{ {\VARIABLES} mem ,\, interface ,\, istate}%
\@pvspace{8.0pt}%
 \@x{ {\INSTANCE} Linearizability {\WITH} ObjValues \.{\leftarrow} MemVals ,\,
 InitObj \.{\leftarrow} InitMem ,\,}%
\@x{\@s{140.48} object \.{\leftarrow} mem}%
\@pvspace{8.0pt}%
\@x{ {\ASSUME} LinearAssumps}%
\@x{}\bottombar\@xx{}%
\end{tlatex}
\caption{Module \emph{LinearSnapshot}.} \label{fig:LinearSnapshot}
\end{figure}

\subsection{The Simplified Afek et al. Snapshot Algorithm}

The snapshot algorithm of Afek et al.\ uses an internal variable $imem$
whose value is an array with $imem[i]$ a pair consisting of the value
of the $i$\tth\ register and an integer whose value is the number of
times the register has been written.  It assumes that the entire pair
can be read and written atomically.

A write operation writes the register value $cmd$ in the obvious way,
the $DoOp(i)$ action setting $imem[i]$ to $<<cmd, imem[i][2]+1>>$.

A read operation first performs the following \emph{scan} procedure:
\begin{display}
It reads all the elements $imem[i]$ once, in any order.  It then reads
them a second time, again in any order.  If it reads the same values
both times for all $i$, it outputs the array of register values it
read.
\end{display}
If the values obtained for each element $imem[i]$ by the two reads are
not all the same, so the scan procedure does not produce an output,
then the procedure is repeated.  The scan procedure is repeated again
and again until it produces an output.

The actual algorithm has an alternative method of producing an output
that can be used when it has read three different values for
$imem[i]$, for some writer~$i$.  By using this method, termination of
the read is assured.  However, for simplicity, we use an algorithm
that keeps performing the scan procedure until it succeeds in
producing an output.  Thus, a read need never terminate, so the
algorithm does not satisfy the liveness requirement of a snapshot
algorithm---namely, it does not satisfy the weak fairness requirement
of the $DoOp(i)$ action for a reader~$i$.  However, it does satisfy
the safety requirement.  The correctness of the complete algorithm
(including liveness) can be verified by essentially the same method
used for our simplified version; but the complete algorithm is more
complicated, so the refinement mapping is more complicated and model
checking takes longer.  We therefore consider only the simplified
algorithm.

To specify the algorithm in \tlaplus, we declare the same constants
$Readers$, $Writers$, $RegVals$, and $InitRegVal$ and make the same
definitions of $MemVals$, $InitMem$, $NotMemVal$, and $NotRegVal$ as
in module $LinearSnapshot$ above.  We also define:
\begin{display}
\begin{notla}
IRegVals == RegVals \X Nat
IMemVals == [Writers -> IRegVals]
InitIMem == [i \in Writers |-> <<InitRegVal, 0>> ]
\end{notla}
\begin{tlatex}
\@x{ IRegVals\@s{6.27} \.{\defeq} RegVals \.{\times} Nat}%
\@x{ IMemVals \.{\defeq} [ Writers \.{\rightarrow} IRegVals ]}%
 \@x{ InitIMem\@s{2.60} \.{\defeq} [ i \.{\in} Writers \.{\mapsto} {\langle}
 InitRegVal ,\, 0 {\rangle} ]}%
\end{tlatex}
\end{display}
We declare five variables, with the following meanings:
\begin{display}
\begin{description}
\item[$interface:$] The same as in $LinearSnapshot$.

\item[$imem:$] Like $mem$ in $LinearSnapshot$, except $imem[i]$ is an
ordered pair in $RegVals\X Nat$, the first component representing
$mem[i]$ and the second the number of times $mem[i]$ has been written.
The initial value of $imem$ is initially $InitIMem$.

\item[$wrNum:$] A function with domain $Writers$, where $wrNum[i]$ is
the number of $BeginWr(i)$ steps that have been taken.

\item[$rdVal1$, $rdVal2:$] They are functions such that $rdVal1[i]$
and $rdVal2[i]$ describe the values read so far by reader $i$ in the
two reads of the \emph{scan} procedure.  Both $rdVal1[i]$ and
$rdVal2[i]$ are functions whose domain is the set of writers $j$ for
which the first or second read of $imem[j]$ has been performed,
mapping each such $j$ to the value read.  They are set initially to
the empty function (the function with empty domain), which we write
$<<\,>>$.
\end{description}
\end{display}
The writer actions are straightforward.  Note that because $wrNum[i]$
counts the number of $BeginWr(i,cmd)$ steps and $imem[i][2]$ is set to
$wrNum[i]$ by the $DoWr(i)$, the $EndWrite(i)$ action should be
enabled and $DoWr(i)$ disabled when $imem[i][2]$ equals $wrNum[i]$.
\begin{display}
\begin{notla}
BeginWr(i, cmd) == /\ interface[i] = NotRegVal
                   /\ wrNum' = [wrNum EXCEPT ![i] = wrNum[i] + 1]
                   /\ interface' = [interface EXCEPT ![i] = cmd]
                   /\ UNCHANGED <<imem, rdVal1, rdVal2>>

DoWr(i) == /\ interface[i] \in RegVals
           /\ imem[i][2] # wrNum[i]
           /\ imem' = [imem EXCEPT ![i] = <<interface[i], wrNum[i]>>]
           /\ UNCHANGED <<interface, wrNum, rdVal1, rdVal2>>
           
EndWr(i) == /\ interface[i] \in RegVals
            /\ imem[i][2] = wrNum[i]
            /\ interface' = [interface EXCEPT ![i] = NotRegVal]
            /\ UNCHANGED <<imem, wrNum, rdVal1, rdVal2>>
        
\end{notla}
\begin{tlatex}
 \@x{ BeginWr ( i ,\, cmd ) \.{\defeq} \.{\land} interface [ i ] \.{=}
 NotRegVal}%
 \@x{\@s{95.48} \.{\land} wrNum \.{'} \.{=} [ wrNum {\EXCEPT} {\bang} [ i ]
 \.{=} wrNum [ i ] \.{+} 1 ]}%
 \@x{\@s{95.48} \.{\land} interface \.{'} \.{=} [ interface {\EXCEPT} {\bang}
 [ i ] \.{=} cmd ]}%
 \@x{\@s{95.48} \.{\land} {\UNCHANGED} {\langle} imem ,\, rdVal1 ,\, rdVal2
 {\rangle}}%
\@pvspace{8.0pt}%
\@x{ DoWr ( i ) \.{\defeq} \.{\land} interface [ i ] \.{\in} RegVals}%
\@x{\@s{58.69} \.{\land} imem [ i ] [ 2 ] \.{\neq} wrNum [ i ]}%
 \@x{\@s{58.69} \.{\land} imem \.{'} \.{=} [ imem {\EXCEPT} {\bang} [ i ]
 \.{=} {\langle} interface [ i ] ,\, wrNum [ i ] {\rangle} ]}%
 \@x{\@s{58.69} \.{\land} {\UNCHANGED} {\langle} interface ,\, wrNum ,\,
 rdVal1 ,\, rdVal2 {\rangle}}%
\@pvspace{8.0pt}%
\@x{ EndWr ( i ) \.{\defeq} \.{\land} interface [ i ] \.{\in} RegVals}%
\@x{\@s{63.55} \.{\land} imem [ i ] [ 2 ] \.{=} wrNum [ i ]}%
 \@x{\@s{63.55} \.{\land} interface \.{'}\@s{3.73} \.{=} [ interface {\EXCEPT}
 {\bang} [ i ] \.{=} NotRegVal ]}%
 \@x{\@s{63.55} \.{\land} {\UNCHANGED} {\langle} imem ,\, wrNum ,\, rdVal1 ,\,
 rdVal2 {\rangle}}%
\@pvspace{8.0pt}%
\end{tlatex}
\end{display}
The $BeginRd(i)$ action is straightforward.  
\begin{display}
\begin{notla}
BeginRd(i) == /\ interface[i] \in MemVals
              /\ interface' = [interface EXCEPT ![i] = NotMemVal]
              /\ UNCHANGED <<imem, wrNum, rdVal1, rdVal2>>
\end{notla}
\begin{tlatex}
\@x{ BeginRd ( i ) \.{\defeq} \.{\land} interface [ i ] \.{\in} MemVals}%
 \@x{\@s{68.09} \.{\land} interface \.{'} \.{=} [ interface {\EXCEPT} {\bang}
 [ i ] \.{=} NotMemVal ]}%
 \@x{\@s{68.09} \.{\land} {\UNCHANGED} {\langle} imem ,\, wrNum ,\, rdVal1 ,\,
 rdVal2 {\rangle}}%
\end{tlatex}
\end{display}
The definitions of the actions that perform the \emph{scan} procedure
use the following definition.  We define $AddToFcn(f, x, v)$  to
be the function $g$ obtained from the function $f$ by adding $x$ to its
domain and defining $g[x]$ to equal $v$.  Using operators defined in
the $TLC$ module, it can be defined to equal $f@@(x :> v)$.  However,
it's easy enough to define it directly as:
\begin{display}
\begin{notla}
AddToFcn(f, x, v) == 
     [y \in (DOMAIN f) \cup {x} |-> IF y = x THEN v ELSE f[y]]
\end{notla}
\begin{tlatex}
\@x{ AddToFcn ( f ,\, x ,\, v ) \.{\defeq}}%
 \@x{\@s{20.5} [ y \.{\in} ( {\DOMAIN} f ) \.{\cup} \{ x \} \.{\mapsto} {\IF}
 y \.{=} x \.{\THEN} v \.{\ELSE} f [ y ] ]}%
\end{tlatex}
\end{display}
Using $AddToFcn$, we define the $Rd1$ action that performs the scan's
first read of $imem$ and the $Rd2$ action that performs its second
read.
%
%
%
\begin{display}
\begin{notla}
Rd1(i) == /\ interface[i] = NotMemVal
          /\ \E j \in Writers \ DOMAIN rdVal1[i] :
                rdVal1' = [rdVal1 EXCEPT 
                            ![i] = AddToFcn(rdVal1[i], j, imem[j])]
          /\ UNCHANGED <<interface, imem, wrNum, rdVal2>>
          
Rd2(i) == /\ interface[i] = NotMemVal
          /\ DOMAIN rdVal1[i] = Writers
          /\ \E j \in Writers \ DOMAIN rdVal2[i] :
                rdVal2' = [rdVal2 EXCEPT 
                            ![i] = AddToFcn(rdVal2[i], j, imem[j])]
          /\ UNCHANGED <<interface, imem, wrNum, rdVal1>>
\end{notla}
\begin{tlatex}
\@x{ Rd1 ( i ) \.{\defeq} \.{\land} interface [ i ] \.{=} NotMemVal}%
 \@x{\@s{48.67} \.{\land} \E\, j \.{\in} Writers \.{\,\backslash\,} {\DOMAIN}
 rdVal1 [ i ] \.{:}}%
\@x{\@s{67.01} rdVal1 \.{'} \.{=} [ rdVal1 {\EXCEPT}}%
 \@x{\@s{119.21} {\bang} [ i ] \.{=} AddToFcn ( rdVal1 [ i ] ,\, j ,\, imem [
 j ] ) ]}%
 \@x{\@s{48.67} \.{\land} {\UNCHANGED} {\langle} interface ,\, imem ,\, wrNum
 ,\, rdVal2 {\rangle}}%
\@pvspace{8.0pt}%
\@x{ Rd2 ( i ) \.{\defeq} \.{\land} interface [ i ] \.{=} NotMemVal}%
\@x{\@s{48.67} \.{\land} {\DOMAIN} rdVal1 [ i ] \.{=} Writers}%
 \@x{\@s{48.67} \.{\land} \E\, j \.{\in} Writers \.{\,\backslash\,} {\DOMAIN}
 rdVal2 [ i ] \.{:}}%
\@x{\@s{67.01} rdVal2 \.{'} \.{=} [ rdVal2 {\EXCEPT}}%
 \@x{\@s{119.21} {\bang} [ i ] \.{=} AddToFcn ( rdVal2 [ i ] ,\, j ,\, imem [
 j ] ) ]}%
 \@x{\@s{48.67} \.{\land} {\UNCHANGED} {\langle} interface ,\, imem ,\, wrNum
 ,\, rdVal1 {\rangle}}%
\end{tlatex}
\end{display}
Finally, we define $TryEndRd(i)$ to be an action that is enabled when
the reader's scan operation has completed.  It compares the values
read by the two sets of reads and, if they are equal, it performs
the $EndOp$ for the read.  Otherwise, it enables the next scan to begin.
\begin{display}
%
%
\begin{notla}
TryEndRd(i) == /\ interface[i] = NotMemVal
               /\ DOMAIN rdVal1[i] = Writers
               /\ DOMAIN rdVal2[i] = Writers
               /\ IF rdVal1[i] = rdVal2[i]
                    THEN interface' = 
                            [interface EXCEPT 
                              ![i] = [j \in Writers |-> rdVal1[i][j][1]] ]
                    ELSE interface' = interface
               /\ rdVal1' = [rdVal1 EXCEPT ![i] = << >>]
               /\ rdVal2' = [rdVal2 EXCEPT ![i] = << >>]
               /\ UNCHANGED <<imem, wrNum>>
\end{notla}
\begin{tlatex}
\@x{ TryEndRd ( i ) \.{\defeq} \.{\land} interface [ i ] \.{=} NotMemVal}%
\@x{\@s{76.65} \.{\land} {\DOMAIN} rdVal1 [ i ] \.{=} Writers}%
\@x{\@s{76.65} \.{\land} {\DOMAIN} rdVal2 [ i ] \.{=} Writers}%
\@x{\@s{76.65} \.{\land} {\IF} rdVal1 [ i ] \.{=} rdVal2 [ i ]}%
\@x{\@s{95.96} \.{\THEN} interface \.{'} \.{=}}%
\@x{\@s{139.57} [ interface {\EXCEPT}}%
 \@x{\@s{146.45} {\bang} [ i ] \.{=} [ j \.{\in} Writers \.{\mapsto} rdVal1 [
 i ] [ j ] [ 1 ] ] ]}%
\@x{\@s{95.96} \.{\ELSE} interface \.{'} \.{=} interface}%
 \@x{\@s{76.65} \.{\land} rdVal1 \.{'} \.{=} [ rdVal1 {\EXCEPT} {\bang} [ i ]
 \.{=} {\langle} {\rangle} ]}%
 \@x{\@s{76.65} \.{\land} rdVal2 \.{'} \.{=} [ rdVal2 {\EXCEPT} {\bang} [ i ]
 \.{=} {\langle} {\rangle} ]}%
\@x{\@s{76.65} \.{\land} {\UNCHANGED} {\langle} imem ,\, wrNum {\rangle}}%
\end{tlatex}
\end{display}
The complete specification is in module $AfekSimplified$, shown in
\lref{targ:AfekSimplified}{Figure~\ref{fig:AfekSimplified}} with the action definitions
above elided.
\begin{figure} \target{targ:AfekSimplified}
\begin{notla}
--------------------------- MODULE AfekSimplified ---------------------------
EXTENDS Integers 

CONSTANTS Readers, Writers, RegVals, InitRegVal

MemVals == [Writers -> RegVals]
InitMem == [i \in Writers |-> InitRegVal]
NotMemVal == CHOOSE v : v \notin MemVals
NotRegVal == CHOOSE v : v \notin RegVals

IRegVals == RegVals \X Nat
IMemVals == [Writers -> IRegVals]
InitIMem == [i \in Writers |-> <<InitRegVal, 0>> ]

VARIABLES imem, interface, wrNum, rdVal1, rdVal2
vars == <<imem, interface, wrNum, rdVal1, rdVal2>>

Init == /\ imem = InitIMem
        /\ interface = [i \in Readers \cup Writers |->
                          IF i \in Readers THEN InitMem ELSE NotRegVal]
        /\ wrNum = [i \in Writers |-> 0]
        /\ rdVal1 = [i \in Readers |-> << >>]
        /\ rdVal2 = [i \in Readers |-> << >>]

BeginWr(i, cmd) == ...
DoWr(i) == ...
EndWr(i) == ...
BeginRd(i) == ...
AddToFcn(f, x, v) == ...
Rd1(i) == ... 
Rd2(i) == ... 
TryEndRd(i) == ...
 
Next == \/ \E i \in Readers : BeginRd(i) \/ Rd1(i) \/ Rd2(i) \/ TryEndRd(i)
        \/ \E i \in Writers : \/ \E cmd \in RegVals : BeginWr(i, cmd)
                              \/ DoWr(i) \/ EndWr(i)  
                              
Spec == Init /\ [][Next]_vars       
=============================================================================
\end{notla}
\begin{tlatex}
\@x{}\moduleLeftDash\@xx{ {\MODULE} AfekSimplified}\moduleRightDash\@xx{}%
\@x{ {\EXTENDS} Integers}%
\@pvspace{8.0pt}%
\@x{ {\CONSTANTS} Readers ,\, Writers ,\, RegVals ,\, InitRegVal}%
\@pvspace{8.0pt}%
\@x{ MemVals \.{\defeq} [ Writers \.{\rightarrow} RegVals ]}%
 \@x{ InitMem\@s{2.60} \.{\defeq} [ i \.{\in} Writers \.{\mapsto} InitRegVal
 ]}%
\@x{ NotMemVal \.{\defeq} {\CHOOSE} v \.{:} v \.{\notin} MemVals}%
\@x{ NotRegVal\@s{6.27} \.{\defeq} {\CHOOSE} v \.{:} v \.{\notin} RegVals}%
\@pvspace{8.0pt}%
\@x{ IRegVals\@s{6.27} \.{\defeq} RegVals \.{\times} Nat}%
\@x{ IMemVals \.{\defeq} [ Writers \.{\rightarrow} IRegVals ]}%
 \@x{ InitIMem\@s{2.60} \.{\defeq} [ i \.{\in} Writers \.{\mapsto} {\langle}
 InitRegVal ,\, 0 {\rangle} ]}%
\@pvspace{8.0pt}%
\@x{ {\VARIABLES} imem ,\, interface ,\, wrNum ,\, rdVal1 ,\, rdVal2}%
 \@x{ vars \.{\defeq} {\langle} imem ,\, interface ,\, wrNum ,\, rdVal1 ,\,
 rdVal2 {\rangle}}%
\@pvspace{8.0pt}%
\@x{ Init\@s{2.02} \.{\defeq} \.{\land} imem \.{=} InitIMem}%
 \@x{\@s{37.72} \.{\land} interface \.{=} [ i \.{\in} Readers \.{\cup} Writers
 \.{\mapsto}}%
 \@x{\@s{107.47} {\IF} i \.{\in} Readers \.{\THEN} InitMem \.{\ELSE} NotRegVal
 ]}%
\@x{\@s{37.72} \.{\land} wrNum \.{=} [ i \.{\in} Writers \.{\mapsto} 0 ]}%
 \@x{\@s{37.72} \.{\land} rdVal1 \.{=} [ i \.{\in} Readers \.{\mapsto}
 {\langle} {\rangle} ]}%
 \@x{\@s{37.72} \.{\land} rdVal2 \.{=} [ i \.{\in} Readers \.{\mapsto}
 {\langle} {\rangle} ]}%
\@pvspace{8.0pt}%
\@x{ BeginWr ( i ,\, cmd ) \.{\defeq} \.{\dots}}%
\@x{ DoWr ( i ) \.{\defeq} \.{\dots}}%
\@x{ EndWr ( i ) \.{\defeq} \.{\dots}}%
\@x{ BeginRd ( i ) \.{\defeq} \.{\dots}}%
\@x{ AddToFcn ( f ,\, x ,\, v ) \.{\defeq} \.{\dots}}%
\@x{ Rd1 ( i ) \.{\defeq} \.{\dots}}%
\@x{ Rd2 ( i ) \.{\defeq} \.{\dots}}%
\@x{ TryEndRd ( i ) \.{\defeq} \.{\dots}}%
\@pvspace{8.0pt}%
 \@x{ Next \.{\defeq} \.{\lor} \E\, i \.{\in} Readers \.{:} BeginRd ( i )
 \.{\lor} Rd1 ( i ) \.{\lor} Rd2 ( i ) \.{\lor} TryEndRd ( i )}%
 \@x{\@s{39.83} \.{\lor} \E\, i \.{\in} Writers\@s{0.49} \.{:} \.{\lor} \E\,
 cmd \.{\in} RegVals \.{:} BeginWr ( i ,\, cmd )}%
\@x{\@s{118.73} \.{\lor} DoWr ( i ) \.{\lor} EndWr ( i )}%
\@pvspace{8.0pt}%
\@x{ Spec\@s{1.46} \.{\defeq} Init \.{\land} {\Box} [ Next ]_{ vars}}%
\@x{}\bottombar\@xx{}%
\end{tlatex}
\caption{Module \emph{AfekSimplified}} \label{fig:AfekSimplified}
\end{figure}

\subsection{Another Snapshot Specification} \label{sec:another-snapshot}

The algorithm in module $AfekSimplified$ satisfies the safety
specification in $LinearSnapshot$, but we now show that it does not
implement that safety specification under any refinement mapping.  Let
$Spec_{A}$ be the algorithm's specification and let $SSpec_{L}$ be the
specification $SafeSpec$ of $LinearSnapshot$.  We assume there is a
refinement mapping $mem<-\ov{mem}$ and $istate <- \ov{istate}$ under
which $Spec_{A}$ implements $SSpec_{L}$ and obtain a contradiction.
Let $\ov{F}$ be the formula obtained from a formula $F$ of module
$LinearSnapshot$ by replacing $mem$ with $\ov{mem}$ and $istate$ with
$\ov{istate}$.  Consider a behavior satisfying $Spec_{A}$ that begins
with the following three sequences of steps.
\begin{enumerate}
\item 
\begin{sloppypar}
Reader $i$ does a $BeginRd(i)$ step, completes its first scan
(so $\DOMAIN rdVal1{[i]}$ equals $Writers$) and begins its second scan by
reading $imem[j]=<<v_{1},0>>$ for some writer $j$ and $v_{1}$ in
$RegVals$ (so $\DOMAIN rdVal2{[i]}$ equals $\{j\}$ and $rdVal2{[i]}[j]$ equals
$<<v_{1},0>>$).
\end{sloppypar}

\item Writer $j$ then does a complete write operation, writing a new
value $v_{2}$ different from $v_{1}$.

\item Reader $i$ completes its second scan, executes its $TryEndRd(i)$
action, finding $rdVal2[i]$ equal to $rdVal1[i]$, and completing the
read operation by setting $interface[i]$ to a value $M$ with
$M[j]=v_{1}$.
\end{enumerate}
The behavior satisfies \ov{SSpec_{L}}, so this sequence of actions
must start with a \ov{BeginRd(i)} step, contain a \ov{DoRd(i)} step,
and end with an \ov{EndRd(i)} step.  The reader has not determined the
value to be output by the read command until it has finished its
second scan, so the \ov{DoRd(i)} step must occur in sequence~3.  The
three steps of the write of $v_{2}$ by writer $j$ occur in sequence~2,
so the \ov{DoWr(j)} step for that write must occur in that sequence,
therefore preceding the \ov{DoRd(i)} step.  Hence, the
$LinearSnapshot$ spec implies that the \ov{DoRd(i)} step must set
$\ov{istate}[i][j]$ to $v_{2}$.  However in the last step of~3, the
reader sets the value of $interface[i][j]$ to $v_{1}$, which implies
that the \ov{DoRd(i)} step set $\ov{istate}[i][j]$ to $v_{1}$.  Since
$v_{1}#v_{2}$, this is a contradiction, showing that the refinement
mapping cannot exist.

This behavior of $Spec_{A}$ is allowed by \tlabox{\EE
mem,istate:SSpec_{L}}, since we can choose values of $mem$ and $istate$
for which $SSpec_{L}$ is satisfied---namely, values for which the
\ov{DoRd(i)} step occurs before the \ov{DoWr(j)} step.  However,
choosing those values requires knowing what steps occur after the
\ov{DoWr(j)} step.  The linearizability specification $SSpec_{L}$
chooses the value returned by a read sooner than it has to.  This
tells us that to find a refinement mapping that shows $Spec_{A}$
implements \tlabox{\EE mem,istate:SSpec_{L}}, we must add a prophecy
variable to $Spec_{A}$.

Instead of adding a prophecy variable to $Spec_{A}$, we write a new
snapshot specification $Spec_{NL}$ that allows the same externally
visible behaviors as specification $Spec_{L}$ of $LinearSnapshot$; and
whose safety specification $SSpec_{NL}$ allows the same visible
behaviors as $SSpec_{L}$.  However, in $Spec_{NL}$ we make a reader
wait as long as possible before choosing its output value.  We can
then find a refinement mapping to show that $Spec_{A}$ implements
$SSpec_{NL}$ without using a prophecy variable.  

We will still need a prophecy variable to show that $SSpec_{NL}$
allows the same externally visible behavior as $SSpec_{L}$.  The
advantage of introducing $Spec_{NL}$ is that the specification of what
an algorithm is supposed to do is generally much simpler than the
algorithm.  Prophecy variables are the most complicated kind of
auxiliary variables, and it is easier to add one to a high-level
specification than to a lower-level algorithm.  (This same idea of
modifying the high-level specification to avoid adding a prophecy
variable to the algorithm can be applied to the queue example of
Herlihy and Wing~\cite{herlihy:axioms}.)

Specification $Spec_{NL}$ records in its internal state all values of
the memory $mem$ that a read operation is allowed to return.  The
$EndRd$ operation nondeterministically chooses one of those values as
its output.  Its internal state therefore remembers much more about
what happened in the past than a reasonable implementation would.  This
means that defining a refinement mapping under which an algorithm
implements $Spec_{NL}$ will require adding a history variable to the
algorithm's spec.  Adding a history variable is much easier than
adding a prophecy variable.

We write specification $Spec_{NL}$ (and $SSpec_{NL}$) in module
$NewLinearSnapshot$.  It has the same declarations of $Readers$,
$Writers$, $RegVals$, and $InitRegVal$ and the same definitions of
$MemVals$, $InitMem$, $NotMemVal$, and $NotRegVal$ as in module
$LinearSnapshot$.  It has the same variables $interface$ and $mem$ as
module $LinearSnapshot$, plus these two internal variables:
\begin{describe}{$wstate$}
\item[$wstate$] A function with domain $Writers$ such that the value
$wstate[i]$ is the same as the value of $istate[i]$ in
$LinearSnapshot$, for each writer $i$.

\item[$rstate$] A function with domain $Readers$ so that, for each
reader $i$ currently executing a read operation, $rstate[i]$ is the
sequence of values that $mem$ has assumed thus far while the operation
has been executing.  The first element of $rstate[i]$ is therefore the
value $mem$ had when the $BeginRd(i)$ step occurred.  The value of
$rstate[i]$ is the empty sequence $<<\,>>$ when $i$ is not executing a
read operation.
\end{describe}
The $BeginWr$ command is essentially the same as in $LinearSnapshot$.
\begin{display}
\begin{notla}
BeginWr(i, cmd) == /\ interface[i] = NotRegVal
                   /\ interface' = [interface EXCEPT ![i] = cmd]
                   /\ wstate' = [wstate EXCEPT ![i] = cmd]
                   /\ UNCHANGED <<mem, rstate>>
\end{notla}
\begin{tlatex}
 \@x{ BeginWr ( i ,\, cmd ) \.{\defeq} \.{\land} interface [ i ] \.{=}
 NotRegVal}%
 \@x{\@s{95.48} \.{\land} interface \.{'} \.{=} [ interface {\EXCEPT} {\bang}
 [ i ] \.{=} cmd ]}%
 \@x{\@s{95.48} \.{\land} wstate \.{'} \.{=} [ wstate {\EXCEPT} {\bang} [ i ]
 \.{=} cmd ]}%
\@x{\@s{95.48} \.{\land} {\UNCHANGED} {\langle} mem ,\, rstate {\rangle}}%
\end{tlatex}
\end{display}
The $BeginRd(i)$ command, which sets $rstate[i]$ to a one-element sequence
containing the current value of $mem$, is:
\begin{display}
\begin{notla}
BeginRd(i) == /\ interface[i] \in MemVals
              /\ interface' = [interface EXCEPT ![i] = NotMemVal]
              /\ rstate' = [rstate EXCEPT ![i] = <<mem>>]
              /\ UNCHANGED <<mem, wstate>> 
\end{notla}
\begin{tlatex}
\@x{ BeginRd ( i ) \.{\defeq} \.{\land} interface [ i ] \.{\in} MemVals}%
 \@x{\@s{68.09} \.{\land} interface \.{'} \.{=} [ interface {\EXCEPT} {\bang}
 [ i ] \.{=} NotMemVal ]}%
 \@x{\@s{68.09} \.{\land} rstate \.{'} \.{=} [ rstate {\EXCEPT} {\bang} [ i ]
 \.{=} {\langle} mem {\rangle} ]}%
\@x{\@s{68.09} \.{\land} {\UNCHANGED} {\langle} mem ,\, wstate {\rangle}}%
\end{tlatex}
\end{display}
The writer executes a $DoWr$ that is the same as in $LinearSnapshot$, except
that it also appends the new value of $mem$ to the end of $rstate[j]$
for every reader $j$ currently executing a read operation.
\begin{display}
\begin{notla}
DoWr(i) == /\ interface[i] \in RegVals
           /\ wstate[i] = interface[i]
           /\ mem' = [mem EXCEPT ![i] = interface[i]]
           /\ wstate' = [wstate EXCEPT ![i] = NotRegVal]
           /\ rstate' = [j \in Readers |-> 
                            IF rstate[j] = << >>
                              THEN << >>
                              ELSE Append(rstate[j], mem')]
           /\ interface' = interface
\end{notla}
\begin{tlatex}
\@x{ DoWr ( i ) \.{\defeq} \.{\land} interface [ i ] \.{\in} RegVals}%
\@x{\@s{58.69} \.{\land} wstate [ i ] \.{=} interface [ i ]}%
 \@x{\@s{58.69} \.{\land} mem \.{'} \.{=} [ mem {\EXCEPT} {\bang} [ i ] \.{=}
 interface [ i ] ]}%
 \@x{\@s{58.69} \.{\land} wstate \.{'} \.{=} [ wstate {\EXCEPT} {\bang} [ i ]
 \.{=} NotRegVal ]}%
 \@x{\@s{58.69} \.{\land} rstate \.{'}\@s{2.42} \.{=} [ j \.{\in} Readers
 \.{\mapsto}}%
\@x{\@s{125.17} {\IF} rstate [ j ] \.{=} {\langle} {\rangle}}%
\@x{\@s{133.37} \.{\THEN} {\langle} {\rangle}}%
\@x{\@s{133.37} \.{\ELSE} Append ( rstate [ j ] ,\, mem \.{'} ) ]}%
\@x{\@s{58.69} \.{\land} interface \.{'} \.{=} interface}%
\end{tlatex}
\end{display}
A reader $i$ has no internal actions, only the externally visible
$BeginRd(i)$ and $EndRd(i)$ actions.  Its $EndRd(i)$ action outputs an
arbitrarily chosen element of $rstate[i]$.
\begin{display}
\begin{notla}
EndRd(i) == /\ interface[i] = NotMemVal
            /\ \E j \in 1..Len(rstate[i]) :
                   interface' = [interface EXCEPT ![i] = rstate[i][j]]
            /\ rstate' = [rstate EXCEPT ![i] = << >>]
            /\ UNCHANGED <<mem, wstate>>
\end{notla}
\begin{tlatex}
\@x{ EndRd ( i ) \.{\defeq} \.{\land} interface [ i ] \.{=} NotMemVal}%
 \@x{\@s{61.19} \.{\land} \E\, j \.{\in} 1 \.{\dotdot} Len ( rstate [ i ] )
 \.{:}}%
 \@x{\@s{83.62} interface \.{'} \.{=} [ interface {\EXCEPT} {\bang} [ i ]
 \.{=} rstate [ i ] [ j ] ]}%
 \@x{\@s{61.19} \.{\land} rstate \.{'} \.{=} [ rstate {\EXCEPT} {\bang} [ i ]
 \.{=} {\langle} {\rangle} ]}%
\@x{\@s{61.19} \.{\land} {\UNCHANGED} {\langle} mem ,\, wstate {\rangle}}%
\end{tlatex}
\end{display}
The writer's $EndWr$ action is essentially the same as in
$LinearSnapshot$.
\begin{display}
\begin{notla}
EndWr(i) == /\ interface[i] \in RegVals
            /\ wstate[i] = NotRegVal
            /\ interface' = [interface EXCEPT ![i] = wstate[i]]
            /\ UNCHANGED <<mem, rstate, wstate>> 
\end{notla}
\begin{tlatex}
\@x{ EndWr ( i ) \.{\defeq} \.{\land} interface [ i ] \.{\in} RegVals}%
\@x{\@s{63.55} \.{\land} wstate [ i ] \.{=} NotRegVal}%
 \@x{\@s{63.55} \.{\land} interface \.{'} \.{=} [ interface {\EXCEPT} {\bang}
 [ i ] \.{=} wstate [ i ] ]}%
 \@x{\@s{63.55} \.{\land} {\UNCHANGED} {\langle} mem ,\, rstate ,\, wstate
 {\rangle}}%
\end{tlatex}
\end{display}
The complete module, minus the action definitions above, is in
\lref{targ:NewLinearSnapshot}{Figure~\ref{fig:NewLinearSnapshot}}.
\begin{figure} \target{targ:NewLinearSnapshot}
\begin{notla}
------------------------- MODULE NewLinearSnapshot -------------------------
EXTENDS Integers, Sequences

CONSTANTS Readers, Writers, RegVals, InitRegVal

ASSUME /\ Readers  \cap Writers = {}
       /\ InitRegVal \in RegVals

InitMem == [i \in Writers |-> InitRegVal]
MemVals == [Writers -> RegVals]
NotMemVal == CHOOSE v : v \notin MemVals
NotRegVal == CHOOSE v : v \notin RegVals

VARIABLES mem, interface, rstate, wstate
vars == <<mem, interface, rstate, wstate>>

Init == /\ mem = InitMem
        /\ interface = [i \in Readers \cup Writers |->
                          IF i \in Readers THEN InitMem ELSE NotRegVal]
        /\ rstate = [i \in Readers |-> << >>]
        /\ wstate = [i \in Writers |-> NotRegVal]

BeginRd(i) == ...
BeginWr(i, cmd) == ...
DoWr(i) == ...
EndRd(i) == ...
EndWr(i) == ...

Next == \/ \E i \in Readers : BeginRd(i) \/ EndRd(i)
        \/ \E i \in Writers : \/ \E cmd \in RegVals : BeginWr(i, cmd)
                              \/ DoWr(i) \/ EndWr(i)

SafeSpec ==  Init /\ [][Next]_vars

Fairness == /\ \A i \in Readers : WF_vars(EndRd(i))
            /\ \A i \in Writers : WF_vars(DoWr(i)) /\ WF_vars(EndWr(i))
                    
Spec == Init /\ [][Next]_vars /\ Fairness   
\end{notla}
\begin{tlatex}
\@x{}\moduleLeftDash\@xx{ {\MODULE} NewLinearSnapshot}\moduleRightDash\@xx{}%
\@x{ {\EXTENDS} Integers ,\, Sequences}%
\@pvspace{8.0pt}%
\@x{ {\CONSTANTS} Readers ,\, Writers ,\, RegVals ,\, InitRegVal}%
\@pvspace{8.0pt}%
\@x{ {\ASSUME} \.{\land} Readers\@s{4.1} \.{\cap} Writers \.{=} \{ \}}%
\@x{\@s{38.24} \.{\land} InitRegVal \.{\in} RegVals}%
\@pvspace{8.0pt}%
 \@x{ InitMem\@s{2.60} \.{\defeq} [ i \.{\in} Writers \.{\mapsto} InitRegVal
 ]}%
\@x{ MemVals \.{\defeq} [ Writers \.{\rightarrow} RegVals ]}%
\@x{ NotMemVal \.{\defeq} {\CHOOSE} v \.{:} v \.{\notin} MemVals}%
\@x{ NotRegVal\@s{6.27} \.{\defeq} {\CHOOSE} v \.{:} v \.{\notin} RegVals}%
\@pvspace{8.0pt}%
\@x{ {\VARIABLES} mem ,\, interface ,\, rstate ,\, wstate}%
 \@x{ vars \.{\defeq} {\langle} mem ,\, interface ,\, rstate ,\, wstate
 {\rangle}}%
\@pvspace{16.0pt}%
\@x{ Init\@s{2.02} \.{\defeq} \.{\land} mem \.{=} InitMem}%
 \@x{\@s{37.72} \.{\land} interface \.{=} [ i \.{\in} Readers \.{\cup} Writers
 \.{\mapsto}}%
 \@x{\@s{107.47} {\IF} i \.{\in} Readers \.{\THEN} InitMem \.{\ELSE} NotRegVal
 ]}%
 \@x{\@s{37.72} \.{\land} rstate\@s{2.42} \.{=} [ i \.{\in} Readers
 \.{\mapsto} {\langle} {\rangle} ]}%
 \@x{\@s{37.72} \.{\land} wstate \.{=} [ i \.{\in} Writers\@s{0.49}
 \.{\mapsto} NotRegVal ]}%
\@pvspace{8.0pt}%
\@x{ BeginRd ( i ) \.{\defeq} \.{\dots}}%
\@x{ BeginWr ( i ,\, cmd ) \.{\defeq} \.{\dots}}%
\@x{ DoWr ( i ) \.{\defeq} \.{\dots}}%
\@x{ EndRd ( i )\@s{2.35} \.{\defeq} \.{\dots}}%
\@x{ EndWr ( i ) \.{\defeq} \.{\dots}}%
\@pvspace{8.0pt}%
 \@x{ Next \.{\defeq} \.{\lor} \E\, i \.{\in} Readers \.{:} BeginRd ( i )
 \.{\lor} EndRd ( i )}%
 \@x{\@s{39.83} \.{\lor} \E\, i \.{\in} Writers\@s{0.49} \.{:} \.{\lor} \E\,
 cmd \.{\in} RegVals \.{:} BeginWr ( i ,\, cmd )}%
\@x{\@s{118.73} \.{\lor} DoWr ( i ) \.{\lor} EndWr ( i )}%
\@pvspace{8.0pt}%
\@x{ SafeSpec \.{\defeq}\@s{4.1} Init \.{\land} {\Box} [ Next ]_{ vars}}%
\@pvspace{8.0pt}%
 \@x{ Fairness\@s{0.49} \.{\defeq} \.{\land} \A\, i \.{\in} Readers \.{:}
 {\WF}_{ vars} ( EndRd ( i ) )}%
 \@x{\@s{56.76} \.{\land} \A\, i \.{\in} Writers\@s{0.49} \.{:} {\WF}_{ vars}
 ( DoWr ( i ) ) \.{\land} {\WF}_{ vars} ( EndWr ( i ) )}%
\@pvspace{8.0pt}%
 \@x{ Spec \.{\defeq} Init \.{\land} {\Box} [ Next ]_{ vars} \.{\land}
 Fairness}%
\end{tlatex}
\caption{Module \emph{NewLinearSnapshot}\label{fig:NewLinearSnapshot}}
\end{figure}

\subsection{\emph{NewLinearSnapshot} Implements \emph{LinearSnapshot}}

For compactness, in the following discussion we let:
\[ \begin{noj3}
   \Sp_{L} & == & \tlabox{\EE mem, istate:Spec_{L}} \V{.2}
   \Sp_{NL} & == & \tlabox{\EE mem, rstate, wstate : Spec_{NL}}
   \end{noj3}\]
Specifications $\Sp_{L}$ and $\Sp_{NL}$ are equivalent.  However, our
goal is to prove that $Spec_{A}$ implements $\Sp_{L}$, for which it
suffices to show that it implements specification $\Sp_{NL}$ and that
$\Sp_{NL}$ implements $\Sp_{L}$.  So, we won't bother showing
equivalence of the two specs; we just show here that $\Sp_{NL}$
implements $\Sp_{L}$.  We show in Section~\ref{sec:Afek-implements}
below that $Spec_{A}$ implements $\Sp_{NL}$.

To show that $\Sp_{NL}$ implements $\Sp_{L}$, we 
add to $Spec_{NL}$ a prophecy variable $p$
then a stuttering variable $s$ to obtain a specification
 $Spec_{NL}^{ps}$ such that 
\tlabox{\EE s, p : Spec_{NL}^{ps}}
is equivalent to $Spec_{NL}$.  We then show that
\tlabox{\EE s, p : Spec_{NL}^{ps}} implements $\Sp_{L}$ by showing
that
 $Spec_{NL}^{ps}$ implements $Spec_{L}$ under
a suitable refinement mapping $mem<-\ov{mem}$, $istate <- \ov{istate}$.

The two auxiliary variables we add to $Spec_{NL}$ have the following
functions:
\begin{describe}{$s$}
\item[$p$] A prophecy variable that predicts for each reader $i$
 which element of the sequence of memory values $rstate[i]$ will
 be chosen as the output.

\item[$s$] A stuttering variable that adds: 
\begin{itemize}

\item A single stuttering step after a $BeginRd(i)$ step if $p[i]$
predicts that the read will return the current value of memory.  The
refinement mapping will be defined so that stuttering step will be a
\ov{DoRd(i)}.

\item Stuttering steps after a $DoWr(i)$ step that will implement
       the \ov{DoRd(j)} step of every current read operation that 
       returns the value of $mem$ immediately after the $DoWr(i)$ step.
\end{itemize}
\end{describe}
Both these variables are added in a single module named
$NewLinearSnapshotPS$.

\subsubsection{Adding the Prophecy Variable}

The prophecy variable $p$ is a prophecy data structure variable as
described in Section~\ref{sec:proph-data-struct}.  Its domain $Dom$
is the set of readers that are currently executing a read, which can be
described as the set of readers $i$ such that $rstate[i]$ is a
nonempty sequence.  The value of $p[i]$ is a positive integer that
predicts which element of the list $rstate[i]$ will be chosen as the
output.  This value can be arbitrarily large, since arbitrarily many
writes can occur during a read operation, so $\Pi$ is the set $Nat :\:
\{0\}$.  Module $NewLinearSnapshotPS$ therefore begins
\begin{display}
\begin{notla}
EXTENDS NewLinearSnapshot

Pi == Nat \ {0}
Dom == {r \in Readers : rstate[r] # << >>}
INSTANCE Prophecy WITH DomPrime <- Dom'
\end{notla}
\begin{tlatex}
\@x{ {\EXTENDS} NewLinearSnapshot}%
\@pvspace{5.0pt}%
\@x{ Pi \.{\defeq} Nat \.{\,\backslash\,} \{ 0 \}}%
 \@x{ Dom \.{\defeq} \{ r \.{\in} Readers \.{:} rstate [ r ] \.{\neq}
 {\langle} {\rangle} \}}%
\@x{ {\INSTANCE} Prophecy {\WITH} DomPrime \.{\leftarrow} Dom \.{'}}%
\end{tlatex}
\end{display}
\begin{sloppypar} \noindent
It is most convenient to define $p$ in terms of a disjunctive representation
in which $EndRd(i)$ is decomposed into 
  \tlabox{\,\E \, j \in 1\dd Len(rstate[i]) : IEndRd(i,j)\,},
where $IEndRd$ can be defined by:
\end{sloppypar}
\begin{display}
\begin{notla}
IEndRd(i, j) == /\ interface[i] = NotMemVal
                /\ interface' = [interface EXCEPT ![i] = rstate[i][j]]
                /\ rstate' = [rstate EXCEPT ![i] = << >>]
                /\ UNCHANGED <<mem, wstate>>
\end{notla}
\begin{tlatex}
\@x{ IEndRd ( i ,\, j ) \.{\defeq} \.{\land} interface [ i ] \.{=} NotMemVal}%
 \@x{\@s{75.67} \.{\land} interface \.{'} \.{=} [ interface {\EXCEPT} {\bang}
 [ i ] \.{=} rstate [ i ] [ j ] ]}%
 \@x{\@s{75.67} \.{\land} rstate \.{'} \.{=} [ rstate {\EXCEPT} {\bang} [ i ]
 \.{=} {\langle} {\rangle} ]}%
\@x{\@s{75.67} \.{\land} {\UNCHANGED} {\langle} mem ,\, wstate {\rangle}}%
\end{tlatex}
\end{display}
We could make the change in our original specification $NewLinearSnapshot$,
but instead we define a new next-state action $Nxt$ that is equivalent
to $Next$:
\begin{display}
\begin{notla}
Nxt == \/ \E i \in Readers : \/ BeginRd(i)  
                             \/ \E j \in 1..Len(rstate[i]) : IEndRd(i,j)
       \/ \E i \in Writers : \/ \E cmd \in RegVals : BeginWr(i, cmd)
                             \/ DoWr(i) \/ EndWr(i)
\end{notla}
\begin{tlatex}
 \@x{ Nxt \.{\defeq} \.{\lor} \E\, i \.{\in} Readers \.{:} \.{\lor} BeginRd (
 i )}%
 \@x{\@s{114.13} \.{\lor} \E\, j \.{\in} 1 \.{\dotdot} Len ( rstate [ i ] )
 \.{:} IEndRd ( i ,\, j )}%
 \@x{\@s{35.23} \.{\lor} \E\, i \.{\in} Writers\@s{0.49} \.{:} \.{\lor} \E\,
 cmd \.{\in} RegVals \.{:} BeginWr ( i ,\, cmd )}%
\@x{\@s{114.13} \.{\lor} DoWr ( i ) \.{\lor} EndWr ( i )}%
\end{tlatex}
\end{display}
It's easy to see that $Nxt$ is equivalent to formula $Next$ of
$NewLinearSnapShot$, and TLAPS can easily check this proof.
\begin{display}
\begin{notla}
THEOREM Next = Nxt
BY DEF Next, Nxt, EndRd, IEndRd
\end{notla}
\begin{tlatex}
\@x{ {\THEOREM} Next \.{=} Nxt}%
\@x{ {\BY} {\DEF} Next ,\, Nxt ,\, EndRd ,\, IEndRd}%
\end{tlatex}
\end{display}
A prediction is made for reader $i$ when the element $i$ is added to
$Dom$, which is done by a $BeginRd(i)$ step.  The prediction is used
by the $IEndRd(i, j)$ action, allowing it to be performed only if
$p[i]$ has predicted that the $j$\tth\ item in the sequence
$rstate[i]$ will be output.  The definitions of $Pred_{A}$, $PredDom_{A}$,
and $DomInj_{A}$ for the subactions $A$ are given along with the
beginning of the module in \lref{targ:NewLinearSnapshotPS1}{Figure~\ref{fig:NewLinearSnapshotPS1}}.
\begin{figure} \target{targ:NewLinearSnapshotPS1}
\begin{notla}
------------------------ MODULE NewLinearSnapshotPS ------------------------
EXTENDS NewLinearSnapshot

Pi == Nat \ {0}
Dom == {r \in Readers : rstate[r] # << >>}
INSTANCE Prophecy WITH DomPrime <- Dom'

IEndRd(i, j) == /\ interface[i] = NotMemVal
                /\ interface' = [interface EXCEPT ![i] = rstate[i][j]]
                /\ rstate' = [rstate EXCEPT ![i] = << >>]
                /\ UNCHANGED <<mem, wstate>>

Nxt == \/ \E i \in Readers : \/ BeginRd(i)  
                             \/ \E j \in 1..Len(rstate[i]) : IEndRd(i,j)
       \/ \E i \in Writers : \/ \E cmd \in RegVals : BeginWr(i, cmd)
                             \/ DoWr(i) \/ EndWr(i)

THEOREM Next = Nxt
BY DEF Next, Nxt, EndRd, IEndRd
                   
PredBeginRd(p) == TRUE
PredDomBeginRd == {}
DomInjBeginRd  == IdFcn(Dom)

PredIEndRd(p, i, j) == j = p[i]
PredDomIEndRd(i) == {i}
DomInjIEndRd  == IdFcn(Dom')

PredBeginWr(p) == TRUE
PredDomBeginWr == {}
DomInjBeginWr  == IdFcn(Dom)

PredDoWr(p) == TRUE
PredDomDoWr == {}
DomInjDoWr  == IdFcn(Dom)

PredEndWr(p) == TRUE
PredDomEndWr == {}
DomInjEndWr  == IdFcn(Dom)
\end{notla}
\begin{tlatex}
 \@x{}\moduleLeftDash\@xx{ {\MODULE}
 NewLinearSnapshotPS}\moduleRightDash\@xx{}%
\@x{ {\EXTENDS} NewLinearSnapshot}%
\@pvspace{8.0pt}%
\@x{ Pi \.{\defeq} Nat \.{\,\backslash\,} \{ 0 \}}%
 \@x{ Dom \.{\defeq} \{ r \.{\in} Readers \.{:} rstate [ r ] \.{\neq}
 {\langle} {\rangle} \}}%
\@x{ {\INSTANCE} Prophecy {\WITH} DomPrime \.{\leftarrow} Dom \.{'}}%
\@pvspace{8.0pt}%
\@x{ IEndRd ( i ,\, j ) \.{\defeq} \.{\land} interface [ i ] \.{=} NotMemVal}%
 \@x{\@s{75.67} \.{\land} interface \.{'} \.{=} [ interface {\EXCEPT} {\bang}
 [ i ] \.{=} rstate [ i ] [ j ] ]}%
 \@x{\@s{75.67} \.{\land} rstate \.{'} \.{=} [ rstate {\EXCEPT} {\bang} [ i ]
 \.{=} {\langle} {\rangle} ]}%
\@x{\@s{75.67} \.{\land} {\UNCHANGED} {\langle} mem ,\, wstate {\rangle}}%
\@pvspace{8.0pt}%
 \@x{ Nxt \.{\defeq} \.{\lor} \E\, i \.{\in} Readers \.{:} \.{\lor} BeginRd (
 i )}%
 \@x{\@s{114.13} \.{\lor} \E\, j \.{\in} 1 \.{\dotdot} Len ( rstate [ i ] )
 \.{:} IEndRd ( i ,\, j )}%
 \@x{\@s{35.23} \.{\lor} \E\, i \.{\in} Writers\@s{0.49} \.{:} \.{\lor} \E\,
 cmd \.{\in} RegVals \.{:} BeginWr ( i ,\, cmd )}%
\@x{\@s{114.13} \.{\lor} DoWr ( i ) \.{\lor} EndWr ( i )}%
\@pvspace{8.0pt}%
\@x{ {\THEOREM} Next \.{=} Nxt}%
\@x{ {\BY} {\DEF} Next ,\, Nxt ,\, EndRd ,\, IEndRd}%
\@pvspace{8.0pt}%
\@x{ PredBeginRd ( p )\@s{7.31} \.{\defeq} {\TRUE}}%
\@x{ PredDomBeginRd \.{\defeq} \{ \}}%
\@x{ DomInjBeginRd\@s{7.14} \.{\defeq} IdFcn ( Dom )}%
\@pvspace{8.0pt}%
\@x{ PredIEndRd ( p ,\, i ,\, j ) \.{\defeq} j \.{=} p [ i ]}%
\@x{ PredDomIEndRd ( i ) \.{\defeq} \{ i \}}%
\@x{ DomInjIEndRd\@s{4.1} \.{\defeq} IdFcn ( Dom \.{'} )}%
\@pvspace{8.0pt}%
\@x{ PredBeginWr ( p )\@s{7.31} \.{\defeq} {\TRUE}}%
\@x{ PredDomBeginWr \.{\defeq} \{ \}}%
\@x{ DomInjBeginWr\@s{7.14} \.{\defeq} IdFcn ( Dom )}%
\@pvspace{8.0pt}%
\@x{ PredDoWr ( p )\@s{7.31} \.{\defeq} {\TRUE}}%
\@x{ PredDomDoWr \.{\defeq} \{ \}}%
\@x{ DomInjDoWr\@s{7.14} \.{\defeq} IdFcn ( Dom )}%
\@pvspace{8.0pt}%
\@x{ PredEndWr ( p )\@s{7.31} \.{\defeq} {\TRUE}}%
\@x{ PredDomEndWr \.{\defeq} \{ \}}%
\@x{ DomInjEndWr\@s{7.14} \.{\defeq} IdFcn ( Dom )}%
\end{tlatex}
\caption{Module \emph{NewLinearSnapshotPS}, part 1.}
\label{fig:NewLinearSnapshotPS1}
\end{figure}
The module next defines the temporal formula $Condition$, which should
be implied by $Spec$.  There follows the definition of the
specification $SpecP$ obtained by adding the prophecy variable $p$ to
$Spec$.  TLC can check that $Condition$ is implied by $Spec$, which
implies that \tlabox{\EE p: SpecP} is equivalent to $Spec$.  These
definitions appear in \lref{targ:NewLinearSnapshotPS2}{Figure~\ref{fig:NewLinearSnapshotPS2}}.
\begin{figure} \target{targ:NewLinearSnapshotPS2}
\begin{notla}
Condition ==
 [][ /\ \A i \in Readers :
          /\ ProphCondition(BeginRd(i), DomInjBeginRd, 
                            PredDomBeginRd, PredBeginRd)
          /\ \A j \in 1..Len(rstate[i]) :
                ProphCondition(IEndRd(i, j), DomInjIEndRd, 
                               PredDomIEndRd(i),
                               LAMBDA p : PredIEndRd(p, i, j))
     /\ \A i \in Writers :
          /\ \A cmd \in RegVals :
                ProphCondition(BeginWr(i, cmd), DomInjBeginWr, 
                               PredDomBeginWr, PredBeginWr)
          /\ ProphCondition(DoWr(i), DomInjDoWr, PredDomDoWr, 
                            PredDoWr)
          /\ ProphCondition(EndWr(i), DomInjEndWr, PredDomEndWr, 
                            PredEndWr)
   ]_vars

VARIABLE p
varsP == <<vars, p>>

InitP == Init /\ (p = EmptyFcn)

BeginRdP(i) == ProphAction(BeginRd(i), p, p', DomInjBeginRd,  
                           PredDomBeginRd, PredBeginRd)

BeginWrP(i, cmd) == ProphAction(BeginWr(i, cmd), p, p', DomInjBeginWr,  
                                PredDomBeginWr, PredBeginWr)
          
DoWrP(i) == ProphAction(DoWr(i), p, p', DomInjDoWr, 
                        PredDomDoWr, PredDoWr)

IEndRdP(i, j) == ProphAction(IEndRd(i, j),  p, p', DomInjIEndRd, 
                             PredDomIEndRd(i), 
                             LAMBDA q : PredIEndRd(q, i, j))

EndWrP(i) == ProphAction(EndWr(i), p, p', DomInjEndWr, 
                         PredDomEndWr, PredEndWr)

NextP == \/ \E i \in Readers : \/ BeginRdP(i)  
                               \/ \E j \in 1..Len(rstate[i]) : IEndRdP(i,j)
         \/ \E i \in Writers : \/ \E cmd \in RegVals : BeginWrP(i, cmd)
                               \/ DoWrP(i) \/ EndWrP(i)

SpecP == InitP /\ [][NextP]_varsP /\ Fairness
\end{notla}
\begin{tlatex}
\@x{ Condition \.{\defeq}}%
\@x{\@s{4.1} {\Box} [ \.{\land} \A\, i \.{\in} Readers \.{:}}%
 \@x{\@s{33.66} \.{\land} ProphCondition ( BeginRd ( i ) ,\, DomInjBeginRd
 ,\,}%
\@x{\@s{118.43} PredDomBeginRd ,\, PredBeginRd )}%
 \@x{\@s{33.66} \.{\land} \A\, j \.{\in} 1 \.{\dotdot} Len ( rstate [ i ] )
 \.{:}}%
\@x{\@s{51.99} ProphCondition ( IEndRd ( i ,\, j ) ,\, DomInjIEndRd ,\,}%
\@x{\@s{125.66} PredDomIEndRd ( i ) ,\,}%
\@x{\@s{125.66} {\LAMBDA} p \.{:} PredIEndRd ( p ,\, i ,\, j ) )}%
\@x{\@s{14.35} \.{\land} \A\, i \.{\in} Writers \.{:}}%
\@x{\@s{33.66} \.{\land} \A\, cmd \.{\in} RegVals \.{:}}%
\@x{\@s{51.99} ProphCondition ( BeginWr ( i ,\, cmd ) ,\, DomInjBeginWr ,\,}%
\@x{\@s{125.66} PredDomBeginWr ,\, PredBeginWr )}%
 \@x{\@s{33.66} \.{\land} ProphCondition ( DoWr ( i ) ,\, DomInjDoWr ,\,
 PredDomDoWr ,\,}%
\@x{\@s{118.43} PredDoWr )}%
 \@x{\@s{33.66} \.{\land} ProphCondition ( EndWr ( i ) ,\, DomInjEndWr ,\,
 PredDomEndWr ,\,}%
\@x{\@s{118.43} PredEndWr )}%
\@x{\@s{11.57} ]_{ vars}}%
\@pvspace{4.0pt}%
\@x{ {\VARIABLE} p}%
\@x{ varsP \.{\defeq} {\langle} vars ,\, p {\rangle}}%
\@pvspace{4.0pt}%
\@x{ InitP\@s{2.14} \.{\defeq} Init \.{\land} ( p \.{=} EmptyFcn )}%
\@pvspace{4.0pt}%
 \@x{ BeginRdP ( i ) \.{\defeq} ProphAction ( BeginRd ( i ) ,\, p ,\, p \.{'}
 ,\, DomInjBeginRd ,\,}%
\@x{\@s{133.99} PredDomBeginRd ,\, PredBeginRd )}%
\@pvspace{4.0pt}%
 \@x{ BeginWrP ( i ,\, cmd ) \.{\defeq} ProphAction ( BeginWr ( i ,\, cmd )
 ,\, p ,\, p \.{'} ,\, DomInjBeginWr ,\,}%
\@x{\@s{161.33} PredDomBeginWr ,\, PredBeginWr )}%
\@pvspace{4.0pt}%
 \@x{ DoWrP ( i ) \.{\defeq} ProphAction ( DoWr ( i ) ,\, p ,\, p \.{'} ,\,
 DomInjDoWr ,\,}%
\@x{\@s{124.54} PredDomDoWr ,\, PredDoWr )}%
\@pvspace{4.0pt}%
 \@x{ IEndRdP ( i ,\, j ) \.{\defeq} ProphAction ( IEndRd ( i ,\, j )
 ,\,\@s{4.1} p ,\, p \.{'} ,\, DomInjIEndRd ,\,}%
\@x{\@s{141.57} PredDomIEndRd ( i ) ,\,}%
\@x{\@s{141.57} {\LAMBDA} q \.{:} PredIEndRd ( q ,\, i ,\, j ) )}%
\@pvspace{4.0pt}%
 \@x{ EndWrP ( i ) \.{\defeq} ProphAction ( EndWr ( i ) ,\, p ,\, p \.{'} ,\,
 DomInjEndWr ,\,}%
\@x{\@s{129.40} PredDomEndWr ,\, PredEndWr )}%
\@pvspace{4.0pt}%
 \@x{ NextP \.{\defeq} \.{\lor} \E\, i \.{\in} Readers \.{:} \.{\lor} BeginRdP
 ( i )}%
 \@x{\@s{125.59} \.{\lor} \E\, j \.{\in} 1 \.{\dotdot} Len ( rstate [ i ] )
 \.{:} IEndRdP ( i ,\, j )}%
 \@x{\@s{46.69} \.{\lor} \E\, i \.{\in} Writers\@s{0.49} \.{:} \.{\lor} \E\,
 cmd \.{\in} RegVals \.{:} BeginWrP ( i ,\, cmd )}%
\@x{\@s{125.59} \.{\lor} DoWrP ( i ) \.{\lor} EndWrP ( i )}%
\@pvspace{4.0pt}%
 \@x{ SpecP\@s{1.08} \.{\defeq} InitP \.{\land} {\Box} [ NextP ]_{ varsP}
 \.{\land} Fairness}%
\end{tlatex}
\caption{Module \emph{NewLinearSnapshotPS}, part 2.}
\label{fig:NewLinearSnapshotPS2}
\end{figure}

\subsubsection{Adding the Stuttering Variable}

The module next adds the stuttering variable $s$ to $SpecS$.  We need
to add a single stuttering step after a $BeginRdP(i)$ step iff the
reader will output the current value of $mem$, which is the case iff
the step sets $p[i]$ to 1.  We also need to add a stuttering step
after ${DoWrP(i)}$ for every currently reading reader $j$ for which
$p[j]$ predicts that the value {of $mem$} that the step appends to
$rstate[j]$ is the one that the read will output.  These steps are
added by letting the values of $s.val$ be subsets of readers, ordered
by the subset relation, with the decrement operation removing an
element from the set chosen with the \textsc{choose} operator.

The specification $SpecPS$ obtained by adding the stuttering variable
$s$ to $SpecP$ is defined in the part of module $NewLinearSnapshotPS$
shown in
\lref{targ:NewLinearSnapshotPS3}{Figure~\ref{fig:NewLinearSnapshotPS3}}.
The two theorems at the beginning are conditions
(\ref{eq:stuttering-cond}) for adding the stuttering steps to
$BeginRdP(i)$ and $DoWrP(i)$ steps.  They can be checked by 
temporarily ending the module immediately after those theorems and
running TLC on a model having $SpecP$ as its specification.  Two
\textsc{assume} statements have been added to check the constant
conditions on the arguments of the $MayPostStutter$ operators used to
add those stuttering steps.

\begin{figure} \target{targ:NewLinearSnapshotPS3}
\begin{notla}
THEOREM SpecP => [][\A i \in Readers : BeginRdP(i) => 
                      (IF p'[i] = 1 THEN 1 ELSE 0) \in {0,1}]_varsP

THEOREM SpecP => [][\A i \in Writers, cmd \in RegVals :
                      DoWrP(i) => 
                        {j \in Readers : (rstate[j] # << >>) 
                                            /\  (p[j] = Len(rstate'[j]))}
                          \in (SUBSET Readers)]_varsP
------------------------------------------------------------------------
VARIABLE s
varsPS == <<vars, p, s>>

INSTANCE Stuttering WITH vars <- varsP 

InitPS == InitP /\ (s = top)

BeginRdPS(i) == MayPostStutter(BeginRdP(i), "BeginRd", i, 0, 
                               IF p'[i] = 1 THEN 1 ELSE 0, 
                               LAMBDA j : j-1)   
ASSUME StutterConstantCondition({0,1}, 0, LAMBDA j : j-1)    

BeginWrPS(i, cmd) == NoStutter(BeginWrP(i, cmd))

DoWrPS(i) == MayPostStutter(DoWrP(i), "DoWr", i, {}, 
                            {j \in Readers : 
                              (rstate[j] # << >>) /\ (p[j] = Len(rstate'[j]))},
                            LAMBDA S : S \ {CHOOSE x \in S : TRUE})
ASSUME StutterConstantCondition(SUBSET Readers, {}, 
                                LAMBDA S : S \ {CHOOSE x \in S : TRUE}) 

IEndRdPS(i, j) == NoStutter(IEndRdP(i, j))

EndWrPS(i) == NoStutter(EndWrP(i))

NextPS == \/ \E i \in Readers : \/ BeginRdPS(i)  
                                \/ \E j \in 1..Len(rstate[i]) : IEndRdPS(i,j)
          \/ \E i \in Writers : \/ \E cmd \in RegVals : BeginWrPS(i, cmd)
                                \/ DoWrPS(i) \/ EndWrPS(i)

SafeSpecPS == InitPS /\ [][NextPS]_varsPS
SpecPS == SafeSpecPS /\ Fairness
\end{notla}
\begin{tlatex}
 \@x{ {\THEOREM} SpecP \.{\implies} {\Box} [ \A\, i \.{\in} Readers \.{:}
 BeginRdP ( i ) \.{\implies}}%
 \@x{\@s{107.54} ( {\IF} p \.{'} [ i ] \.{=} 1 \.{\THEN} 1 \.{\ELSE} 0 )
 \.{\in} \{ 0 ,\, 1 \} ]_{ varsP}}%
\@pvspace{8.0pt}%
 \@x{ {\THEOREM} SpecP \.{\implies} {\Box} [ \A\, i \.{\in} Writers ,\, cmd
 \.{\in} RegVals \.{:}}%
\@x{\@s{107.54} DoWrP ( i ) \.{\implies}}%
 \@x{\@s{115.74} \{ j \.{\in} Readers \.{:} ( rstate [ j ] \.{\neq} {\langle}
 {\rangle} )}%
 \@x{\@s{193.83} \.{\land}\@s{6.10} ( p [ j ] \.{=} Len ( rstate \.{'} [ j ] )
 ) \}}%
\@x{\@s{124.84} \.{\in} ( {\SUBSET} Readers ) ]_{ varsP}}%
\@x{}\midbar\@xx{}%
\@x{ {\VARIABLE} s}%
\@x{ varsPS \.{\defeq} {\langle} vars ,\, p ,\, s {\rangle}}%
\@pvspace{8.0pt}%
\@x{ {\INSTANCE} Stuttering {\WITH} vars \.{\leftarrow} varsP}%
\@pvspace{8.0pt}%
\@x{ InitPS \.{\defeq} InitP \.{\land} ( s \.{=} top )}%
\@pvspace{8.0pt}%
 \@x{ BeginRdPS ( i ) \.{\defeq} MayPostStutter ( BeginRdP ( i )
 ,\,\@w{BeginRd} ,\, i ,\, 0 ,\,}%
\@x{\@s{153.64} {\IF} p \.{'} [ i ] \.{=} 1 \.{\THEN} 1 \.{\ELSE} 0 ,\,}%
\@x{\@s{153.64} {\LAMBDA} j \.{:} j \.{-} 1 )}%
 \@x{ {\ASSUME} StutterConstantCondition ( \{ 0 ,\, 1 \} ,\, 0 ,\, {\LAMBDA} j
 \.{:} j \.{-} 1 )}%
\@pvspace{8.0pt}%
\@x{ BeginWrPS ( i ,\, cmd ) \.{\defeq} NoStutter ( BeginWrP ( i ,\, cmd ) )}%
\@pvspace{8.0pt}%
 \@x{ DoWrPS ( i ) \.{\defeq} MayPostStutter ( DoWrP ( i ) ,\,\@w{DoWr} ,\, i
 ,\, \{ \} ,\,}%
\@x{\@s{144.19} \{ j \.{\in} Readers \.{:}}%
 \@x{\@s{153.29} ( rstate [ j ] \.{\neq} {\langle} {\rangle} ) \.{\land} ( p [
 j ] \.{=} Len ( rstate \.{'} [ j ] ) ) \} ,\,}%
 \@x{\@s{144.19} {\LAMBDA} S \.{:} S \.{\,\backslash\,} \{ {\CHOOSE} x \.{\in}
 S \.{:} {\TRUE} \} )}%
\@x{ {\ASSUME} StutterConstantCondition ( {\SUBSET} Readers ,\, \{ \} ,\,}%
 \@x{\@s{155.21} {\LAMBDA} S \.{:} S \.{\,\backslash\,} \{ {\CHOOSE} x \.{\in}
 S \.{:} {\TRUE} \} )}%
\@pvspace{8.0pt}%
\@x{ IEndRdPS ( i ,\, j ) \.{\defeq} NoStutter ( IEndRdP ( i ,\, j ) )}%
\@pvspace{8.0pt}%
\@x{ EndWrPS ( i ) \.{\defeq} NoStutter ( EndWrP ( i ) )}%
\@pvspace{8.0pt}%
 \@x{ NextPS \.{\defeq} \.{\lor} \E\, i \.{\in} Readers \.{:} \.{\lor}
 BeginRdPS ( i )}%
 \@x{\@s{131.39} \.{\lor} \E\, j \.{\in} 1 \.{\dotdot} Len ( rstate [ i ] )
 \.{:} IEndRdPS ( i ,\, j )}%
 \@x{\@s{52.48} \.{\lor} \E\, i \.{\in} Writers\@s{0.49} \.{:} \.{\lor} \E\,
 cmd \.{\in} RegVals \.{:} BeginWrPS ( i ,\, cmd )}%
\@x{\@s{131.39} \.{\lor} DoWrPS ( i ) \.{\lor} EndWrPS ( i )}%
\@pvspace{8.0pt}%
 \@x{ SafeSpecPS \.{\defeq} InitPS\@s{3.04} \.{\land} {\Box} [ NextPS ]_{
 varsPS}}%
\@x{ SpecPS \.{\defeq} SafeSpecPS \.{\land} Fairness}%
\end{tlatex}
\caption{Module \emph{NewLinearSnapshotPS}, part 3.}
\label{fig:NewLinearSnapshotPS3}
\end{figure}

\subsubsection{The Refinement Mapping}

Let's again use the abbreviations $Spec_{L}$ for formula $Spec$ of
$LinearSnapshot$ and $Spec_{PS}$ for formula $SpecPS$ of module
$NewLinearSnapshot$.  We now define the state functions \ov{mem} and
\ov{istate} such that $Spec_{PS}$ implements $Spec_{L}$ under the
refinement mapping 
 $mem<-\ov{mem}$, $istate <- \ov{istate}$.  
We want writer actions of $Spec_{L}$ to be simulated
by the corresponding writer actions of $Spec_{PS}$.  Hence, we
let \ov{mem} equal $mem$ and we let \ov{istate[i]} equal $wstate[i]$
for every writer~$i$.  The problem is defining \ov{istate[i]} 
for readers~$i$.

In $Spec_{L}$, for any process $i$ not executing a read or write,
$istate[i]$ equals $interface[i]$.  Hence, we can define
\ov{istate[i]} to equal $interface[i]$ for any reader $i$ not
currently reading.  We now consider the case when $i$ is currently
reading, which is true iff $rstate[i]#<<\,>>$, which implies $p[i]$
is a positive integer.  There are two possibilities:
%
%
\begin{describe}{$p[i]=1$}
\item[{$p[i]=1$}] In this case, the $DoRd(i)$ step of $Spec_{L}$ is 
simulated by the stuttering step added to $BeginRd(i)$.  The $DoRd(i)$
step changes $istate[i]$ from $NotMemVal$ to the memory value to be
output, so \ov{istate[i]} should equal $rstate[i][1]$ when
$rstate[i]#<<\,>>$, except after the $BeginRd(i)$ step and before the
stuttering step added immediately after it.

\item[{$p[i]>1$}] In this case, the $DoRd(i)$ step of $Spec_{L}$ is
simulated by one of the stuttering steps added to the $DoWr(j)$
step for the writer that appends the $p[i]$\tth\ element to
$rstate[i]$.  We let it be the stuttering step that removes $i$ from
$s.val$, so \ov{istate[i]} equals $NotMemVal$ until $p[i]\leq
Len(rstate[i])$ and it's not the case that $i$ is an element of
$s.val$ while some writer is performing a stuttering step added after
its $DoWr$ step.
%
\end{describe}
The definition of \ov{istate}, under the name $istateBar$,
appears near the end of the module, shown in
\lref{targ:NewLinearSnapshotPS4}{Figure~\ref{fig:NewLinearSnapshotPS4}}.
\begin{figure} \target{targ:NewLinearSnapshotPS4}
\begin{notla}
istateBar == [i \in Readers \cup Writers |->
                IF i \in Writers 
                  THEN wstate[i]
                  ELSE IF rstate[i] = << >> 
                         THEN interface[i]
                         ELSE IF p[i] = 1 
                                THEN IF /\ s # top 
                                        /\ s.id  = "BeginRd"
                                        /\ s.ctxt = i
                                       THEN NotMemVal
                                       ELSE rstate[i][1]
                                ELSE IF \/ p[i] > Len(rstate[i])
                                        \/ /\ s # top
                                           /\ s.id = "DoWr"
                                           /\ i \in s.val
                                       THEN NotMemVal
                                       ELSE rstate[i][p[i]] ]    

LS == INSTANCE LinearSnapshot WITH istate <- istateBar   

THEOREM SafeSpecPS => LS!SafeSpec

THEOREM SpecPS => LS!Spec                              
=============================================================================
\end{notla}
\begin{tlatex}
\@x{ istateBar \.{\defeq} [ i \.{\in} Readers \.{\cup} Writers \.{\mapsto}}%
\@x{\@s{66.70} {\IF} i \.{\in} Writers}%
\@x{\@s{74.90} \.{\THEN} wstate [ i ]}%
\@x{\@s{74.90} \.{\ELSE} {\IF} rstate [ i ] \.{=} {\langle} {\rangle}}%
\@x{\@s{114.41} \.{\THEN} interface [ i ]}%
\@x{\@s{114.41} \.{\ELSE} {\IF} p [ i ] \.{=} 1}%
\@x{\@s{153.92} \.{\THEN} {\IF} \.{\land} s \.{\neq} top}%
\@x{\@s{197.39} \.{\land} s . id\@s{4.1} \.{=}\@w{BeginRd}}%
\@x{\@s{197.39} \.{\land} s . ctxt \.{=} i}%
\@x{\@s{193.43} \.{\THEN} NotMemVal}%
\@x{\@s{193.43} \.{\ELSE} rstate [ i ] [ 1 ]}%
 \@x{\@s{153.92} \.{\ELSE} {\IF} \.{\lor} p [ i ]\@s{0.63} \.{>} Len ( rstate
 [ i ] )}%
\@x{\@s{197.39} \.{\lor} \.{\land} s \.{\neq} top}%
\@x{\@s{208.50} \.{\land} s . id \.{=}\@w{DoWr}}%
\@x{\@s{208.50} \.{\land} i \.{\in} s . val}%
\@x{\@s{193.43} \.{\THEN} NotMemVal}%
\@x{\@s{193.43} \.{\ELSE} rstate [ i ] [ p [ i ] ] ]}%
\@pvspace{8.0pt}%
 \@x{ LS \.{\defeq} {\INSTANCE} LinearSnapshot {\WITH} istate \.{\leftarrow}
 istateBar}%
\@pvspace{8.0pt}%
\@x{ {\THEOREM} SafeSpecPS \.{\implies} LS {\bang} SafeSpec}%
\@pvspace{4.0pt}%
\@x{ {\THEOREM} SpecPS \.{\implies} LS {\bang} Spec}%
\@x{}\bottombar\@xx{}%
\end{tlatex}
\caption{Module \emph{NewLinearSnapshotPS}, part 4.}
\label{fig:NewLinearSnapshotPS4}
\end{figure}
The theorems at the end of the module can be checked with TLC. In
fact, TLC checks that $SpecPS$ satisfies property $LS!Spec$ by
checking that the safety part of $SpecPS$ satisfies both (a)~the
safety part of $LS!Spec$ and (b)~the property that the liveness part
of $SpecPS$ implies the liveness part of $LS!Spec$.  Therefore, having
TLC check that $SpecPS$ satisfies $LS!Spec$ checks both theorems.

\subsection{\emph{AfekSimplified} Implements \emph{NewLinearSnapshot}}
\label{sec:Afek-implements}

We now finish checking the correctness of the algorithm $Spec_{A}$ of
$AfekSimplified$ by showing that it implements 
 \tlabox{\EE mem, rstate, wstate : SSpec_{NL}},
where $SSpec_{NL}$ is the safety specification of $NewLinearSnapshot$.
As we suggested in Section~\ref{sec:another-snapshot}, finding a
refinement mapping to show this requires adding a history variable to
$Spec_{A}$ that captures the information remembered by $SSpec_{NL}$ in
the variable $rstate$.  This is straightforward.  We just add a
history variable $h$ such that $h[i]$ is changed by $BeginRd(i)$,
$DoWr(i)$, and an ending $TryEndRdH(i)$ action (one executed with
$rdVal1[i] = rdVal2[i]$) the same way $rstate[i]$ is changed by the
corresponding $BeginRd(i)$, $DoWr(i)$, and $EndRd(i)$ action of
$SSpec_{NL}$.

The specification $SpecH$, obtained by adding the
history variable $h$ to specification $Spec$ of module $AfekSimplified$,
is defined in module $AfekSimplified$ as shown in
\lref{targ:AfekSimplifiedH-beg}{Figure~\ref{fig:AfekSimplifiedH-beg}}.
\begin{figure} \target{targ:AfekSimplifiedH-beg}
\begin{notla}
-------------------------- MODULE AfekSimplifiedH --------------------------
EXTENDS AfekSimplified, Sequences

VARIABLE h
varsH == <<vars, h>>

InitH == Init /\ (h = [i \in Readers |-> << >>])

memBar == [i \in Writers |-> imem[i][1]]

BeginWrH(i, cmd) == BeginWr(i, cmd) /\ (h' = h)

DoWrH(i) == /\ DoWr(i) 
            /\ h' = [j \in Readers |-> 
                            IF h[j] = << >>
                              THEN << >>
                              ELSE Append(h[j], memBar')]

EndWrH(i) == EndWr(i) /\ (h' = h)

BeginRdH(i) == /\ BeginRd(i) 
               /\ h' = [h EXCEPT ![i] = <<memBar>>]

Rd1H(i) == Rd1(i) /\ (h' = h)

Rd2H(i) == Rd2(i) /\ (h' = h)

TryEndRdH(i) == /\ TryEndRd(i) 
                /\ h' = IF rdVal1[i] = rdVal2[i]
                          THEN [h EXCEPT ![i] = << >>]
                          ELSE h

NextH == 
  \/ \E i \in Readers : BeginRdH(i) \/ Rd1H(i) \/ Rd2H(i) \/ TryEndRdH(i)
  \/ \E i \in Writers : \/ \E cmd \in RegVals : BeginWrH(i, cmd)
                        \/ DoWrH(i) \/ EndWrH(i)  
                        
SpecH == InitH /\ [][NextH]_varsH                        
\end{notla}
\begin{tlatex}
\@x{}\moduleLeftDash\@xx{ {\MODULE} AfekSimplifiedH}\moduleRightDash\@xx{}%
\@x{ {\EXTENDS} AfekSimplified ,\, Sequences}%
\@pvspace{8.0pt}%
\@x{ {\VARIABLE} h}%
\@x{ varsH \.{\defeq} {\langle} vars ,\, h {\rangle}}%
\@pvspace{8.0pt}%
 \@x{ InitH\@s{2.14} \.{\defeq} Init \.{\land} ( h \.{=} [ i \.{\in} Readers
 \.{\mapsto} {\langle} {\rangle} ] )}%
\@pvspace{8.0pt}%
\@x{ memBar \.{\defeq} [ i \.{\in} Writers \.{\mapsto} imem [ i ] [ 1 ] ]}%
\@pvspace{8.0pt}%
 \@x{ BeginWrH ( i ,\, cmd ) \.{\defeq} BeginWr ( i ,\, cmd ) \.{\land} ( h
 \.{'} \.{=} h )}%
\@pvspace{8.0pt}%
\@x{ DoWrH ( i ) \.{\defeq} \.{\land} DoWr ( i )}%
\@x{\@s{66.69} \.{\land} h \.{'} \.{=} [ j \.{\in} Readers \.{\mapsto}}%
\@x{\@s{124.54} {\IF} h [ j ] \.{=} {\langle} {\rangle}}%
\@x{\@s{132.74} \.{\THEN} {\langle} {\rangle}}%
\@x{\@s{132.74} \.{\ELSE} Append ( h [ j ] ,\, memBar \.{'} ) ]}%
\@pvspace{8.0pt}%
\@x{ EndWrH ( i ) \.{\defeq} EndWr ( i ) \.{\land} ( h \.{'} \.{=} h )}%
\@pvspace{8.0pt}%
\@x{ BeginRdH ( i ) \.{\defeq} \.{\land} BeginRd ( i )}%
 \@x{\@s{76.13} \.{\land} h \.{'} \.{=} [ h {\EXCEPT} {\bang} [ i ] \.{=}
 {\langle} memBar {\rangle} ]}%
\@pvspace{8.0pt}%
\@x{ Rd1H ( i ) \.{\defeq} Rd1 ( i ) \.{\land} ( h \.{'} \.{=} h )}%
\@pvspace{8.0pt}%
\@x{ Rd2H ( i ) \.{\defeq} Rd2 ( i ) \.{\land} ( h \.{'} \.{=} h )}%
\@pvspace{8.0pt}%
\@x{ TryEndRdH ( i ) \.{\defeq} \.{\land} TryEndRd ( i )}%
\@x{\@s{84.69} \.{\land} h \.{'} \.{=} {\IF} rdVal1 [ i ] \.{=} rdVal2 [ i ]}%
 \@x{\@s{126.02} \.{\THEN} [ h {\EXCEPT} {\bang} [ i ] \.{=} {\langle}
 {\rangle} ]}%
\@x{\@s{126.02} \.{\ELSE} h}%
\@pvspace{8.0pt}%
\@x{ NextH \.{\defeq}}%
 \@x{\@s{8.2} \.{\lor} \E\, i \.{\in} Readers \.{:} BeginRdH ( i ) \.{\lor}
 Rd1H ( i ) \.{\lor} Rd2H ( i ) \.{\lor} TryEndRdH ( i )}%
 \@x{\@s{8.2} \.{\lor} \E\, i \.{\in} Writers\@s{0.49} \.{:} \.{\lor} \E\, cmd
 \.{\in} RegVals \.{:} BeginWrH ( i ,\, cmd )}%
\@x{\@s{87.10} \.{\lor} DoWrH ( i ) \.{\lor} EndWrH ( i )}%
\@pvspace{8.0pt}%
\@x{ SpecH \.{\defeq} InitH \.{\land} {\Box} [ NextH ]_{ varsH}}%
\end{tlatex}

\caption{Beginning of module \emph{AfekSimplifiedH}.}
\label{fig:AfekSimplifiedH-beg}
\end{figure}
The definition should be easy to understand by comparing the
definition of the initial predicate $InitH$ and of the actions in the
module to the corresponding definitions in module $NewLinearSnapshot$.
Note that the action definitions use $memBar$ where the corresponding
action definitions in $NewLinearSnapshot$ use $mem$.  The module
defines $memBar$ to equal the memory value obtained from $imem$ in the
obvious way, by letting $memBar[i]$ equal the first element of
$imem[i]$.  The expression $memBar$ is, of course, the value substituted
for $mem$ by the refinement mapping.

The rest of the refinement mapping is defined at the end of the module,
\begin{figure} \target{targ:AfekSimplifiedH-end}
\begin{notla}
wstateBar == [i \in Writers |-> 
               IF (interface[i] = NotRegVal) \/ (wrNum[i] = imem[i][2])
                 THEN NotRegVal
                 ELSE interface[i]]
                 
NLS == INSTANCE NewLinearSnapshot
         WITH mem <- memBar, rstate <- h, wstate <- wstateBar
               
THEOREM SpecH => NLS!SafeSpec
=============================================================================
\end{notla}
\begin{tlatex}
\@x{ wstateBar \.{\defeq} [ i \.{\in} Writers \.{\mapsto}}%
 \@x{\@s{70.29} {\IF} ( interface [ i ] \.{=} NotRegVal ) \.{\lor} ( wrNum [ i
 ] \.{=} imem [ i ] [ 2 ] )}%
\@x{\@s{78.49} \.{\THEN} NotRegVal}%
\@x{\@s{78.49} \.{\ELSE} interface [ i ] ]}%
\@pvspace{8.0pt}%
\@x{ NLS \.{\defeq} {\INSTANCE} NewLinearSnapshot}%
 \@x{\@s{47.61} {\WITH} mem \.{\leftarrow} memBar ,\, rstate \.{\leftarrow} h
 ,\, wstate \.{\leftarrow} wstateBar}%
\@pvspace{8.0pt}%
\@x{ {\THEOREM} SpecH \.{\implies} NLS {\bang} SafeSpec}%
\@x{}\bottombar\@xx{}%
\end{tlatex}
\caption{End of module \emph{AfekSimplifiedH}.}
\label{fig:AfekSimplifiedH-end}
\end{figure}
shown in
\lref{targ:AfekSimplifiedH-end}{Figure~\ref{fig:AfekSimplifiedH-end}}.
It substitutes $h$ for $rstate$.  The expression $wstateBar$ is
substituted for $wstate$.  To understand it, remember that in
$SSpec_{NL}$, the value of $wstate[i]$ for a writer $i$ is $NotRegVal$
except between a $BeginWr(i,cmd)$ action and a $DoWr(i)$ action, when
it equals $interface[i]$ (which equals $cmd$).  

The theorem was checked by TLC in about 10 hours on a circa 2012
laptop, using a model with: two readers, two writers, and two register
values; symmetry of readers and writers (register values aren't
symmetric because $IntRegVal$ equals one of them); a state constraint
limiting each writer to at most three writes; and two worker threads.

\clearpage

\addcontentsline{toc}{section}{References}
\bibliography{auxiliary}

\begin{thebibliography}{1}

\bibitem{abadi:undo}
Mart\'{\i}n Abadi.
\newblock The prophecy of undo.
\newblock In Alexander Egyed and Ina Schaefer, editors, {\em Fundamental
  Approaches to Software Engineering}, volume 9033 of {\em Lecture Notes in
  Computer Science}, pages 347--361, Berlin Heidelberg, 2015. Springer.

\bibitem{abadi:existence}
Mart\'{\i}n Abadi and Leslie Lamport.
\newblock The existence of refinement mappings.
\newblock {\em Theoretical Computer Science}, 82(2):253--284, May 1991.

\bibitem{afek:atomic-snap}
Yehuda Afek, Hagit Attiya, Danny Dolev, Eli Gafni, Michael Merritt, and Nir
  Shavit.
\newblock Atomic snapshots of shared memory.
\newblock {\em Journal of the ACM}, 40(4):873--890, September 1993.

\bibitem{herlihy:axioms}
M.P. Herlihy and J.M. Wing.
\newblock Axioms for concurrent objects.
\newblock In {\em Proceedings of the Fourteenth Annual ACM SIGPLAN-SIGACT
  Symposium on Principles of Programming Languages}, pages 13--26, Munich,
  January 1987. ACM.

\bibitem{hoare:proof}
C.~A.~R. Hoare.
\newblock Proof of correctness of data representations.
\newblock {\em Acta Informatica}, 1:271--281, 1972.

\bibitem{lamport:auxiliary-variables-web}
Leslie Lamport.
\newblock Auxiliary variables in {TLA}+.
\newblock Web page at URL
  \url{http://research.microsoft.com/en-us/um/people/lamport/tla/auxiliary/auxiliary.html}.

\bibitem{lamport:tla-webpage}
Leslie Lamport.
\newblock {TLA}---temporal logic of actions.
\newblock A web page, a link to which can be found at {URL}
  \url{http://lamport.org}. The page can also be found by searching the Web for
  the 21-letter string formed by concatenating {\tt uid} and {\tt
  lamporttlahomepage}.

\bibitem{lamport:tla-book}
Leslie Lamport.
\newblock {\em Specifying Systems}.
\newblock Addison-Wesley, Boston, 2003.
\newblock A link to an electronic copy can be found at
  \url{http://lamport.org}.

\bibitem{owicki:verifying}
Susan Owicki and David Gries.
\newblock Verifying properties of parallel programs: An axiomatic approach.
\newblock {\em Communications of the ACM}, 19(5):279--284, May 1976.

\end{thebibliography}
\bibliographystyle{plain}

\end{document}